\documentclass[manuscript,numberedappendix,appendixfloats]{emulateapj}

\newcommand{\as}{$^{\prime\prime}$}
\newcommand{\ci}{[C{\sc{I}}]}
\newcommand{\cii}{[C{\sc{II}}]}
\newcommand{\htwo}{H$_2$}
\newcommand{\nii}{[N{\sc{II}}]}
\newcommand{\ls}{L$_{\odot}$}
\newcommand{\ms}{M$_{\odot}$}
\newcommand{\kms}{$\,\rm km\,s^{-1}$}

\newcommand{\unit}[1]{\ensuremath{\, \mathrm{#1}}}
\newcommand{\eqq}{\!=\!}  
\newcommand{\too}{\!\rightarrow\!} 
\newcommand{\jone}{{$\rm J\eqq 1\too0$}}
\newcommand{\jtwo}{{$ \rm J\eqq2\too1$}}
\newcommand{\jthree}{{$\rm J\eqq3\too2$}}
\newcommand{\jfour}{{$\rm J\eqq4\too3$}}
\newcommand{\jfive}{{$\rm J\eqq5\too4$}}
\newcommand{\jsix}{{$\rm J\eqq6\too5$}}
\newcommand{\jseven}{{$\rm J\eqq7\too6$}}
\newcommand{\jeight}{{$\rm J\eqq8\too7$}}
\newcommand{\jnine}{{$\rm J\eqq9\too8$}}
\newcommand{\jten}{{$\rm J\eqq10\too9$}}
\newcommand{\jeleven}{{$\rm J\eqq11\too10$}}
\newcommand{\jtwelve}{{$\rm J\eqq12\too11$}}
\newcommand{\jthirteen}{{$\rm J\eqq13\too12$}}
\newcommand{\jfourteen}{{$\rm J\eqq14\too13$}}
\newcommand{\jtwentyfour}{{$\rm J\eqq24\too23$}}
\newcommand{\jthirty}{{$\rm J\eqq30\too29$}}
\newcommand{\lfir}{$L_{\rm FIR}$}
\newcommand{\mhtwo}{$M_{\rm H_2}$}

\slugcomment{}

\shorttitle{Survey of ISM Properties}
\shortauthors{Kamenetzky et al.}

\usepackage{color}
\usepackage{natbib}
\usepackage{bm}
\usepackage{hyperref}
\usepackage{listings}
\begin{document}

\title{A Survey of the Molecular ISM Properties of Nearby Galaxies using the Herschel FTS}

\author{J. Kamenetzky\altaffilmark{1}, N. Rangwala\altaffilmark{2}, J. Glenn, P. R. Maloney, A. Conley}
\affil{University of Colorado at Boulder \\
       Center for Astrophysics and Space Astronomy \\
       389-UCB, Boulder, CO, USA}
\altaffiltext{1}{Steward Observatory, University of Arizona, 933 North Cherry Avenue, Tucson, AZ 85721}
\altaffiltext{2}{Visiting Scientist, Space Science and Astrobiology Division, NASA Ames Research Center}
\email{jkamenetzky@as.arizona.edu}

\begin{abstract}
The $^{12}$CO \jfour\ to \jthirteen\ lines of the interstellar medium from nearby galaxies, 
newly observable with the {\it Herschel} SPIRE Fourier Transform Spectrometer (FTS), 
offer an opportunity to study warmer, more luminous molecular gas than that traced by $^{12}$CO \jone.
Here we present a survey of 17 nearby infrared-luminous galaxy systems (21 pointings).
In addition to photometric modeling of dust, 
we modeled full $^{12}$CO spectral line energy distributions from \jone\ to \jthirteen\ with 
two components of warm and cool CO gas, and included LTE analysis of \ci, \cii, \nii, and \htwo\ lines.
CO is emitted from a low-pressure/high-mass component traced by the low-J lines and a 
high-pressure/low-mass component which dominates the luminosity.  
We found that, on average, the ratios of the warm/cool pressure, mass, and $^{12}$CO luminosity 
are $60 \pm 30$, $0.11 \pm 0.02$, and $15.6 \pm 2.7$.  The gas-to-dust-mass ratios are $< 120$ throughout the sample.  
The $^{12}$CO luminosity is dominated by the high-J lines and is $4 \times 10^{-4}$ \lfir\ on average.
We discuss systematic effects of single-component and multi-component CO modeling
(e.g., single-component J $\le$ 3 models overestimate gas pressure by $\sim$ 0.5 dex), as 
well as compare to Galactic star-forming regions.
With this comparison, we show the molecular interstellar medium of starburst galaxies is not simply an ensemble of Galactic-type GMCs.
The warm gas emission is likely dominated by regions resembling the  
warm extended cloud of Sgr B2.
\end{abstract}


\section{Introduction}\label{sec:intro}

Cool molecular gas in the interstellar medium (ISM) is the raw material out of which stars will form.  
The carbon monoxide molecule ($^{12}$CO, henceforth CO) is known to be an excellent tracer of the total molecular hydrogen in the ISM, 
especially at the ground-state rotational transition of \jone.
For many prominent nearby galaxies, such as M82, Arp~220, and NGC~1068, the emission from the higher-J lines of CO has proven to be far more luminous than would 
have been predicted from ground-based observations restricted to low-J lines \citep[e.g.,][]{Panuzzo2010,Kamenetzky2012,Rangwala2011,Spinoglio2012,Rigopoulou2013,
Pereira-Santaella2013}.  
It is already well established that the ISM is comprised of a 
multitude of constituents, both in composition (ionized, atomic, molecular) and physical conditions (temperatures, densities).  High-J lines 
of CO offer a new opportunity to study the relatively warm (compared to low-J CO), yet still molecular, ISM. 
This warmer CO is notable because of its much larger luminosity, representing a much greater role in the total energy budget 
of the molecular gas.  Therefore, to study the ongoing questions regarding the feedback interactions between 
different energy sources (cosmic rays, ultraviolet light from stars, X-rays from AGN, or mechanical heating from turbulence, winds, shocks, 
etc.), one must specifically examine the warmer CO via high-J transitions.

In general, due to atmospheric water 
absorption, only the lowest-J lines of CO can be observed from the ground.  However, the launch 
of the {\it Herschel} Space Observatory \citep{Pilbratt2010} 
offered a unique opportunity to observe at higher frequencies.\footnote{Herschel is an ESA space observatory with science instruments provided by 
European-led Principal Investigator consortia and with important participation from NASA.}
  The Spectral and Photometric Imaging REceiver (SPIRE) instrument \citep{Griffin2010} 
consisted of a three-band imaging photometer and an imaging Fourier transform spectrometer (FTS).
The FTS simultaneously observed spectra from 447-1550 GHz, which for nearby galaxies, encompasses the $^{12}$CO \jfour\ to \jthirteen\ 
transitions, among other molecular and atomic fine structure lines.

{\it Herschel's} mission has come to an end because of its finite supply of cryogens, but it has left behind an impressive legacy  
of observations.  Approximately 300 galaxies have been observed in point-source mode with the FTS, 
with spectra of varying quality.  In this 
paper, we establish a uniform pipeline for analysis of FTS spectra of galaxies, from the raw observations to the determination of the physical 
parameters of the cool and warm emitting CO gas.  We present initial results for 21 pointings of 17 unique 
galaxy systems, most of which have been well studied in the past.  The motivation for this survey is to understand 
the physical parameters (e.g., pressure and mass) which describe warm CO emission and to determine how the parameters 
vary with galaxy type, total infrared luminosity, etc.  
We sought to answer many questions brought to light by this new data.
Is the highly luminous warm component of molecular gas found in early {\it Herschel} SPIRE FTS studies generically present in high star-formation rate galaxies, 
and how does its presence modify what we already knew about cool molecular gas?
Given the broad range of CO lines we now have with {\it Herschel}, can we reassess the CO luminosity to mass conversion factor, and dust-to-gas mass ratios?
Accounting for the warm gas component luminosities, what $L_{\rm CO}$/\lfir\ are observed, and what other properties vary with e.g., \lfir?
What are the relative column densities of C$^+$, C, and CO in nearby starburst galaxies; in other words, where is the carbon not locked up in grains? 
Finally, how do the molecular gas pressures of the galaxies in our sample compare to Galactic giant molecular clouds?
Studying the physical parameters of the gas 
provides necessary information to interpret the energy budget and physical processes acting on the gas.
Additionally, it informs the analysis of high-redshift galaxies with fewer observed lines. 
This paper represents a sub-sample of a planned archival study of the molecular gas and dust in as many galaxies 
as possible observed with the FTS.  

\begin{figure*} 
\includegraphics[width=\textwidth]{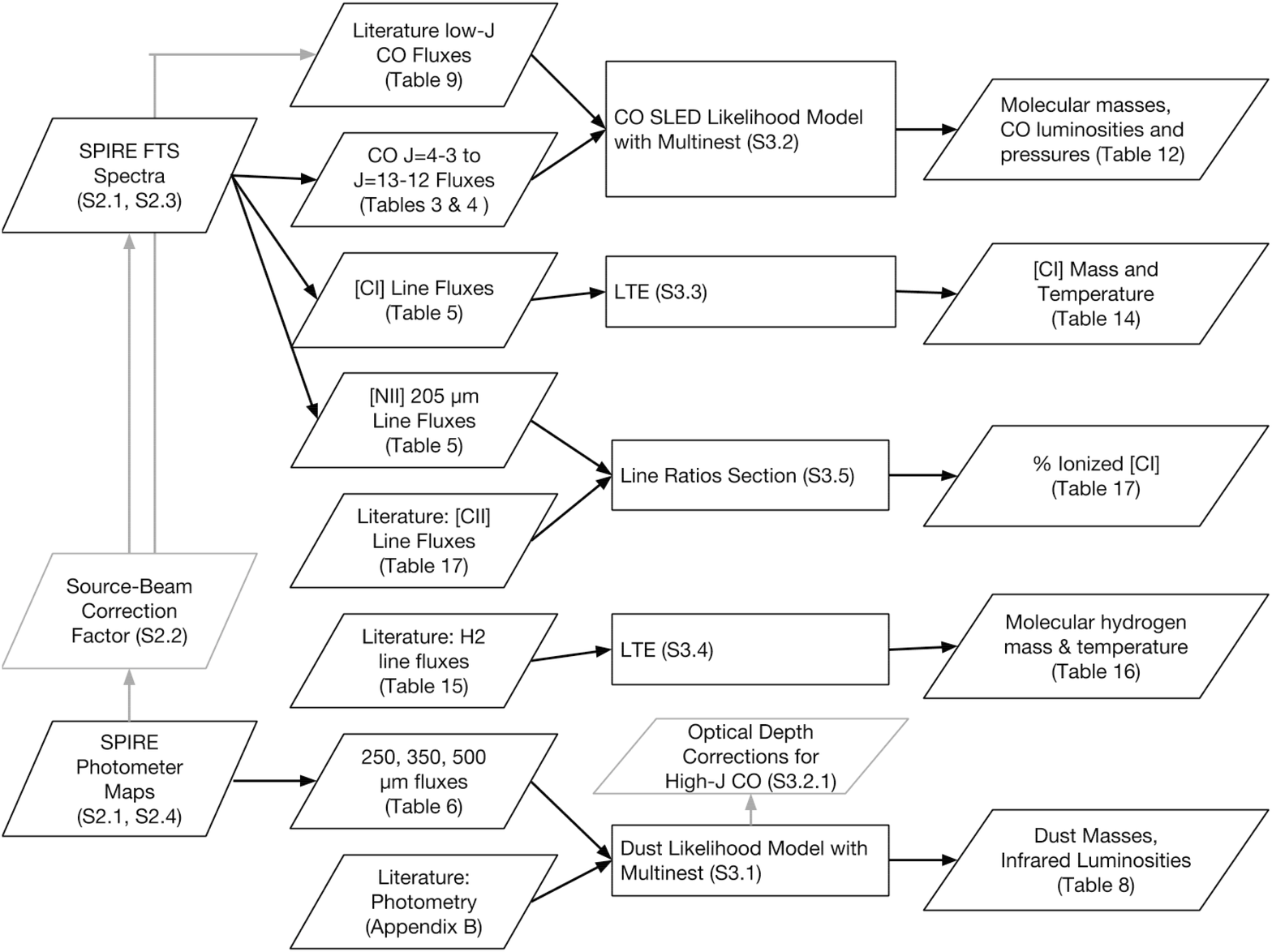} 
\caption[Flowchart of Data Analysis]{Flowchart of Data Analysis.
\label{fig:flowchart}}
\end{figure*}

Section \ref{sec:obs} describes how the {\it Herschel} observations were utilized, 
including sample selection, source-beam coupling correction and spectral line fitting.  
Section \ref{sec:analysis} details our modeling methodology: dust and CO likelihood analysis with Multinest \citep{Feroz2009},  
and LTE analysis of \htwo, \ci, \cii, and \nii.  
 A flowchart in Figure \ref{fig:flowchart} should help 
the reader understand the many types of data and modeling utilized in this work.  
Section \ref{sec:disc} summarizes our findings for this sample of galaxies, including 
a discussion of systematic effects of two-component likelihood modeling of physical properties, 
the calculation of the CO luminosity-to-mass conversion factor and gas-to-dust mass, and comparisons 
among our galaxies and Galactic star-forming regions.  Finally, we present 
conclusions and future plans in Section \ref{sec:conclusions}.

\begin{deluxetable*}{llrrccrrr}
\tabletypesize{\scriptsize}
\tablecaption{{\it Herschel}-SPIRE Observation Numbers for Galaxies in Sample\label{tbl:obs}}
\tablehead{
\colhead{FTS Name} & \colhead{Type} & \colhead{FTS RA} & \colhead{FTS Dec} & 
\colhead{FTS ObsID} & \colhead{Phot ObsID} & \colhead{$D_L$} & \colhead{$L_{FIR}$} & \colhead{Order}
}
\startdata
NGC 253              & SB & 0h47m33.12s & -25d17m17.6s & 1342210847 & 1342199424 &     3.4 & 10.52 &  19\\
NGC 1068             & AGN & 2h42m40.71s & -0d00m47.8s & 1342213445 & 1342189440 &    16.1 & 11.40 &  11\\
NGC 1222             & Early & 3h08m56.74s & -2d57m18.5s & 1342239354 & 1342239232 &    34.6 & 10.66 &  17\\
NGC 1266             & Early & 3h16m00.70s & -2d25m38.0s & 1342239338 & 1342189440 &    30.6 & 10.44 &  20\\
NGC 1365-SW          & AGN & 3h33m35.90s & -36d08m35.0s & 1342204021 & 1342201472 &    20.7 & 11.11 &  12\\
NGC 1365-NE          & AGN & 3h33m36.60s & -36d08m20.0s & 1342204020 & 1342201472 &    20.7 & 11.11 &  13\\
IRAS 09022-3615      & ULIRG & 9h04m12.72s & -36d27m01.3s & 1342231063 & 1342230784 &   261.7 & 12.21 &   3\\
UGC 05101            & AGN & 9h35m51.65s & +61d21m11.3s & 1342209278 & 1342204928 &   176.4 & 11.95 &   6\\
M82                  & SB & 9h55m52.22s & +69d40m46.9s & 1342208389 & 1342185600 &     3.7 & 10.79 &  16\\
Arp 299-B            & SB & 11h28m31.00s & +58d33m41.0s & 1342199249 & 1342199296 &    49.3 & 11.74 &   9\\
Arp 299-C            & SB & 11h28m31.00s & +58d33m50.0s & 1342199250 & 1342199296 &    49.3 & 11.74 &   8\\
Arp 299-A            & SB & 11h28m33.63s & +58d33m47.0s & 1342199248 & 1342199296 &    49.3 & 11.74 &  10\\
NGC 4038             & SB & 12h01m53.00s & -18d52m01.0s & 1342210860 & 1342188672 &    23.0 & 10.90 &  14\\
NGC 4038 (Overlap)   & SB & 12h01m54.90s & -18d52m46.0s & 1342210859 & 1342188672 &    23.0 & 10.90 &  15\\
Mrk 231              & AGN & 12h56m14.23s & +56d52m25.2s & 1342210493 & 1342201216 &   187.9 & 12.41 &   1\\
Cen A                & AGN & 13h25m27.61s & -43d01m08.8s & 1342204037 & 1342188672 &     3.6 &  9.93 &  21\\
M83                  & SB & 13h37m00.92s & -29d51m56.7s & 1342212345 & 1342188672 &     6.1 & 10.53 &  18\\
Mrk 273              & AGN & 13h44m42.11s & +55d53m12.7s & 1342209850 & 1342201216 &   167.6 & 12.13 &   5\\
Arp 220              & ULIRG & 15h34m57.12s & +23d30m11.5s & 1342190674 & 1342188672 &    81.4 & 12.14 &   4\\
NGC 6240             & SB & 16h52m58.89s & +2d24m03.4s & 1342214831 & 1342203648 &   108.0 & 11.83 &   7\\
IRAS F17207-0014     & ULIRG & 17h23m21.96s & -0d17m00.9s & 1342192829 & 1342203648 &   189.8 & 12.34 &   2
\enddata
\tablecomments{In this table only, sources are sorted by RA; in subsequent tables, sources are sorted by \lfir, in the order given in the rightmost column.  For example, Mrk~231 has the highest \lfir, so it is first in subsequent tables, IRAS~F17207-0014 is second, etc.  FTS/Phot ObsID are the observation IDs for the FTS spectrometer/photometer. $D_L$ is the luminosity distance in Mpc and \lfir\ is in log(\ls), both from HyperLeda\footnote{\url{http://leda.univ-lyon1.fr/}}.  
The category given in the Type column is not necessarily exclusive to other types (SB = Starburst, Early = early type).}
\end{deluxetable*}

\section{Observations}\label{sec:obs}

\subsection{Sample Selection}

This paper utilizes publicly available data from both the SPIRE spectrometer (FTS) and photometer, downloaded from the Herschel Science Archive.  
The 21 pointings in Table \ref{tbl:obs} represent 17 unique galaxy systems.  
This table lists both the observation IDs, coordinates and some basic 
properties of the sample galaxies; they are presented in order by RA in this table only, and will subsequently be presented in order of far-infrared luminosity, \lfir\ (the last column, ``Order," can help the reader find each galaxy in 
subsequent tables and figures).\footnote{Throughout this work, we use \lfir\ to refer to the integrated flux from 8-1000 $\mu$m; definitions vary in the literature.}
Some of the interacting galaxies with separated nuclei had 
FTS observations centered at each nucleus: there are two separate pointings within the Antennae 
(NGC~4038 and the Overlap region), two within NGC~1365, and three within Arp~299.
Appendix \ref{appendix:individual} includes additional information on how these and 
other extended galaxies were handled.

These specific galaxies were chosen 
because they are relatively nearby (10 within 50 Mpc, all within $\sim$ 260 Mpc) and their spectra showed measurable, bright CO 
emission.  Six are identified as AGN (\object{UGC~05101}, \object{NGC~1068}, \object{Cen A}, \object{NGC~1365}, \object{Mrk~273}, 
\object{Mrk~231}), three as ULIRGS (\object{Arp~220} 
[which may also have an AGN, e.g., \citet{Rangwala2011}], \object{IRAS~F17207-0014}, \object{IRAS~09022-3615}), two as early-type galaxies 
(\object{NGC~1266}, \object{NGC~1222}), and the remaining six as starbursts (\object{NGC~4038/4039}, \object{M82}, \object{M83}, 
\object{NGC~6240}, \object{Arp~299}, \object{NGC~253}).  The aforementioned categories are not mutually exclusive or necessarily complete. 

The SPIRE photometer bands are 250, 350, and 500 $\mu$m for the Photometer 
Short/Medium/Long Wave (PSW, PMW, PLW), respectively.  
The photometer maps were all used as downloaded, as processed with Standard Product Generation (SPG) 
v8.2.1 and calibration \texttt{spire\_cal\_8\_1}.
The Fourier transform spectrometer (FTS) contains 
two arrays of detectors: the lowest frequencies are captured by the Spectrometer Long Wave (SLW, 303-671 $\mu$m) and the higher 
frequencies by the Spectrometer Short Wave (SSW, 194-313 $\mu$m).
The FTS spectra were, in general, single(sparse)-pointed observations downloaded as 
SPG v6.1.0 and 
reprocessed with HIPE  v9
and \texttt{spire\_cal\_9\_1}.\footnote{HCSS, HSpot, and HIPE are joint developments by the 
Herschel Science Ground Segment Consortium, consisting of ESA, the NASA Herschel Science Center, and the HIFI, PACS and SPIRE consortia.}  
We used smooth off-axis background subtraction for UGC~05101 and Mrk~231 (see Appendix \ref{appendix:individual}) and 
daily dark background subtraction for NGC~1222, IRAS~09022-3615 (resulting in the notable reduction in the flux from the early calibration), and NGC~6240.
The exceptions to the HIPE v9 reprocessing were NGC~4038 and its Overlap region, which were extracted from a mapping 
observation \citep{Schirm2014}, as was M83 \citep{Wu2014} and provided via private communication - see Appendix \ref{appendix:individual} for more information on 
these extended galaxies.  IRAS~F17207-0014 and Mrk~273 were reprocessed using HIPE v10 because the detector temperatures were abnormally low for those particular 
observations and standard telescope background subtraction available in v9 was inadequate.

\begin{figure*}
\includegraphics[width=\textwidth]{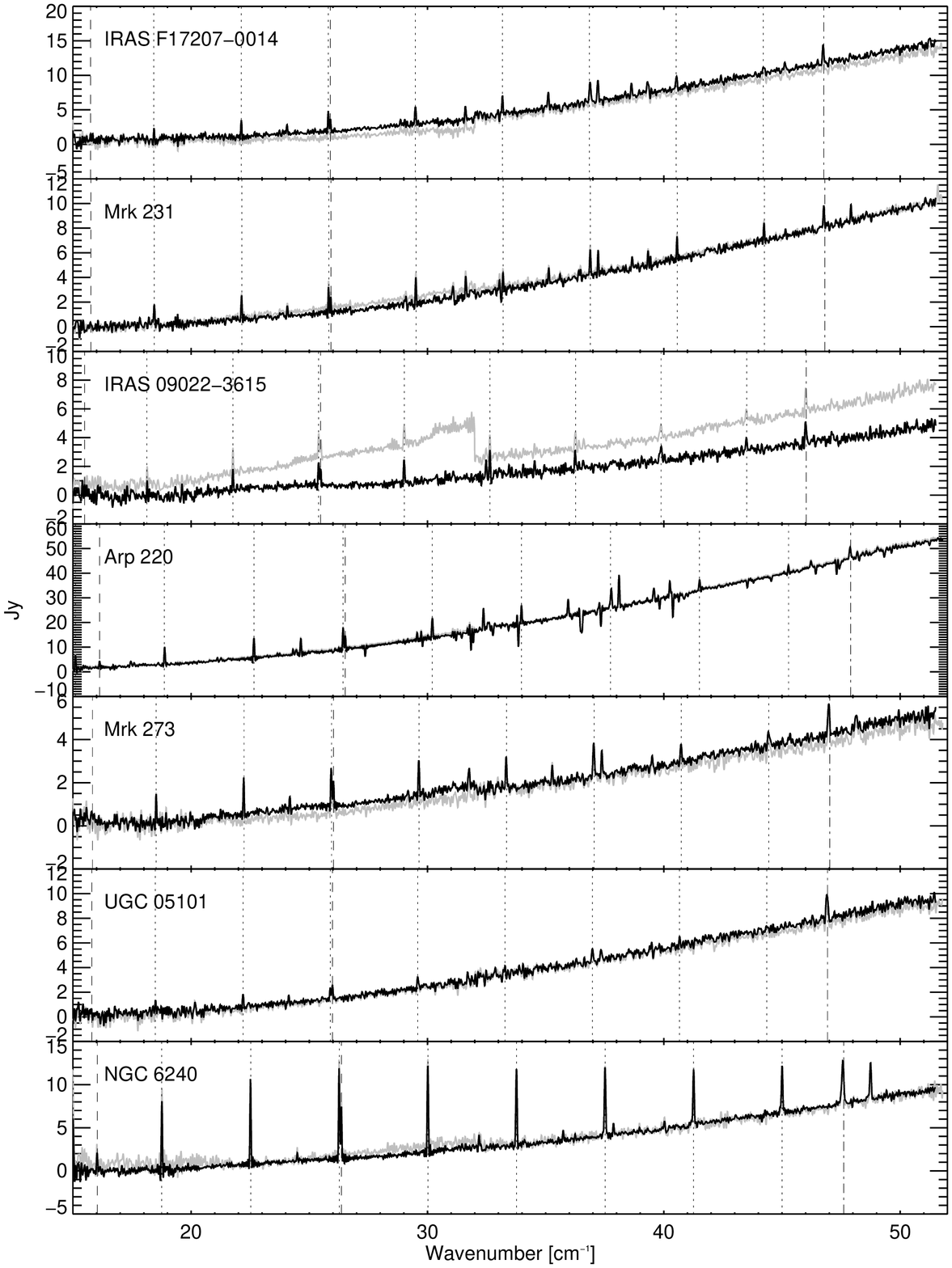} 
\caption[Source-Beam Coupling Corrected FTS Spectra 1 of 3]{Source-Beam Coupling Corrected FTS Spectra 1 of 3.  
The galaxies, starting with this figure, are in order of decreasing far-infrared luminosity.
The uncorrected spectra are shown in gray, the corrected spectra (as described in Section \ref{sec:sourcebeam}) are plotted over top in black.  For point-source-like 
spectra, e.g. UGC~05101, the correction may be unnoticeable.  The redshifted locations of the $^{12}$CO, \ci, and \nii\ lines are shown with 
dotted, dashed, and dash-dotted lines, respectively.\label{fig:spectra_a}}
\end{figure*}

\begin{figure*} 
\includegraphics[width=\textwidth]{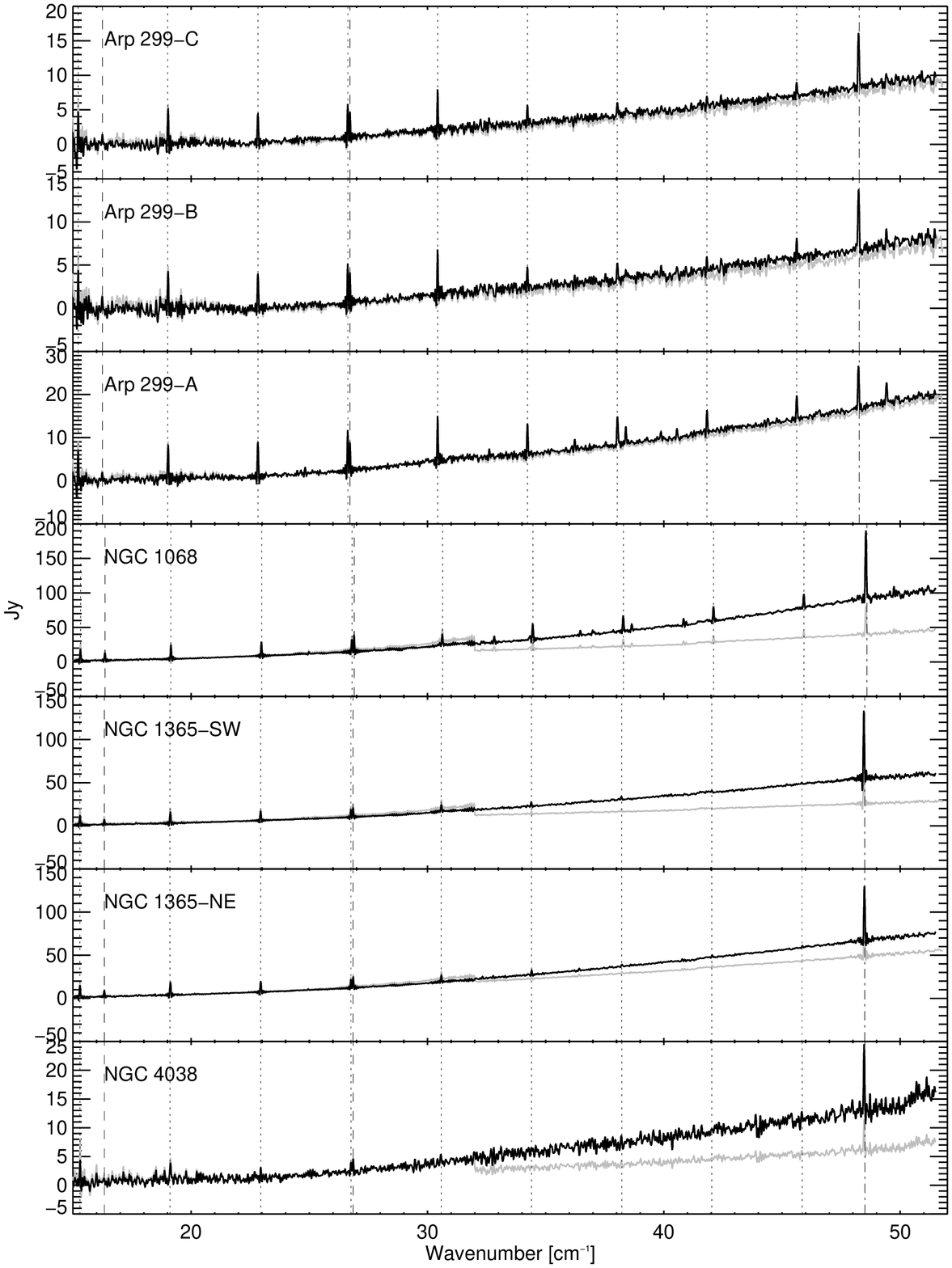}
\caption[Source-Beam Coupling Corrected FTS Spectra 2 of 3]{Source-Beam Coupling Corrected FTS Spectra 2 of 3.  See Figure \ref{fig:spectra_a} caption for  more information.\label{fig:spectra_b}}
\end{figure*}

\begin{figure*} 
\includegraphics[width=\textwidth]{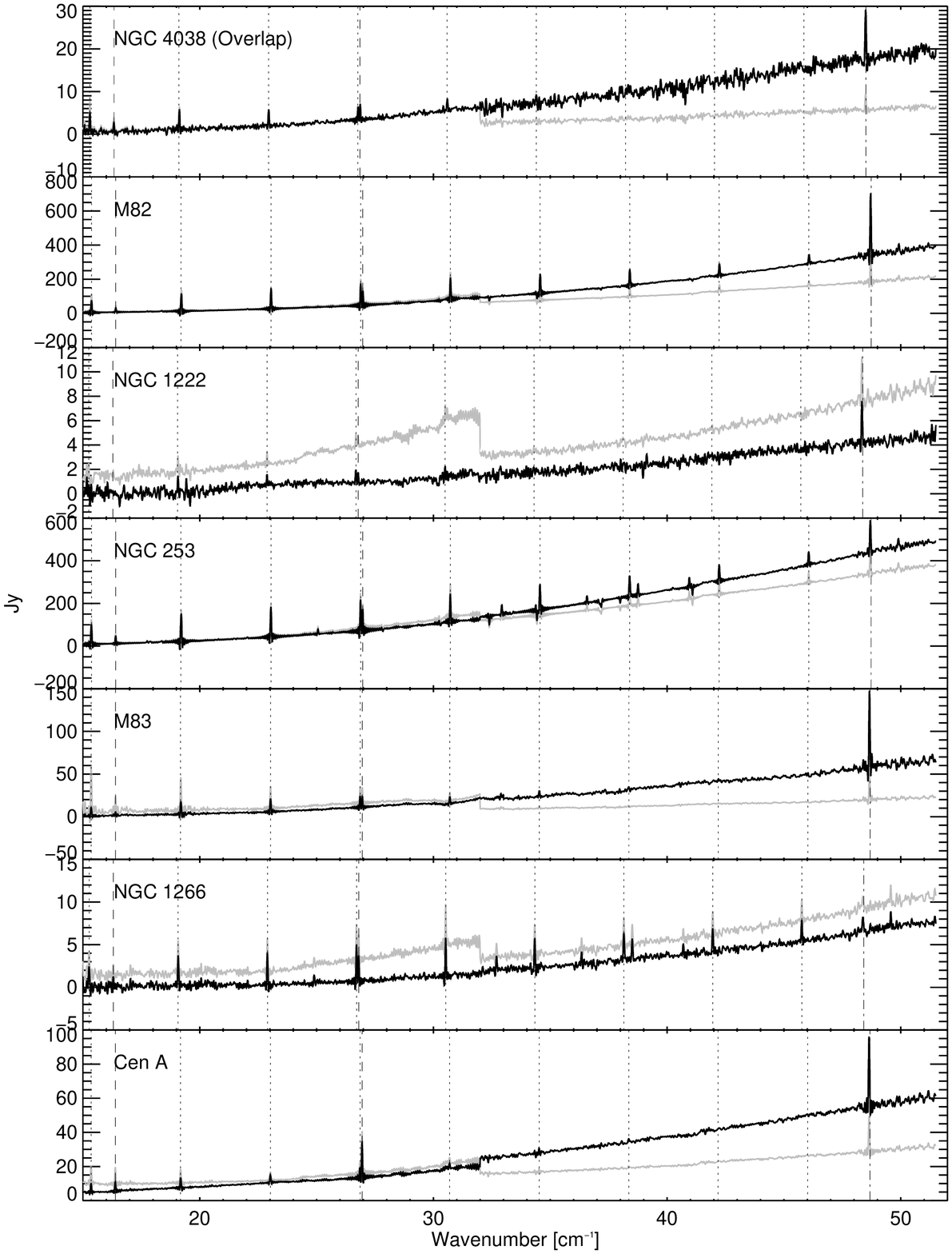}
\caption[Source-Beam Coupling Corrected FTS Spectra 3 of 3]{Source-Beam Coupling Corrected FTS Spectra 3 of 3.  See Figure \ref{fig:spectra_a} caption for more information.\label{fig:spectra_c}}
\end{figure*}

\subsection{Source-Beam Correction of Semi-Extended Sources}\label{sec:sourcebeam}

\begin{deluxetable}{lrrrrcc}
\tabletypesize{\scriptsize}
\tablecaption{Source/Beam Correction Parameters\label{tbl:corr}}
\tablehead{
\colhead{FTS Name} & \multicolumn{4}{c}{$\eta(\Omega,43.5) = a + b\Omega+c\Omega^2+d\Omega^3$} & \colhead{$\eta(15,43.5)$} \\
 \colhead{}         & \colhead{a} & \colhead{b} & \colhead{c} & \colhead{d} & \colhead{}\\
 \colhead{}         & \colhead{$\times 10^1$} & \colhead{$\times 10^2$} & \colhead{$\times 10^4$} & \colhead{$\times 10^6$} & \colhead{}
}
\startdata
Mrk 231              & 7.3\phantom{00} & 1.3\phantom{0} & -2.4\phantom{0} & 1.9 &  0.88\\
IRAS F17207-0014     & 8.1\phantom{00} & 0.92 & -1.8\phantom{0} & 1.5 &  0.91\\
IRAS 09022-3615      & 7.0\phantom{00} & 1.6\phantom{0} & -2.9\phantom{0} & 2.1 &  0.88\\
Arp 220              & 6.6\phantom{00} & 1.6\phantom{0} & -2.9\phantom{0} & 2.2 &  0.85\\
Mrk 273              & 7.8\phantom{00} & 1.1\phantom{0} & -2.0\phantom{0} & 1.7 &  0.90\\
UGC 05101            & 8.2\phantom{00} & 0.88 & -1.7\phantom{0} & 1.5 &  0.92\\
NGC 6240             & 4.3\phantom{00} & 2.5\phantom{0} & -4.0\phantom{0} & 2.7 &  0.73\\
Arp 299-C            & -3.7\phantom{00} & 4.7\phantom{0} & -4.6\phantom{0} & 2.1 &  0.25\\
Arp 299-B            & -4.0\phantom{00} & 4.7\phantom{0} & -4.1\phantom{0} & 1.7 &  0.21\\
Arp 299-A            & 0.45\phantom{0} & 3.8\phantom{0} & -4.9\phantom{0} & 2.9 &  0.51\\
NGC 1068             & -4.7\phantom{00} & 5.7\phantom{0} & -7.3\phantom{0} & 4.2 &  0.24\\
NGC 1365-SW          & -4.7\phantom{00} & 5.8\phantom{0} & -7.4\phantom{0} & 4.2 &  0.25\\
NGC 1365-NE          & -0.042 & 4.4\phantom{0} & -6.6\phantom{0} & 4.1 &  0.52\\
NGC 4038             & 1.8\phantom{00} & 0.54 & 3.7\phantom{0} & -1.4 &  0.34\\
NGC 4038 (Overlap)   & -2.7\phantom{00} & 4.3\phantom{0} & -4.6\phantom{0} & 3.4 &  0.28\\
M82                  & -2.9\phantom{00} & 5.2\phantom{0} & -6.9\phantom{0} & 4.1 &  0.35\\
NGC 1222             & 1.3\phantom{00} & 3.9\phantom{0} & -6.1\phantom{0} & 3.8 &  0.60\\
M83                  & -1.7\phantom{00} & 4.4\phantom{0} & -5.5\phantom{0} & 3.6 &  0.38\\
NGC 253              & 1.9\phantom{00} & 3.4\phantom{0} & -5.0\phantom{0} & 3.5 &  0.60\\
NGC 1266             & 5.2\phantom{00} & 2.3\phantom{0} & -3.8\phantom{0} & 2.7 &  0.78\\
Cen A                & -0.012 & 2.2\phantom{0} & -0.34 & 1.2 &  0.33
\enddata
\end{deluxetable}

The emission we measured in the FTS ($F^\prime$) was produced by the multiplication of the (generally non-Gaussian) 
source and beam.  The beam size, $\Omega(\nu)_{b}$, varied from $b$ (effective Gaussian FWHM) $\sim 45$\as\ to 17\as\ across both 
bands ($\Omega_b = 1.133 b^2$).  For a point source ($\Omega_{b} \gg \Omega_{s}$), the FTS measured the total 
flux in Jy at all wavelengths.  For a uniformly extended source ($\Omega_{b} \ll \Omega_{s}$), 
the flux density scaled with $\Omega_{b}$, i.e. 
$(F^\prime_{\Omega_1}/F^\prime_{\Omega_2}) = \Omega_1/\Omega_2$.  Many of the sources in this sample are semi-extended, 
meaning their source size is comparable to the beam size of the FTS.  In this case, 
$(F^\prime_{\Omega_1}/F^\prime_{\Omega_2}) = \eta_{1,2}$, where $\eta_{1,2}$ will be between $\Omega_1/\Omega_2$ 
(uniform extended) and 1 (point source).  We use a two-step procedure to correct the spectra for source-beam coupling 
effects: 1) derive $\eta_{1,2}$ from photometry maps and use it to scale all fluxes to a 43\farcs5 beam (that of the CO \jfour\ line), and 
2) scale the resulting spectrum to match the total photometer fluxes.  The final spectrum is as if it were observed by an instrument 
with a constant beam size.

Proper modeling of $\eta_{1,2}$ is necessary to compare flux densities measured at different frequencies of 
the FTS band as well as those measured from other telescopes.  For this we follow a similar procedure to that used 
in \citet{Panuzzo2010} using the Photometer Short Wave (PSW, 250 $\mu$m) maps.  
Because $\Omega_{PSW} = 423$ arcsec$^2$, the effective FWHM of the PSW beam is 19\farcs32.
To determine the light distribution at a given beam size 
$\Omega_{b}$ ($> \Omega_{PSW}$), we convolve the PSW map with a Gaussian kernel of size $\Omega_{kernel} = \Omega_{b} - \Omega_{PSW}$.  This procedure 
thus relies on two assumptions: that the distribution of the molecular gas follows the same spatial distribution 
as the 250 $\mu$m dust emission, and that the beams are roughly Gaussian.

We wish to obtain the spectrum measured within a 43\farcs5 beam, as this is the largest size we require, 
at the CO \jfour\ line.  We divide the flux convolved to $\Omega_{b}$ by that convolved to $\Omega_{43.5}$ to 
find $\eta(\Omega_b,\Omega_{43.5})$ for various beams from 20 to 43 arcseconds.  We fit the resultant curve as a 
third order polynomial, the parameters for which are given in Table \ref{tbl:corr}.  The last column shows 
$\eta_{15,43.5}$ as an example.  For M82, a telescope with a 15\as\ beam (pointed at the same location as the FTS) 
would only measure 35\% of the emission measured in a 43\farcs5 beam.  UGC~05101, in contrast, is extremely point like; a 
15\as\ beam would measure 91\% of the flux despite having only 12\% of the beam area.  Knowing the beam size at each 
frequency, we divide the entire spectrum by the appropriate $\eta(\nu)$ for those galaxies whose $\eta_{20,43.5}$ 
is less than 90\% (where we expect to be measuring real signal above a conservative 10\% calibration error).  
This step removes the discontinuity between SLW and SSW, which indicates that we have correctly stitched together the flux 
over the discontinuity in beam sizes.  

The second step is an additional calibration to match the absolute flux to that of the SPIRE photometer maps.  We convolve each 
map by a kernel $\Omega_{kernel} = \Omega_{43.5} - \Omega_{phot}$ to match 43\farcs5.  $\Omega_{phot}$ is 423, 751, 
and 1587 square arcseconds for PSW, PMW, and PLW, respectively.  We then measure the flux in Jy/Beam$_{43.5}$ at 
the point where the FTS beam is pointing, which is F$^{\prime\prime}$(PSW), for example.  We compare this flux to 
that from the FTS spectrum ($S(\nu)$), considering the (unweighted) photometer response function, $R(\nu)$, which is 
$\hat{F}(\hat{\lambda_j}) = \frac{\int_0^\infty R(\nu) S(\nu) d\nu}{\int_0^\infty R(\nu) d\nu}$.  For the SSW band, 
we multiply by the ratio $X_{SSW} = F''(PSW)/\hat{F}(PSW)$.  
There are two photometer bands (PMW and PLW) which overlap with the SLW band, so we define a line that connects those two ratios, 
thus dividing by $X_{SLW}(\nu)$.  Often, the first step ($\eta$) 
overestimates the total flux in the SSW especially, and the second step (absolute flux calibration) reduces the overestimation, and 
can result in lower spectra (e.g. NGC~1222, NGC~1266).

Therefore the final corrected spectrum for SLW/SSW is $F_c(\nu) = X_{\rm SLW/SSW} F'(\nu) / \eta(\nu)$.  The spectra are 
shown in Figures \ref{fig:spectra_a}, \ref{fig:spectra_b}, and \ref{fig:spectra_c}.  The empirical fits for 
$\eta(\Omega_b,\Omega_{43.5})$ from Table \ref{tbl:corr} were also used to correct fluxes from the literature 
for CO modeling (see Section \ref{sec:coground}).

\subsection{FTS Spectral Line Fitting}\label{sec:FTFit}
 
\begin{deluxetable*}{lrrrrrrrrrr}
\tabletypesize{\scriptsize}
\tablecaption{Integrated Fluxes in 10$^2$ Jy km s$^{-1}$: CO\label{tbl:flux1}}
\tablehead{
\colhead{FTS Name}  & \multicolumn{2}{c}{\jfour} & \multicolumn{2}{c}{\jfive} & \multicolumn{2}{c}{\jsix} & \multicolumn{2}{c}{\jseven} & \multicolumn{2}{c}{\jeight} \\
\colhead{} & \colhead{$I \Delta v$} & \colhead{$\sigma$} & \colhead{$I \Delta v$} & \colhead{$\sigma$} & \colhead{$I \Delta v$} & \colhead{$\sigma$} & \colhead{$I \Delta v$} & \colhead{$\sigma$} & \colhead{$I \Delta v$} & \colhead{$\sigma$} 
}
\startdata
Mrk 231              &\ldots & \ldots & 10.0\phantom{0} & 0.96 & 9.77 & 0.49 & 9.04 & 0.29 & 8.15 & 0.36 \\
IRAS F17207-0014     &\ldots & \ldots & 9.71 & 1.5\phantom{0} & 12.1\phantom{0} & 0.70 & 12.4\phantom{0} & 0.45 & 10.4\phantom{0} & 0.49 \\
IRAS 09022-3615      &\ldots & \ldots & 5.99 & 0.98 & 7.27 & 0.55 & 8.19 & 0.34 & 6.55 & 0.36 \\
Arp 220              &\ldots & \ldots & 41.0\phantom{0} & 4.7\phantom{0} & 43.1\phantom{0} & 1.8\phantom{0} & 38.0\phantom{0} & 1.7\phantom{0} & 33.7\phantom{0} & 1.8\phantom{0} \\
Mrk 273              &\ldots & \ldots & 7.77 & 0.81 & 8.53 & 0.38 & 8.33 & 0.21 & 6.17 & 0.30 \\
UGC 05101            &\ldots & \ldots & 6.31 & 0.91 & 5.84 & 0.43 & 4.33 & 0.36 & 3.66 & 0.35 \\
NGC 6240             &48.6\phantom{0} & 2.2 & 48.7\phantom{0} & 0.90 & 63.1\phantom{0} & 1.4\phantom{0} & 64.0\phantom{0} & 1.3\phantom{0} & 59.3\phantom{0} & 1.4\phantom{0} \\
Arp 299-C            &34.2\phantom{0} & 3.2 & 32.9\phantom{0} & 1.3\phantom{0} & 22.1\phantom{0} & 0.58 & 20.1\phantom{0} & 0.46 & 20.9\phantom{0} & 0.72 \\
Arp 299-B            &31.4\phantom{0} & 3.3 & 27.9\phantom{0} & 1.4\phantom{0} & 21.4\phantom{0} & 0.56 & 19.0\phantom{0} & 0.49 & 18.5\phantom{0} & 0.83 \\
Arp 299-A            &51.5\phantom{0} & 4.1 & 48.8\phantom{0} & 1.8\phantom{0} & 42.5\phantom{0} & 0.62 & 38.6\phantom{0} & 0.57 & 36.4\phantom{0} & 0.89 \\
NGC 1068             &129\phantom{.00} & 8.6 & 111\phantom{.00} & 5.2\phantom{0} & 99.9\phantom{0} & 2.3\phantom{0} & 69.3\phantom{0} & 1.3\phantom{0} & 59.7\phantom{0} & 2.0\phantom{0} \\
NGC 1365-SW          &86.3\phantom{0} & 3.9 & 77.1\phantom{0} & 2.2\phantom{0} & 58.6\phantom{0} & 0.94 & 34.4\phantom{0} & 0.69 & 32.3\phantom{0} & 1.1\phantom{0} \\
NGC 1365-NE          &104\phantom{.00} & 4.8 & 90.2\phantom{0} & 2.4\phantom{0} & 64.0\phantom{0} & 0.89 & 38.6\phantom{0} & 0.65 & 29.2\phantom{0} & 1.0\phantom{0} \\
NGC 4038             &28.3\phantom{0} & 3.5 & 19.0\phantom{0} & 1.8\phantom{0} & 8.85 & 1.0\phantom{0} & 6.51 & 0.81 & 5.47 & 0.74 \\
NGC 4038 (Overlap)   &35.5\phantom{0} & 1.8 & 28.5\phantom{0} & 1.4\phantom{0} & 19.5\phantom{0} & 0.62 & 12.3\phantom{0} & 0.46 & 10.1\phantom{0} & 0.68 \\
M82                  &583\phantom{.00} & 22\phantom{.0} & 625\phantom{.00} & 12\phantom{.00} & 607\phantom{.00} & 5.1\phantom{0} & 537\phantom{.00} & 4.3\phantom{0} & 482\phantom{.00} & 5.1\phantom{0} \\
NGC 1222             &6.89 & 2.5 & 6.79 & 1.2\phantom{0} & 4.41 & 0.47 & 4.15 & 0.29 & 3.43 & 0.50 \\
M83                  &74.9\phantom{0} & 4.6 & 101\phantom{.00} & 3.0\phantom{0} & 80.6\phantom{0} & 1.3\phantom{0} & 53.9\phantom{0} & 0.85 & 25.8\phantom{0} & 1.1\phantom{0} \\
NGC 253              &750.\phantom{00} & 34\phantom{.0} & 842\phantom{.00} & 19\phantom{.00} & 725\phantom{.00} & 7.3\phantom{0} & 605\phantom{.00} & 7.0\phantom{0} & 495\phantom{.00} & 7.6\phantom{0} \\
NGC 1266             &17.8\phantom{0} & 2.4 & 21.9\phantom{0} & 0.99 & 18.7\phantom{0} & 0.42 & 17.9\phantom{0} & 0.32 & 16.7\phantom{0} & 0.51 \\
Cen A                &39.4\phantom{0} & 3.0 & 37.0\phantom{0} & 2.1\phantom{0} & 25.2\phantom{0} & 0.93 & 17.4\phantom{0} & 0.54 & 13.5\phantom{0} & 1.0\phantom{0}
\enddata
\tablecomments{Some galaxies are missing measurements for \jfour\ because the line falls outside the band or very close to the edge. 
Errors are one sigma; lines with S/N $< 3$ are used as upper limits in CO analysis.
Lines were fit as unresolved, using instrumental line sinc profiles, unless otherwise specified in Section \ref{sec:FTFit}.}
\end{deluxetable*}
\begin{deluxetable*}{lrrrrrrrrrr}
\tabletypesize{\scriptsize}
\tablecaption{Integrated Fluxes in 10$^2$ Jy km s$^{-1}$: CO, Continued\label{tbl:flux2}}
\tablehead{
\colhead{FTS Name}  & \multicolumn{2}{c}{\jnine} & \multicolumn{2}{c}{\jten} & \multicolumn{2}{c}{\jeleven} & \multicolumn{2}{c}{\jtwelve} & \multicolumn{2}{c}{\jthirteen} \\
\colhead{} & \colhead{$I \Delta v$} & \colhead{$\sigma$} & \colhead{$I \Delta v$} & \colhead{$\sigma$} & \colhead{$I \Delta v$} & \colhead{$\sigma$} & \colhead{$I \Delta v$} & \colhead{$\sigma$} & \colhead{$I \Delta v$} & \colhead{$\sigma$} 
}
\startdata
Mrk 231              &5.00 & 0.48 & 6.53\phantom{0} & 0.35 & 5.14 & 0.33 & 3.80\phantom{0} & 0.27 & 3.31 & 0.27 \\
IRAS F17207-0014     &9.45 & 0.87 & 9.04\phantom{0} & 0.53 & 4.77 & 0.57 & 2.39\phantom{0} & 0.45 & 1.65 & 0.44 \\
IRAS 09022-3615      &6.27 & 0.50 & 4.11\phantom{0} & 0.30 & 2.80 & 0.28 & 2.45\phantom{0} & 0.27 & 1.42 & 0.24 \\
Arp 220              &25.6\phantom{0} & 2.6\phantom{0} & 24.4\phantom{00} & 2.1\phantom{0} & 12.6\phantom{0} & 2.2\phantom{0} & 7.23\phantom{0} & 1.8\phantom{0} & 5.47 & 1.7\phantom{0} \\
Mrk 273              &7.85 & 0.95 & 9.33\phantom{0} & 0.68 & 4.19 & 0.67 & 4.23\phantom{0} & 0.65 & 4.42 & 0.66 \\
UGC 05101            &2.51 & 0.45 & 3.08\phantom{0} & 0.31 & 2.13 & 0.34 & 0.734 & 0.27 & 1.10 & 0.25 \\
NGC 6240             &49.2\phantom{0} & 1.3\phantom{0} & 43.5\phantom{00} & 0.92 & 31.2\phantom{0} & 0.95 & 24.4\phantom{00} & 0.78 & 20.2\phantom{0} & 0.81 \\
Arp 299-C            &9.20 & 0.60 & 5.74\phantom{0} & 0.48 & 3.93 & 0.49 & 4.79\phantom{0} & 0.41 & 3.15 & 0.40 \\
Arp 299-B            &8.15 & 0.60 & 5.85\phantom{0} & 0.49 & 4.15 & 0.46 & 5.88\phantom{0} & 0.42 & 4.41 & 0.43 \\
Arp 299-A            &24.5\phantom{0} & 0.78 & 19.8\phantom{00} & 0.71 & 14.2\phantom{0} & 0.69 & 13.6\phantom{00} & 0.60 & 10.8\phantom{0} & 0.64 \\
NGC 1068             &75.2\phantom{0} & 2.5\phantom{0} & 69.1\phantom{00} & 2.2\phantom{0} & 56.8\phantom{0} & 2.1\phantom{0} & 50.8\phantom{00} & 2.0\phantom{0} & 27.7\phantom{0} & 2.1\phantom{0} \\
NGC 1365-SW          &15.3\phantom{0} & 1.6\phantom{0} & 10.2\phantom{00} & 1.5\phantom{0} & 4.83 & 1.5\phantom{0} & 3.94\phantom{0} & 1.3\phantom{0} & \ldots & \ldots \\
NGC 1365-NE          &16.6\phantom{0} & 1.6\phantom{0} & 6.58\phantom{0} & 1.5\phantom{0} & 7.13 & 1.3\phantom{0} & 4.33\phantom{0} & 1.0\phantom{0} & 5.52 & 1.4\phantom{0} \\
NGC 4038             &2.30 & 1.2\phantom{0} & 3.52\phantom{0} & 0.93 & 1.46 & 0.99 & 3.36\phantom{0} & 0.89 & 3.29 & 0.97 \\
NGC 4038 (Overlap)   &6.68 & 1.3\phantom{0} & 3.58\phantom{0} & 1.4\phantom{0} & \ldots & \ldots & \ldots & \ldots & \ldots & \ldots \\
M82                  &394\phantom{.00} & 6.4\phantom{0} & 289\phantom{.000} & 6.9\phantom{0} & 188\phantom{.00} & 7.6\phantom{0} & 136\phantom{.000} & 6.4\phantom{0} & 88.1\phantom{0} & 6.2\phantom{0} \\
NGC 1222             &1.56 & 0.40 & 0.861 & 0.38 & 1.02 & 0.37 & 0.277 & 0.29 & 1.05 & 0.33 \\
M83                  &21.8\phantom{0} & 2.7\phantom{0} & 10.0\phantom{00} & 2.7\phantom{0} & 11.4\phantom{0} & 2.8\phantom{0} & \ldots & \ldots & \ldots & \ldots \\
NGC 253              &416\phantom{.00} & 11\phantom{.00} & 291\phantom{.000} & 12\phantom{.00} & 216\phantom{.00} & 14\phantom{.00} & 155\phantom{.000} & 15\phantom{.00} & 91.3\phantom{0} & 17\phantom{.00} \\
NGC 1266             &11.9\phantom{0} & 0.48 & 9.62\phantom{0} & 0.42 & 7.47 & 0.41 & 6.32\phantom{0} & 0.36 & 4.35 & 0.38 \\
Cen A                &9.20 & 1.2\phantom{0} & 5.81\phantom{0} & 1.1\phantom{0} & 2.13 & 1.2\phantom{0} & 2.90\phantom{0} & 0.97 & 5.62 & 1.2\phantom{0}
\enddata
\tablecomments{See Table \ref{tbl:flux1} for more information.}
\end{deluxetable*}
\begin{deluxetable*}{lrrrrrr}
\tabletypesize{\scriptsize}
\tablecaption{Integrated Fluxes in 10$^2$ Jy km s$^{-1}$: [C {\sc{I}}] and [N {\sc{II}}] \label{tbl:flux3}}
\tablehead{
\colhead{FTS Name}  & \multicolumn{2}{c}{\ci\ 1-0} & \multicolumn{2}{c}{\ci\ 2-1} & \multicolumn{2}{c}{\nii} \\
\colhead{} & \colhead{$I \Delta v$} & \colhead{$\sigma$} & \colhead{$I \Delta v$} & \colhead{$\sigma$} & \colhead{$I \Delta v$} & \colhead{$\sigma$} 
}
\startdata
Mrk 231              &\ldots & \ldots & 5.00 & 0.30 & 6.42 & 0.83 \\
IRAS F17207-0014     &6.17 & 1.9\phantom{0} & 6.88 & 0.45 & 6.67 & 0.49 \\
IRAS 09022-3615      &7.46 & 1.8\phantom{0} & 5.65 & 0.30 & 7.26 & 0.74 \\
Arp 220              &18.9\phantom{0} & 5.8\phantom{0} & 19.5\phantom{0} & 1.6\phantom{0} & 23.6\phantom{0} & 5.1\phantom{0} \\
Mrk 273              &5.05 & 0.97 & 5.03 & 0.22 & 7.88 & 0.58 \\
UGC 05101            &\ldots & \ldots & 5.44 & 0.34 & 12.3\phantom{0} & 0.89 \\
NGC 6240             &16.7\phantom{0} & 1.4\phantom{0} & 34.9\phantom{0} & 1.3\phantom{0} & 33.2\phantom{0} & 0.86 \\
Arp 299-C            &10.6\phantom{0} & 2.0\phantom{0} & 14.3\phantom{0} & 0.41 & 30.7\phantom{0} & 1.2\phantom{0} \\
Arp 299-B            &10.6\phantom{0} & 2.0\phantom{0} & 13.5\phantom{0} & 0.38 & 28.6\phantom{0} & 1.2\phantom{0} \\
Arp 299-A            &13.4\phantom{0} & 2.5\phantom{0} & 23.9\phantom{0} & 0.54 & 36.2\phantom{0} & 2.0\phantom{0} \\
NGC 1068             &81.1\phantom{0} & 5.0\phantom{0} & 101\phantom{.00} & 1.4\phantom{0} & 358\phantom{.00} & 7.7\phantom{0} \\
NGC 1365-SW          &39.9\phantom{0} & 2.3\phantom{0} & 43.1\phantom{0} & 0.70 & 192\phantom{.00} & 1.4\phantom{0} \\
NGC 1365-NE          &47.2\phantom{0} & 3.0\phantom{0} & 48.4\phantom{0} & 0.68 & 156\phantom{.00} & 1.3\phantom{0} \\
NGC 4038             &5.44 & 1.8\phantom{0} & 9.02 & 0.81 & 27.6\phantom{0} & 0.96 \\
NGC 4038 (Overlap)   &15.8\phantom{0} & 1.3\phantom{0} & 14.9\phantom{0} & 0.51 & 27.5\phantom{0} & 1.8\phantom{0} \\
M82                  &175\phantom{.00} & 16\phantom{.00} & 315\phantom{.00} & 3.7\phantom{0} & 867\phantom{.00} & 6.2\phantom{0} \\
NGC 1222             &\ldots & \ldots & 2.96 & 0.30 & 8.34 & 0.30 \\
M83                  &27.7\phantom{0} & 2.7\phantom{0} & 51.9\phantom{0} & 1.2\phantom{0} & 219\phantom{.00} & 2.2\phantom{0} \\
NGC 253              &262\phantom{.00} & 21\phantom{.00} & 387\phantom{.00} & 5.8\phantom{0} & 357\phantom{.00} & 18\phantom{.00} \\
NGC 1266             &7.97 & 1.4\phantom{0} & 11.4\phantom{0} & 0.31 & 7.51 & 1.0\phantom{0} \\
Cen A                &39.4\phantom{0} & 2.1\phantom{0} & 93.8\phantom{0} & 0.57 & 102\phantom{.00} & 1.0\phantom{0}
\enddata
\tablecomments{See Table \ref{tbl:flux1} for more information.}
\end{deluxetable*}

To fit the CO, \ci, and \nii\ lines, we used the FTFitter code from the University of 
Lethbridge.\footnote{\url{https://www.uleth.ca/phy/naylor/index.php?page=ftfitter}}  
Some of our spectra contain many more lines \citep[for example, Arp~220, ][]{Rangwala2011}, 
but we do not fit them here in the interest of establishing a consistent pipeline for the 
brightest lines in all galaxies.  
For one detector (SLW or SSW) at a time, 
we first determine a polynomial fit to the baseline and then fit unresolved lines at the expected frequencies given 
the known redshifts.  
The code utilizes the instrumental line profile to determine the area underneath each line in Jy cm$^{-1}$.

Most of the lines in this sample are unresolved; however, the velocity resolution improves at the highest 
frequencies, and the highest-J CO lines and \nii\ line are sometimes resolved.  These lines do not show the 
characteristic ringing of the sinc function, which is the expected line profile for a Fourier 
transform spectrometer.
By visual inspection, we determined which lines were clearly 
resolved and refit them as a Gaussian convolved with the instrumental line profile.  Though the code cannot 
break the degeneracy between the Gaussian amplitude and width, the area is well constrained.  The \nii\ line, the 
highest frequency line in our spectrum, was resolved in the following 11 galaxies: UGC~05101, NGC~1068, Arp~220, Mrk~273, Mrk~231,
NGC~6240, Arp~299-A, -B, -C, NGC~1266, and IRAS~09022-3615.  Two of these galaxies showed resolved line structure for multiple lines.
All lines at and above CO \jnine\ were resolved for Mrk~273, and all lines at and above CO \jsix\ were resolved for NGC~6240.
Additionally, the CO \jseven\ and \ci\ \jtwo\ lines are very close to one another; for three of the spectra (UGC~05101, Cen A, Arp~220), we 
manually fitted variable-width sinc functions to the lines when the FTFitter code could not properly fit the two.

The integrated fluxes are shown in Tables \ref{tbl:flux1}, \ref{tbl:flux2}, \ref{tbl:flux3}.  
Our most luminous galaxies are distant enough that the CO \jfour\ line is either completely redshifted out 
of the band, or extremely close to the edge.  Fluxes are not reported in those cases.

\begin{figure*} 
\includegraphics[height=\textwidth,angle=270]{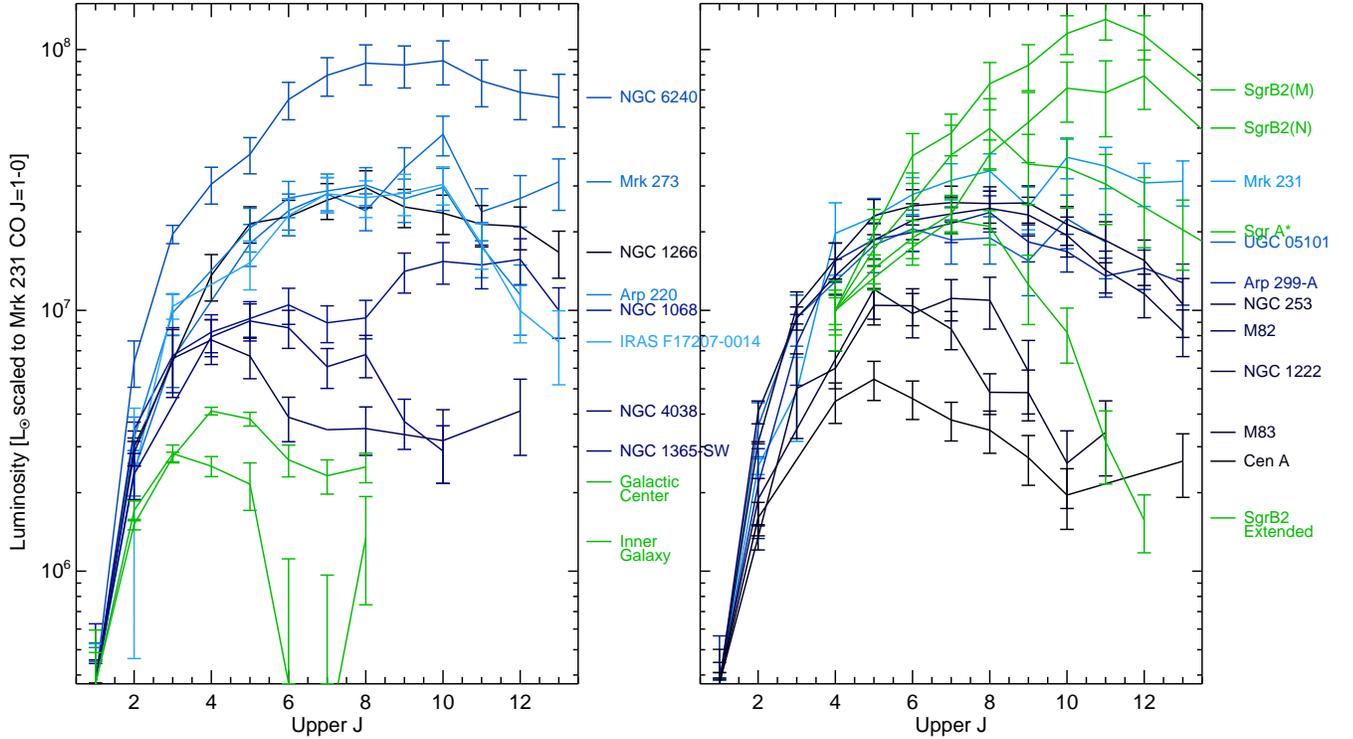} 
\caption[Normalized CO SLEDs]{Normalized CO SLEDS. 
All \jone\ luminosities are scaled to match that of Mrk~231 (3.7 $\times 10^5$ \ls).
SLEDs are colored to indicate increasing \lfir\ with increasing lightness.
Placement in the left or right panel is for clarity only.
On the left panel, the SLEDs of the Galactic center ($|l| <$ 2\fdg{5}) and the Inner Galaxy (2\fdg{5} $< |l| <$  32\fdg{5}), 
also normalized, are shown in green for comparison \citep{Fixsen1999}.
On the right panel, we show the SLEDs of two star-forming cores and the extended envelope of Sgr B2 \citep{Etxaluze2013} and 
that of Sgr A* \citep{Goicoechea2013}.  Because these SLEDs begin at \jfour, we scale \jfour\ to $10^7$ \ls\ for visual comparison 
of the shapes.  
None of the SLEDs are corrected for dust extinction.
\label{fig:slednorm}}
\end{figure*}

In Figure \ref{fig:slednorm}, we show the CO spectral line energy distributions (SLEDs), all scaled such that 
the luminosity of \jone\ line matches that of Mrk~231 (see Section \ref{sec:coground} for how the low-J line fluxes 
were acquired from the literature).  This figure illustrates just how varied the shapes of the CO SLEDs become at higher-J, 
and includes comparisons to Galactic sources, which will be discussed in Section \ref{sec:disc:galactic}.

\subsection{Photometry}\label{sec:aperture}

\begin{deluxetable*}{lrrrrrr}
\tabletypesize{\scriptsize}
\tablecaption{Photometry Flux Densities in Jy: SPIRE Photometer \label{tbl:photflux_spire}}
\tablehead{
\colhead{Galaxy} & \colhead{250 $\mu$m $F_\nu$} & \colhead{$\sigma_{tot}$}  & \colhead{350 $\mu$m $F_\nu$} & \colhead{$\sigma_{tot}$}  & \colhead{500 $\mu$m $F_\nu$} & \colhead{$\sigma_{tot}$}
}
\startdata
Mrk 231 & 5.40 & 0.878 & 1.92\phantom{0} & 0.354 & 0.567 & 0.238\\
IRAS F17207-0014 & 7.67 & 1.29\phantom{0} & 2.84\phantom{0} & 0.451 & 0.944 & 0.284\\
IRAS 09022-3615 & 2.48 & 0.399 & 0.732 & 0.470 & 0.197 & 0.321\\
Arp 220 & 32.6\phantom{0} & 3.76\phantom{0} & 11.7\phantom{00} & 0.955 & 4.01\phantom{0} & 0.421\\
Mrk 273 & 4.27 & 0.822 & 1.49\phantom{0} & 0.225 & 0.491 & 0.139\\
UGC 05101 & 5.77 & 0.941 & 2.23\phantom{0} & 0.322 & 0.687 & 0.166\\
NGC 6240 & 6.82 & 0.877 & 2.64\phantom{0} & 0.370 & 0.815 & 0.246\\
Arp 299-A & 22.0\phantom{0} & 1.96\phantom{0} & 7.60\phantom{0} & 0.721 & 2.36\phantom{0} & 0.325\\
NGC 1068 & 103\phantom{.00} & 6.12\phantom{0} & 41.3\phantom{00} & 2.69\phantom{0} & 14.0\phantom{00} & 1.04\phantom{0}\\
NGC 4038 (Overlap) & 37.8\phantom{0} & 2.50\phantom{0} & 14.9\phantom{00} & 0.984 & 5.04\phantom{0} & 0.464\\
NGC 1222 & 3.65 & 0.387 & 1.43\phantom{0} & 0.319 & 0.446 & 0.202\\
NGC 1266 & 3.71 & 0.522 & 1.30\phantom{0} & 0.201 & 0.411 & 0.147\\
Cen A & 271\phantom{.00} & 60.0\phantom{00} & 110.\phantom{000} & 17.6\phantom{00} & 44.9\phantom{00} & 5.12\phantom{0}
\enddata
\tablecomments{All are from this work, see Section \ref{sec:aperture}.}
\end{deluxetable*}

In addition to using the SPIRE photometry maps to determine the source-beam coupling correction, 
we also performed aperture photometry to obtain galaxy-integrated fluxes to supplement the dust modeling.
Results are in Table \ref{tbl:photflux_spire}.  
A circular aperture was centered on the FTS pointing location of the dust map, and 
a radius ($r_a$) was chosen to encompass all of the measured flux.  The sky background is calculated from an annulus with inner radius $r_a$ to 
outer radius $r_a$ + 10 pixels.  The sky is calculated as the median value of the data (Jy/beam) divided by (beam/pixel) times the area contained in the annulus (in pixels).  For all points within $r_a$, the flux is the total of the data divided by beam/pixel minus the sky.  

\section{Analysis}\label{sec:analysis}

For both dust and CO modeling,\footnote{In addition to the methods presented here, 
we attempted to use MAGPHYS \citep{daCunha2008} to compare galaxies by star formation rate (SFR) and specific star formation rate (sSFR), 
which requires stellar masses (M$_*$).  However, due to the difficulties of SED modeling, the assumptions made in the models, and 
the small sample size here, we found the results to be highly uncertain, and they did not particularly add to the conclusions presented here.
Interested readers may wish to see the results in \citet{Kamenetzky2014}, including compiled $< 10 \mu$m photometry for this sample of galaxies.
We consider star-formation rates (SFRs) to be directly proportional to \lfir, from the Kennicutt relation \citep{Kennicutt1998}.} we utilize the nested sampling algorithm MultiNest \citep{Feroz2009} and its Python wrapper, PyMultiNest \citep{Buchner2014}.\footnote{\url{https://github.com/JohannesBuchner/PyMultiNest}}
  As stated in \citet{Feroz2009}, nested sampling ``is a Monte Carlo technique aimed at efficient evaluation of the Bayesian evidence, but also produces posterior inferences as a by-product."  The evidence in 
 this context is the average likelihood of a model over its prior probability space.
 The algorithm ``nests inwards" to subsequently smaller regions of high-likelihood parameter space.  Unlike calculating the likelihood using a grid method, the algorithm can focus on high likelihood regions and thus estimate parameter constraints more efficiently.
 
In both cases (dust and CO), described in their respective sections (\ref{sec:analysis_dust} and \ref{sec:comodel}), we have a set of measurements $\bm{x}$ with errors $\bm{\sigma}$, 
a model described by parameters $\bm{p}$ with predicted fluxes $I(\bm{p})$.  
For a Gaussian probability, the natural log of the likelihood is $-0.5 (\bm{x} - \bm{I}(\bm{p}))^{T} {\rm Cov}^{(-1)} (\bm{x} - \bm{I}(\bm{p}))$.
The covariance matrix is fully described for the dust modeling in Section \ref{sec:analysis_dust}.  For the CO modeling, 
we use zero covariance between data points, simplifying the probability to $\prod_i (2 \pi \sigma_i^2)^{-2} \unit{exp}(-0.5 (x_i-I_i(\bm{p}))^2 \sigma_i^{-2})$, where $i$ represents 
each data point (integrated flux).  The natural log of the probability is $-\sum_i 0.5 {\rm ln}(2 \pi) + {\rm ln}(\sigma_i) + 0.5 (x_i-I_i(\bm{p}))^2 \sigma_i^{-2}$.

We can determine the probability distribution for any one parameter by marginalizing the full distribution over all other parameters.
There are different statistics that can be used to describe a parameter.  The best-fit set of parameters is the combination that produced the 
highest likelihood.  In the case of a very simple parameter space, where the solution is clustered in a Gaussian fashion in one area, the best-fit will 
likely coincide with the statistical mean of each parameter.  This was the case for the dust modeling.  However, the parameter 
space may not be so simply described, which we will demonstrate for the CO modeling.  This mode of parameter estimation is designed to focus on the 
whole probability density function (PDF), not to refine the best fit.  We present the best-fit in our tables, as it is used in the plots of best-fit spectral (line) 
energy distributions, but we focus our discussion on the marginalized likelihoods.  

In the case of a complicated parameter space, there can be multiple ``modes," or islands 
in parameter space, as was sometimes the case for the CO modeling.  The MultiNest algorithm partitions the posterior likelihood space into ellipsoids, which may overlap.
Non-overlapping ellipsoids can be separated into separate modes, with a separate ``local" evidence.  In our cases, all posterior distributions 
with multiple likelihoods had one mode stand out as containing more posterior mass than others; we focus our parameter estimation on this most-likely 
mode, and present its statistical mean and standard deviation.  This is as opposed to the mean that one would calculate considering the entire
posterior distribution, which would be weighted towards other, less-likely modes.  The extent of that weighting would depend on the ratio of the 
different local evidence.

\subsection{Dust Modeling Likelihood with Multinest}\label{sec:analysis_dust}

In addition to SPIRE photometer observations (250, 350, 500 $\mu$m), we also used fluxes from the literature, those 
listed in Appendix \ref{appendix:lit} above 10 $\mu$m.  These are galaxy-integrated fluxes; we only 
modeled once per galaxy with multiple FTS pointings (NGC~4038, NGC~1365, Arp~299), because fluxes 
separated into individual components were often not available.  
We used the dust model as in \citet{Casey2012}, which is the sum of a greybody and a powerlaw with exponential drop-off:

\begin{equation}\label{eqn:greybody}
S(\lambda) = N_{bb} \frac{(1-e^{-(\lambda_0/\lambda)^\beta})(c/\lambda)^3}{e^{hc/(\lambda kT)}-1} +N_{pl} \lambda^\alpha e^{-(\lambda/\lambda_c)^2}.
\end{equation}

In Equation \ref{eqn:greybody}, the free parameters are $T$ (temperature, K),  $\beta$ (emissivity index), 
$\lambda_0$ (wavelength at which optical depth is unity, $\mu$m), and $\alpha$ (slope of the mid-IR powerlaw).
$N_{bb}$, the normalization in Jy for the greybody component, is fixed at the best-fit value for any 
given combination of the previous parameters.  $\lambda_c$ and $N_{pl}$ are tied to the other parameters as in 
\citet[][Table 1]{Casey2012}.  In calculating the likelihood of the dust parameters, we assume that calibration errors are 50\% correlated 
between measurements from the same instruments.  We expect some degree of correlation, but too far above 50\% in the 
covariance matrix can drive the best fit very far away from the data points of other instruments.

\begin{figure*} 
\centering
\includegraphics[width=\textwidth]{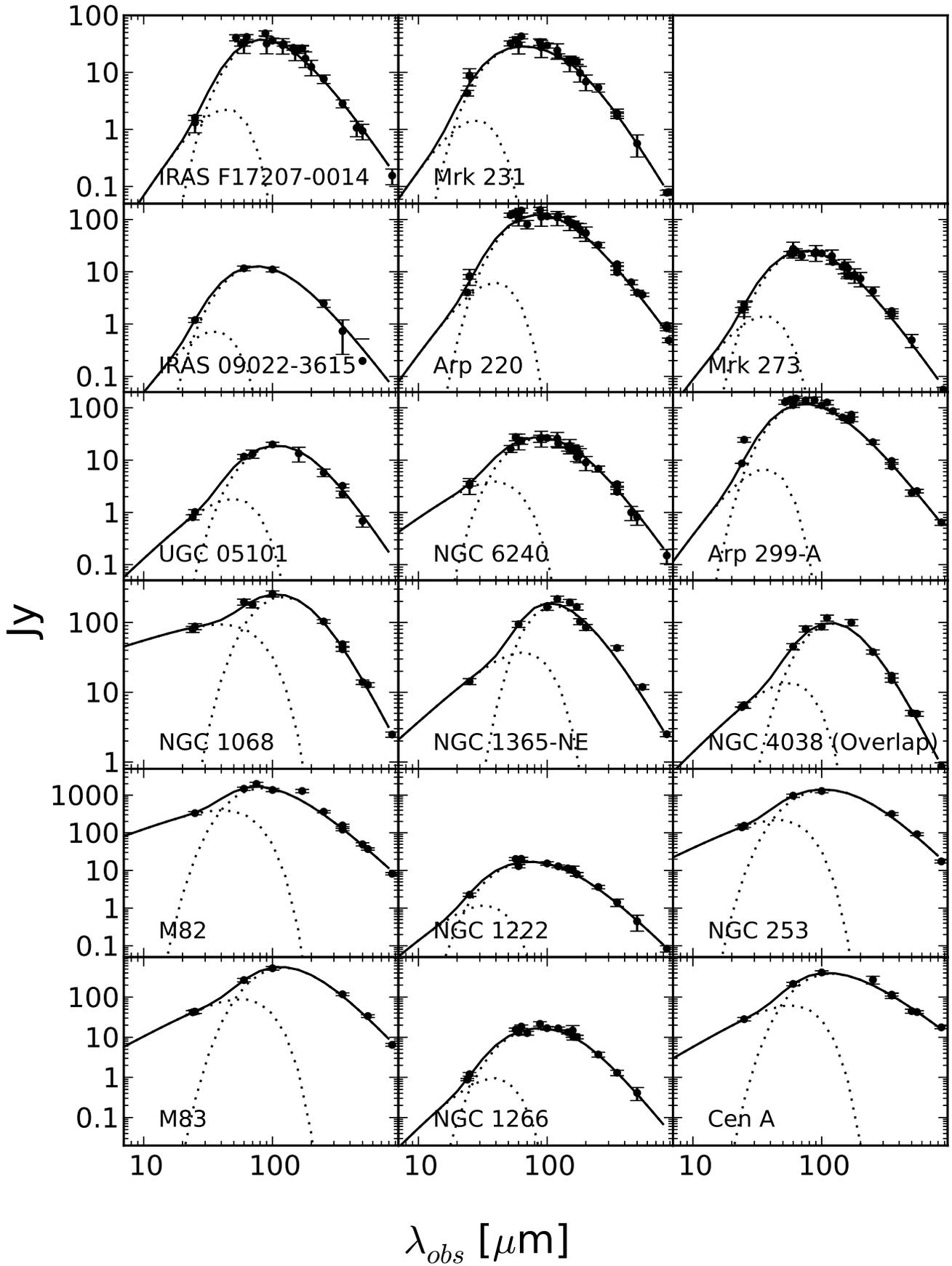}
\caption[Dust Modeling Spectral Energy Distributions]{Dust modeling spectral energy distributions.  
The maximum likelihood solution (Equation \ref{eqn:greybody}) is shown as a solid line.
Dotted lines break up the 
model into the powerlaw (lower wavelength) and greybody (higher wavelength) components.\label{fig:dustsed}}
\end{figure*}

\begin{figure} 
\plotone{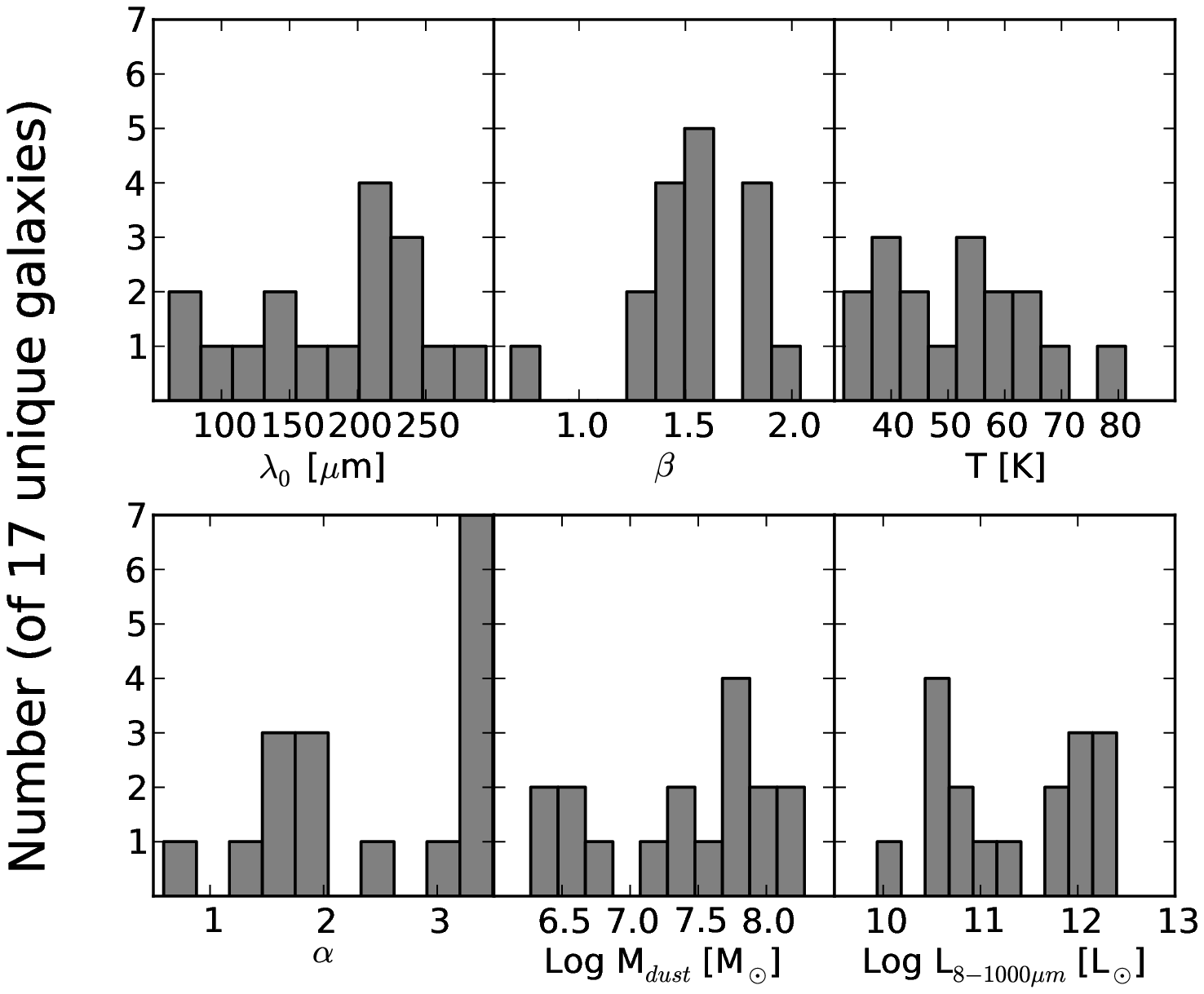} 
\caption[Dust Modeling Histogram Results]{Dust modeling histogram results.  See Section \ref{sec:analysis_dust}.\label{fig:dusthist}}
\end{figure}

The results are shown in Figures \ref{fig:dustsed} (best-fit SEDs) and \ref{fig:dusthist} (histogram of best-fit parameters).
The individual parameter results are in Table \ref{tbl:dust1}.  In all cases, only one mode in likelihood space was found, and the resulting likelihood 
distributions were very well defined by Gaussians (the best-fit and the mode and median of the resulting marginalized parameters all aligned).

Table \ref{tbl:dust2} also lists the results for parameters which can be derived from the model above: the optical depth at 100 $\mu$m ($\tau_{100}$), 
the dust mass (M$_{dust}$), and the far-infrared luminosity from 8 to 1000 $\mu$m (L$_{8-1000 \mu m}$ or \lfir).  
To calculate the dust mass, we utilize $\kappa_{125 \mu m} = 2.64 \unit{m^2/kg}$ \citep{Dunne2003} and $M_d = \frac{S_\nu D_L^2}{\kappa_\nu B_\nu(T)}$ (statistical 
errors in Table \ref{tbl:dust2} do not include uncertainty in $\kappa$).    
The values of \lfir\ we find when modeling the SED ($L = 4 \pi D_L^2 \int_{8 \mu m}^{1000 \mu m} S_\nu d\nu$) are slightly higher than those derived 
from utilizing only the 60 and 100 $\mu$m fluxes (e.g. those presented in Table \ref{tbl:obs} from Hyperleda), by about a factor of 1.7 $\pm$ 0.5.  

\begin{deluxetable*}{l rrr | rrr | rrr | rrr}
\tabletypesize{\scriptsize}
\tablecaption{Dust Fitting Results: Model Parameters \label{tbl:dust1}}
\tablehead{
\colhead{FTS Name}  & \multicolumn{3}{c}{$\lambda_0$ [$\mu$m]} & \multicolumn{3}{c}{$\beta$} & \multicolumn{3}{c}{T [K]} & \multicolumn{3}{c}{$\alpha$} \\
\colhead{}    & \colhead{Mean} & \colhead{$\sigma$} & \colhead{Best}  & \colhead{Mean} & \colhead{$\sigma$} & \colhead{Best}  & \colhead{Mean} & \colhead{$\sigma$} & \colhead{Best}  & \colhead{Mean} & \colhead{$\sigma$} & \colhead{Best} 
}
\startdata
Mrk 231              &     231 &      24 &     242 &    1.82 &    0.12 &    1.85 &    79.6 &     2.3 &    81.3 &    3.07 &    0.34 &    3.45\\
IRAS F17207-0014     &      96 &      24 &     120 &    1.48 &    0.11 &    1.46 &    48.8 &     3.0 &    52.3 &    3.27 &    0.16 &    3.49\\
IRAS 09022-3615      &     176 &      53 &     161 &    1.69 &    0.57 &    1.37 &    62.6 &     4.7 &    65.2 &    2.91 &    0.41 &    3.32\\
Arp 220              &     187 &      27 &     193 &    1.54 &    0.09 &    1.54 &    55.5 &     1.7 &    56.4 &    3.38 &    0.11 &    3.49\\
Mrk 273              &     115 &      25 &     136 &    1.58 &    0.07 &    1.60 &    58.5 &     5.0 &    63.5 &    2.89 &    0.36 &    3.41\\
UGC 05101            &     225 &      50 &     225 &    1.94 &    0.42 &    1.87 &    46.2 &     3.8 &    45.7 &    2.43 &    0.16 &    2.36\\
NGC 6240             &     226 &      46 &     212 &    1.65 &    0.25 &    1.60 &    57.6 &     4.8 &    56.0 &    1.98 &    0.31 &    1.80\\
Arp 299-A            &     120 &      30 &     140 &    1.26 &    0.08 &    1.28 &    58.6 &     2.7 &    60.9 &    3.23 &    0.20 &    3.48\\
NGC 1068             &     244 &      34 &     251 &    1.80 &    0.09 &    1.82 &    39.5 &     3.0 &    40.1 &    0.60 &    0.05 &    0.59\\
NGC 1365-NE          &     181 &      64 &      61 &    1.45 &    0.08 &    1.39 &    38.6 &     3.9 &    31.6 &    1.76 &    0.09 &    1.66\\
NGC 4038 (Overlap)   &     219 &      31 &     220 &    1.87 &    0.11 &    1.88 &    38.0 &     2.3 &    37.9 &    1.85 &    0.05 &    1.85\\
M82                  &     137 &      44 &     100 &    1.30 &    0.06 &    1.30 &    55.6 &     7.3 &    49.3 &    1.48 &    0.37 &    1.25\\
NGC 1222             &     205 &      49 &     226 &    1.48 &    0.17 &    1.49 &    64.8 &     6.0 &    68.1 &    2.65 &    0.53 &    2.93\\
M83                  &     213 &      49 &     211 &    1.59 &    0.09 &    1.60 &    37.1 &     2.8 &    36.9 &    1.70 &    0.07 &    1.71\\
NGC 253              &     267 &      28 &     294 &    1.46 &    0.07 &    1.50 &    45.0 &     2.5 &    46.3 &    1.64 &    0.10 &    1.66\\
NGC 1266             &     206 &      25 &     214 &    2.00 &    0.29 &    2.04 &    56.1 &     2.8 &    57.4 &    3.08 &    0.26 &    3.24\\
Cen A                &     187 &      66 &      64 &    0.71 &    0.10 &    0.68 &    38.8 &     1.8 &    36.5 &    1.91 &    0.09 &    1.88
\enddata
\end{deluxetable*}
\begin{deluxetable*}{l rrr | rrr | rrr}
\tabletypesize{\scriptsize}
\tablecaption{Dust Fitting Results: Derived Parameters \label{tbl:dust2}}
\tablehead{
\colhead{FTS Name}  & \multicolumn{3}{c}{$\tau_{100}$} & \multicolumn{3}{c}{Log $M_{\rm dust}$ [\ms]} & \multicolumn{3}{c}{Log $L_{8-1000 \mu m}$ [\ls]} \\
\colhead{}    & \colhead{Mean} & \colhead{$\sigma$} & \colhead{Best}  & \colhead{Mean} & \colhead{$\sigma$} & \colhead{Best}  & \colhead{Mean} & \colhead{$\sigma$} & \colhead{Best} 
}
\startdata
Mrk 231              &    5.18 &    1.46 &    5.57 &    7.80 &    0.02 &    7.80 &   12.41 &    0.02 &   12.40\\
IRAS F17207-0014     &    1.04 &    0.42 &    1.39 &    8.32 &    0.05 &    8.28 &   12.34 &    0.01 &   12.34\\
IRAS 09022-3615      &    4.08 &    3.61 &    2.09 &    7.96 &    0.07 &    7.90 &   12.21 &    0.02 &   12.23\\
Arp 220              &    2.80 &    0.79 &    2.85 &    8.08 &    0.02 &    8.07 &   12.14 &    0.02 &   12.14\\
Mrk 273              &    1.36 &    0.47 &    1.75 &    7.77 &    0.06 &    7.71 &   12.13 &    0.01 &   12.13\\
UGC 05101            &    6.78 &    4.58 &    4.91 &    8.27 &    0.07 &    8.27 &   11.95 &    0.02 &   11.96\\
NGC 6240             &    4.55 &    2.34 &    3.49 &    7.69 &    0.06 &    7.71 &   11.83 &    0.02 &   11.84\\
Arp 299-A            &    1.31 &    0.45 &    1.56 &    7.46 &    0.03 &    7.44 &   11.74 &    0.02 &   11.74\\
NGC 1068             &    5.22 &    1.46 &    5.41 &    7.49 &    0.07 &    7.48 &   11.40 &    0.02 &   11.41\\
NGC 1365-NE          &    2.56 &    1.34 &    0.51 &    7.56 &    0.09 &    7.72 &   11.11 &    0.02 &   11.12\\
NGC 4038 (Overlap)   &    4.59 &    1.45 &    4.47 &    7.43 &    0.06 &    7.43 &   10.90 &    0.02 &   10.90\\
M82                  &    1.56 &    0.68 &    1.00 &    6.45 &    0.09 &    6.53 &   10.79 &    0.03 &   10.80\\
NGC 1222             &    3.18 &    1.38 &    3.45 &    6.31 &    0.06 &    6.27 &   10.66 &    0.02 &   10.65\\
M83                  &    3.52 &    1.35 &    3.31 &    7.07 &    0.07 &    7.08 &   10.53 &    0.02 &   10.53\\
NGC 253              &    4.28 &    0.77 &    5.05 &    6.78 &    0.06 &    6.77 &   10.52 &    0.02 &   10.52\\
NGC 1266             &    4.69 &    1.94 &    4.82 &    6.34 &    0.04 &    6.33 &   10.44 &    0.01 &   10.45\\
Cen A                &    1.57 &    0.46 &    0.75 &    6.51 &    0.05 &    6.56 &    9.93 &    0.02 &    9.94
\enddata
\tablecomments{The statistical error on $M_{dust}$ does not include the uncertainty in the dust emissivity.}
\end{deluxetable*}

\citet{Casey2012} modeled 65 local LIRGS and ULIRGS, fixing $\lambda_0 = 200$ $\mu$m and finding a mean $\beta$ = 1.60 $\pm$ 0.38 and $\alpha = 2.0 \pm 0.5$. We find, in Figure \ref{fig:dusthist} that $\beta$ and $\alpha$ can vary significantly, but cluster around similar values.  When left to vary, $\lambda_0$ can often be higher than 200 $\mu$m. 

\subsection{CO Modeling Likelihood with Multinest}\label{sec:comodel}

\subsubsection{Measurements from the Literature and Dust Optical Depth Correction}\label{sec:coground}

As discussed in Section \ref{sec:intro}, previous work has shown that the high-J 
lines detected by the FTS are often emitted by a warmer component of gas than the low-J lines.  To accurately model 
both components of gas, we supplemented our FTS line fluxes with measurements from the ground of \jone, \jtwo, and \jthree.
The measurements used for this survey are presented in Table \ref{tbl:coground}.  We first utilized large surveys for consistent data; some 
galaxies had multiple measurements for the same line because of overlaps of the surveys.  When surveys alone did not have enough lines 
for a particular galaxy, we sought out individual measurements in the literature.  For the cases of semi-extended sources which may 
have multiple pointings, see Appendix \ref{appendix:individual} for more information on how these were handled.  
No ground-based measurements for IRAS~09022-3615 were available, so only the warm, high-J component of gas was modeled for this galaxy.

\begin{deluxetable*}{crrll|crrll|crrll}
\tabletypesize{\scriptsize}
\tablecaption{CO Additional Line Measurements in Jy km s$^{-1}$ \label{tbl:coground}}
\tablehead{
\colhead{J$_{up}$} & \colhead{$I \Delta v$} & \colhead{$\sigma$} & \colhead{$\eta(\Omega)$} & \colhead{Ref}  & \colhead{J$_{up}$} & \colhead{$I \Delta v$} & \colhead{$\sigma$} & \colhead{$\eta(\Omega)$} & \colhead{Ref}  & \colhead{J$_{up}$} & \colhead{$I \Delta v$} & \colhead{$\sigma$} & \colhead{$\eta(\Omega)$} & \colhead{Ref} 
}
\startdata
\multicolumn{  5}{c}{\bf{UGC 05101           }} & 
\multicolumn{  5}{c}{\bf{Arp 220             }} & 
\multicolumn{  5}{c}{\bf{NGC 6240            }}\\
 1 & 73.5 & 15 &  1.00 &  1 &   1 & 403\phantom{.0} & 22 &  1.00 &  7 & 
 1 & 239\phantom{.0} & 29 &  1.01 &  7\\
 1 & 70.3 & 18 &  1.00 &  2 &   1 & 609\phantom{.0} & 120 &  0.90 &  1 & 
 1 & 413\phantom{.0} & 82 &  0.82 &  1\\
 1 & 73.6 & 15 &  1.00 &  3 &   1 & 283\phantom{.0} & 56 &  1.04 & 14 & 
 1 & 261\phantom{.0} & 52 &  1.07 & 14\\
 2 & 350.\phantom{0} & 81 &  1.00 &  2 & 
 1 & 520.\phantom{0} & 57 &  1.04 & 15 & 
 2 & 2210\phantom{.0} & 430 &  0.81 & 16\\
 3 & 609\phantom{.0} & 130 &  1.00 &  2 & 
 2 & 1780\phantom{.0} & 350 &  0.89 & 16 & 
 3 & 4710\phantom{.0} & 900 &  0.72 & 16\\
\multicolumn{  5}{c}{\bf{NGC 4038 (Overlap)  }} & 
 3 & 3970\phantom{.0} & 760 &  0.84 & 16 & 
 3 & 4210\phantom{.0} & 1200 &  0.82 &  6\\
 1 & 851\phantom{.0} & 170 &  0.99 &  4 & 
 3 & 3050\phantom{.0} & 880 &  0.90 &  6 & 
\multicolumn{  5}{c}{\bf{IRAS F17207-0014    }}\\
 2 & 2930\phantom{.0} & 320 &  0.70 &  5 & 
\multicolumn{  5}{c}{\bf{NGC 1365-NE         }} &   1 & 93.2 & 19 &  1.00 &  1\\
 3 & 5380\phantom{.0} & 780 &  0.99 &  4 & 
 1 & 1840\phantom{.0} & 360 &  1.11 & 17 & 
 1 & 161\phantom{.0} & 29 &  1.00 &  2\\
 3 & 8190\phantom{.0} & 2400 &  0.49 &  6 & 
 1 & 2330\phantom{.0} & 460 &  1.00 & 18 & 
 1 & 178\phantom{.0} & 36 &  1.00 &  3\\
 3 & 5610\phantom{.0} & 630 &  0.49 &  5 & 
 1 & 2210\phantom{.0} & 35 &  1.00 & 19 & 
 2 & 695\phantom{.0} & 150 &  1.00 &  2\\
\multicolumn{  5}{c}{\bf{NGC 4038            }} & 
 1 & 1440\phantom{.0} & 140 &  1.11 & 15 & 
 2 & 330.\phantom{0} & 66 &  1.00 &  3\\
 1 & 468\phantom{.0} & 93 &  0.98 &  4 & 
 1 & 2160\phantom{.0} & 440 &  1.01 &  3 & 
 3 & 1220\phantom{.0} & 260 &  1.00 &  2\\
 1 & 651\phantom{.0} & 72 &  1.04 &  7 & 
 2 & 6670\phantom{.0} & 1300 &  0.75 & 18 & 
 3 & 1050\phantom{.0} & 310 &  1.00 &  6\\
 2 & 2230\phantom{.0} & 230 &  0.65 &  5 & 
 3 & 12100\phantom{.0} & 1100 &  0.52 & 20 & 
\multicolumn{  5}{c}{\bf{Arp 299-A           }}\\
 3 & 2830\phantom{.0} & 410 &  0.98 &  4 & 
\multicolumn{  5}{c}{\bf{NGC 1365-SW         }} & 
 1 & 785\phantom{.0} & 160 &  1.12 & 14\\
 3 & 4600\phantom{.0} & 1300 &  0.46 &  6 & 
 1 & 1500\phantom{.0} & 300 &  1.18 & 17 & 
 1 & 442\phantom{.0} & 88 &  0.65 &  3\\
 3 & 3870\phantom{.0} & 410 &  0.46 &  5 & 
 1 & 1860\phantom{.0} & 370 &  1.01 & 18 & 
 2 & 1990\phantom{.0} & 290 &  1.00 & 22\\
\multicolumn{  5}{c}{\bf{M82                 }} & 
 1 & 2140\phantom{.0} & 430 &  1.02 &  3 & 
 2 & 1200\phantom{.0} & 230 &  0.46 &  3\\
 1 & 7060\phantom{.0} & 700 &  0.62 &  8 & 
 2 & 5550\phantom{.0} & 1100 &  0.58 & 18 & 
 3 & 3480\phantom{.0} & 1000 &  0.67 &  6\\
 1 & 6600\phantom{.0} & 650 &  1.02 &  7 & 
\multicolumn{  5}{c}{\bf{Mrk 273             }} & 
\multicolumn{  5}{c}{\bf{Arp 299-B           }}\\
 2 & 37400\phantom{.0} & 3600 &  0.62 &  8 &   1 & 90.0 & 18 &  1.00 &  1 & 
 1 & 394\phantom{.0} & 86 &  0.75 &  3\\
 3 & 57200\phantom{.0} & 5500 &  0.62 &  8 &   1 & 68.1 & 14 &  1.00 & 14 & 
 2 & 1690\phantom{.0} & 250 &  1.00 & 22\\
 6 & 68200\phantom{.0} & 7800 &  0.19 &  5 &   1 & 82.3 & 15 &  1.00 &  2 & 
 3 & 3930\phantom{.0} & 1200 &  0.44 &  6\\
\multicolumn{  5}{c}{\bf{NGC 1068            }} & 
 1 & 112\phantom{.0} & 28 &  1.00 & 21 & 
\multicolumn{  5}{c}{\bf{Arp 299-C           }}\\
 1 & 2260\phantom{.0} & 330 &  1.03 &  7 &   1 & 97.6 & 19 &  1.00 &  3 & 
 1 & 385\phantom{.0} & 84 &  0.77 &  3\\
 1 & 4240\phantom{.0} & 840 &  0.45 &  3 & 
 2 & 273\phantom{.0} & 54 &  1.00 &  2 & 
 2 & 1710\phantom{.0} & 250 &  1.00 & 22\\
 2 & 11700\phantom{.0} & 1100 &  0.74 &  9 & 
 3 & 491\phantom{.0} & 110 &  1.00 &  2 & 
\multicolumn{  5}{c}{\bf{NGC 253             }}\\
 2 & 12600\phantom{.0} & 2500 &  0.16 &  3 & 
\multicolumn{  5}{c}{\bf{Mrk 231             }} & 
 1 & 10800\phantom{.0} & 1100 &  1.01 &  7\\
 3 & 17800\phantom{.0} & 3000 &  1.00 & 10 &   1 & 90.4 & 13 &  1.00 &  7 & 
 1 & 7200\phantom{.0} & 640 &  0.75 & 23\\
 3 & 11800\phantom{.0} & 3400 &  0.48 &  6 & 
 1 & 103\phantom{.0} & 20. &  1.00 &  1 & 
 2 & 33800\phantom{.0} & 3200 &  0.89 & 24\\
\multicolumn{  5}{c}{\bf{Cen A               }} & 
 1 & 88.5 & 16 &  1.00 &  2 &   2 & 34300\phantom{.0} & 3600 &  0.75 & 23\\
 1 & 1540\phantom{.0} & 290 &  1.01 & 11 &   1 & 97.0 & 24 &  1.00 & 21 & 
 3 & 75300\phantom{.0} & 9700 &  0.75 & 23\\
 1 & 1620\phantom{.0} & 170 &  1.04 & 12 &   1 & 83.7 & 12 &  1.00 & 15 & 
\multicolumn{  5}{c}{\bf{NGC 1266            }}\\
 2 & 3440\phantom{.0} & 430 &  0.48 & 12 & 
 1 & 104\phantom{.0} & 21 &  1.00 &  3 & 
 1 & 204\phantom{.0} & 33 &  0.86 & 25\\
\multicolumn{  5}{c}{\bf{M83                 }} & 
 2 & 321\phantom{.0} & 56 &  1.00 &  2 & 
 1 & 204\phantom{.0} & 41 &  0.86 & 26\\
 1 & 2020\phantom{.0} & 200 &  1.02 &  7 & 
 2 & 292\phantom{.0} & 72 &  1.00 & 21 & 
 2 & 906\phantom{.0} & 140 &  0.74 & 25\\
 2 & 3660\phantom{.0} & 370 &  0.76 &  5 & 
 3 & 592\phantom{.0} & 120 &  1.00 &  2 & 
 2 & 727\phantom{.0} & 140 &  0.72 & 26\\
 3 & 10500\phantom{.0} & 2300 &  0.54 & 13 & 
 3 & 354\phantom{.0} & 110 &  1.00 &  6 & 
\multicolumn{  5}{c}{\bf{NGC 1222            }}\\
 3 & 13400\phantom{.0} & 3800 &  0.57 &  6 & 
 3 & 376\phantom{.0} & 46 &  1.00 &  5 & 
 1 & 119\phantom{.0} & 24 &  0.73 & 26\\
 3 & 8680\phantom{.0} & 830 &  0.57 &  5 & 
 4 & 1210\phantom{.0} & 350 &  1.00 &  2 & 
 2 & 301\phantom{.0} & 59 &  0.49 & 26
\enddata
\tablecomments{Calibration errors, from references or assumed, are included in $\sigma$.  {\bf References.} ( 1) \citet{Solomon1997}; ( 2) \citet{Papadopoulos2012}; ( 3) \citet{Baan2008}; ( 4) \citet{Schirm2014}; ( 5) \citet{Bayet2006}; ( 6) \citet{Mao2010}; ( 7) \citet{Young1995}; ( 8) \citet{Ward2003}; ( 9) \citet{Kamenetzky2011}; (10) \citet{Spinoglio2012}; (11) \citet{Wild2000}; (12) \citet{Eckart1990}; (13) \citet{Mauersberger1999}; (14) \citet{Sanders1991}; (15) \citet{Maiolino1997}; (16) \citet{Greve2009}; (17) \citet{Papadopoulos1998}; (18) \citet{Sandqvist1995}; (19) \citet{Elfhag1996}; (20) \citet{Sandqvist1999}; (21) \citet{Albrecht2007}; (22) \citet{Sliwa2013}; (23) \citet{Harrison1999}; (24) Z-Spec; (25) \citet{Alatalo2011}; (26) \citet{Young2011}}
\end{deluxetable*}

The low-J line measurements came from a variety of telescopes with different beam sizes.  We divided all flux densities in Jy km/s by 
the appropriate $\eta(\Omega_b,\Omega_{43.5})$ from Table \ref{tbl:corr} to refer all fluxes to our beam size of 43\farcs5 (see Section \ref{sec:sourcebeam}).  The specific value used 
for each line flux in the last column of Table \ref{tbl:coground}.  In same cases, we gathered multiple transitions of the same line, and 
only discarded measurements which were wildly discrepant from the rest of the SLED.  

We additionally corrected the line fluxes for obscuration by dust.  At higher frequencies, the optical depth is not negligible.  
We corrected the lines by dividing by this factor for a mixed dust model, in which the line-emitting gas and the dust are assumed to be spatially co-extensive: 
$(1-e^{-\tau_\lambda})/\tau_\lambda$, \noindent where $\tau_\lambda$ for a given line was calculated by $(\lambda_0/\lambda)^\beta$, 
using the mean $\lambda_0$ and $\beta$ from the dust model 
results in Table \ref{tbl:dust1}.  
We propagated the uncertainties in both parameters into the correction and resulting line fluxes.  
For a given galaxy, the correction factor decreases approximately linearly from $\sim 1$ at the low-J lines to a median of 0.64 (standard deviation of 0.1) at \jthirteen; in other words, we model \jthirteen\ lines a factor of about 1.6 times higher than measured.  The correction factor at CO \jthirteen\ ranged from $0.51 \pm 0.05$ (NGC253) to $0.84 \pm 0.04$ (IRAS~F17207-0014).  For comparison, at mid-J of CO \jfive, the median correction was 0.90 $\pm$ 0.04.

The correction did not significantly modify the resulting likelihood distributions, 
other than slight increases in the warm gas pressures and luminosities, as one would expect by changing the shape of the SLED in this fashion.  To be explicit, the best-fit solution often moves, but the marginalized parameter distributions shift an insignificant amount compared to their large uncertainties.  This is further discussed 
in Section \ref{sec:disc:priors}.

\subsubsection{Model Description}

To model the CO fluxes, we use a custom version of the non-LTE code RADEX described in \citet{VanderTak2007}.  The input parameters to this model 
are kinetic temperature, $T$ [K], column density of CO per unit linewidth, $N_{CO}/\Delta V$ [cm$^{-2}$ / (km s$^{-1}$)], and density of the colliding partner (molecular hydrogen, $n_{H2}$ [cm$^{-3}$]).
Additionally, we allowed the resultant fluxes to scale uniformly lower by an area filling factor $\Phi_A \le 1$.  

RADEX uses an escape probability method to perform statistical equilibrium calculations, 
first populating the rotational levels in the optically thin limit, then calculating the optical depths for the lines. 
The code continues calculating new level populations using new optical depth values until the two converge on a consistent solution. The line intensities are output as background-subtracted Rayleigh-Jeans equivalent radiation temperatures.
We use the CMB (2.73 K at $z=0$) as the radiation background, because we have found that the solution 
is generally not sensitive to the radiation background temperature for these types of galaxies \citep{Kamenetzky2011,Kamenetzky2012}. 
Previous work has shown that the high-J 
lines detected by the FTS are often emitted by a warmer component of gas which is typically 10\% or less of the total CO mass.  
Therefore we simultaneously 
model two components of gas, described by eight parameters, ${\bf p} = (n_1,T_1,N_1/\Delta V,\Phi_1,n_2,T_2,N_2/\Delta V,\Phi_2)$, four for each component.

In addition to these basic parameters which fully describe the model, we calculated other properties and their likelihoods, 
some of which were better constrained than the formal parameters.  
The product of the temperature and density is the thermal pressure $(P)$, and the product of the column density and filling factor ($<N_{CO}>$) is 
proportional to the total molecular gas mass,

\begin{equation}\label{eqn:mass}
M = A \Phi_A N_{CO} 1.4 \unit{m_{H_2}} X_{^{12}\unit{CO}}^{-1}
\end{equation}

\noindent where A is the area of the region, calculated from the luminosity distance, where $A = \frac{\pi}{4} s^2$ pc$^2$ and 
$s$ is a diameter given by $s=2 \times 10^6 \sqrt{\Omega_s/\pi} D_L(1+z)^{-2}$ \ [pc].  
Using the beam size $\Omega_b = 5.04 \times 10^{-8}$ sr corresponds to areas of 5.04$\times 10^4 (D_L(1+z)^{-2})^2$ pc$^2$.
Note that both A and $\Phi_A$, the 
area filling factor, are present in this equation, which accounts for beam dilution for the many sources that are less than 43\farcs5 across.
The factor of 1.4 accounts for helium and other heavy elements, and $X_{^{12}CO}$ is the abundance ratio of $^{12}$CO to H$_{2}$: we use $2 \times 10^{-4}$. 
We also calculated the probabilities of the total CO 
luminosity in each component from RADEX, which may include 
contributions from higher-J lines other than those modeled here.  Finally, we also determined the likelihood 
for the ratio of warm to cold properties (e.g. $M_1/M_2$).

The temperature and density are degenerate; even when one might not individually be well determined, their product (thermal pressure, henceforth simply ``pressure") 
is better constrained, because the two properties are anti-correlated.  Likewise, the column density and filling factor are degenerate, and it is their product (proportional to mass) that is better determined.  However, it is important to note that these two parameters are not perfectly degenerate.  Pressure is largely responsible for the shape of the SLED, and 
mass for the absolute flux scaling.  
The filling factor linearly scales the absolute values (i.e., the normalization) of the peak temperatures in the CO SLED, while varying the column density changes the absolute values as well as the SLED shape, as a result of optical depth effects. Thus the model SLEDs depend on both quantities independently, and so column density and filling factor are separate parameters. We adopted $\Omega_s$ = 1.133 $\times 43.5^2$ sq arcsec, allowing $\Phi_A$ to account for emission smaller than the extent of the beam.

\subsubsection{Model Constraints and Priors}\label{sec:copriors}

\begin{deluxetable}{lrrr}
\tabletypesize{\scriptsize}
\tablecaption{CO Likelihood Parameters \label{tbl:likeparams}}
\tablehead{
\colhead{FTS Name}  & \colhead{Linewidth} & \colhead{Length Max} & \colhead{Mass Max} \\
\colhead{        }  & \colhead{[\kms]}    & \colhead{[pc]}       & \colhead{[$10^{8}$ \ms]}
}
\startdata
Mrk 231              &    198 &   1415 &     65\\
IRAS F17207-0014     &    373 &   5384 &    869\\
IRAS 09022-3615      &    547 &   5650 &   1961\\
Arp 220              &    428 &   4739 &   1008\\
Mrk 273              &    265 &   5090 &    417\\
UGC 05101            &    350 &   3955 &    562\\
NGC 6240             &    370 &   8770 &   1399\\
Arp 299-C            &     80 &   6351 &     47\\
Arp 299-B            &    155 &   6353 &    177\\
Arp 299-A            &    282 &   6349 &    586\\
NGC 1068             &    254 &   3454 &    258\\
NGC 1365-SW          &    250 &   2567 &    136\\
NGC 1365-NE          &    250 &   2564 &    135\\
NGC 4038             &    133 &   4900 &    101\\
NGC 4038 (Overlap)   &    166 &   5922 &    189\\
M82                  &    174 &    850 &     30\\
NGC 1222             &     80 &   2703 &     20\\
M83                  &    102 &   1169 &     14\\
NGC 253              &    220 &    402 &     23\\
NGC 1266             &    239 &    576 &     38\\
Cen A                &    150 &   1998 &     52
\enddata
\tablecomments{See Section \ref{sec:copriors}.  The CO likelihood also depends on the luminosity distance, given in Table \ref{tbl:obs}, and the beam area.  All are normalized to $\Omega_b = 5.04 \times 10^{-8}$ sr, see Section \ref{sec:sourcebeam}. Linewidths are the average of those reported from references in Table \ref{tbl:coground}. No linewidths were available for NGC~1365-NE, NGC~1365-SW, NGC~1222, and IRAS~09022-3615; the $v_{rot}$ from Hyperleda was used instead for the first three, and the [OIII] linewidth was used for the last from \citet{ChulLee2011}.}
\end{deluxetable}

RADEX calculates antenna temperatures in K, so we
 converted our data from Jy km/s to K km/s by multiplying by 646$\nu^{-2}_{\unit{GHz}}$ (assuming
the aforementioned 43\farcs5 beam). We then divided our integrated flux data into per-unit-linewidth 
units of temperature (K instead of K km/s) to directly compare to RADEX, assuming fixed linewidths.
This is because the actual parameter we were testing was column density per unit linewidth; the modeled emission was simply 
 multiplied by linewidth.  Because the FTS did not resolve the widths of the lines, we relied on ground-based CO measurements 
for linewidths.  The values used are in Table \ref{tbl:likeparams}; many are medians of multiple CO 
linewidths from the literature, if more than one measurement was available.  Our reported masses and luminosities 
scale linearly with linewidth.

Given the size of the eight-dimensional parameter space, it was important to apply some physical constraints to 
limit the potential solutions to those that are physically meaningful.  Some of these constraints were on 
the relationship between the two components: we required the first component to be cooler 
than the second (henceforth referred to as the ``cool/cold" and ``warm" components), and the cool component 
to contain more mass than the warm.  

Two additional priors were used based on known physical constraints: 
the sum of the two components' mass could not exceed the dynamical mass of the 
galaxy, nor could either component's line-of-sight length be greater than the extent of the galaxy in the plane of the sky.  
The mass as a function of our parameters is that given in Equation \ref{eqn:mass}; the dynamic mass 
sets a limit on the product of $\Phi_A \times N_{CO}$.  The length as a function of 
our parameters is $N_{CO} (\sqrt{\Phi_A} n_{H2} X_{^{12}CO})^{-1}$.  

The dynamical mass and length limits are presented in Table \ref{tbl:likeparams}.  
The length upper limits were calculated by fitting a two dimensional Gaussian to the SPIRE PSW maps.  This Gaussian 
was the convolution of the intrinsic source size and the PSW beam (FWHM = 19\farcs3).  We used the longest length 
of the Gaussian ($g$) and found the source size, $s = \sqrt{g^2 - 19.3^2}$.  
We utilized the largest-side-of-a-Gaussian approximation because we sought only an upper limit.
The dynamical mass limit was determined from the linewidth ($\Delta V$) and maximum 
length limit $L$, such that $M_{max} = \Delta V^2 L/G$.  If any of these constraints were violated, the likelihood for that set 
of parameters was not included.  

Furthermore, any one line was not counted in the likelihood if its modeled 
optical depth was less than -0.9 or greater than 100.  
The upper limit of 100 was because the concept of a one-zone model breaks down at this point.
The line center optical depths are so large that the measured excitation temperatures will vary strongly 
across the line profile, causing self-absorption.  In other words, the 
escape probability method becomes invalid. We allowed a slightly negative lower limit because we found that, even 
given normal ISM conditions, the lowest population levels may be slightly inverted (resulting in negative optical depth).
Again, the escape probability method can no longer be used with a strong maser.  

\begin{deluxetable*}{l rrr | rrr | rrr | rrr}
\tabletypesize{\scriptsize}
\tablecaption{CO Fitting Results: Model Parameters \label{tbl:co1}}
\tablehead{
\colhead{FTS Name}  & \multicolumn{3}{c}{Log $n_{H_2}$ [cm$^{-3}$]} & \multicolumn{3}{c}{Log T$_{kin}$ [K]} & \multicolumn{3}{c}{Log $N_{CO}$ [cm$^{-2}$]} & \multicolumn{3}{c}{Log $\Phi_A$} \\
\colhead{}    & \colhead{Mean} & \colhead{$\sigma$} & \colhead{Best}  & \colhead{Mean} & \colhead{$\sigma$} & \colhead{Best}  & \colhead{Mean} & \colhead{$\sigma$} & \colhead{Best}  & \colhead{Mean} & \colhead{$\sigma$} & \colhead{Best} 
}
\startdata
\multicolumn{13}{l}{\bf{Cool Component}} \\
Mrk 231              &     3.6 &     1.3 &     5.4 &     1.4 &     0.6 &     0.8 &    17.4 &     0.8 &    16.4 &    -1.3 &     0.7 &    -0.3\\
IRAS F17207-0014     &     4.4 &     1.2 &     4.1 &     1.2 &     0.4 &     1.1 &    19.5 &     0.5 &    20.1 &    -2.6 &     0.3 &    -2.7\\

Arp 220              &     3.4 &     1.1 &     2.6 &     1.7 &     0.6 &     1.8 &    19.1 &     1.0 &    20.3 &    -2.0 &     0.7 &    -2.8\\
Mrk 273              &     3.5 &     1.3 &     5.4 &     1.5 &     0.6 &     0.8 &    18.4 &     1.1 &    18.8 &    -2.0 &     0.8 &    -2.2\\
UGC 05101            &     3.6 &     1.2 &     5.6 &     1.4 &     0.6 &     1.1 &    18.0 &     1.1 &    19.3 &    -1.9 &     0.9 &    -2.8\\
NGC 6240             &     4.3 &     1.0 &     5.8 &     1.7 &     0.5 &     1.2 &    17.8 &     0.8 &    18.0 &    -1.2 &     0.7 &    -1.4\\
Arp 299-C            &     3.6 &     1.3 &     5.9 &     1.5 &     0.6 &     1.1 &    19.0 &     0.5 &    18.9 &    -2.0 &     0.5 &    -1.7\\
Arp 299-B            &     3.6 &     1.3 &     5.8 &     1.1 &     0.5 &     1.1 &    18.7 &     0.8 &    19.4 &    -1.7 &     0.7 &    -2.1\\
Arp 299-A            &     3.2 &     1.1 &     3.8 &     1.2 &     0.5 &     0.5 &    18.2 &     0.9 &    17.6 &    -1.0 &     0.6 &    -0.3\\
NGC 1068             &     2.7 &     0.3 &     2.7 &     2.4 &     0.3 &     2.7 &    18.7 &     0.8 &    18.3 &    -1.0 &     0.5 &    -0.9\\
NGC 1365-SW          &     3.4 &     1.2 &     2.4 &     1.3 &     0.6 &     1.9 &    18.6 &     0.7 &    18.0 &    -0.9 &     0.5 &    -0.7\\
NGC 1365-NE          &     2.3 &     0.3 &     2.6 &     1.5 &     0.2 &     1.5 &    20.2 &     0.1 &    20.3 &    -1.8 &     0.1 &    -2.0\\
NGC 4038             &     2.6 &     0.9 &     2.1 &     2.1 &     0.5 &     2.6 &    19.0 &     0.5 &    19.1 &    -1.8 &     0.3 &    -1.8\\
NGC 4038 (Overlap)   &     3.5 &     1.2 &     3.8 &     1.6 &     0.7 &     0.9 &    18.7 &     0.6 &    19.4 &    -1.5 &     0.5 &    -1.4\\
M82                  &     3.4 &     1.0 &     2.8 &     1.8 &     0.6 &     2.4 &    18.7 &     0.7 &    18.2 &    -0.6 &     0.5 &    -0.3\\
NGC 1222             &     3.6 &     1.3 &     5.5 &     1.2 &     0.5 &     1.0 &    17.9 &     0.9 &    19.4 &    -1.4 &     0.7 &    -2.0\\
M83                  &     3.8 &     1.3 &     2.9 &     1.3 &     0.7 &     0.8 &    19.2 &     0.5 &    19.4 &    -1.7 &     0.5 &    -0.9\\
NGC 253              &     3.5 &     1.2 &     2.6 &     1.6 &     0.7 &     2.1 &    19.6 &     0.6 &    19.9 &    -1.2 &     0.5 &    -1.3\\
NGC 1266             &     3.4 &     1.2 &     2.2 &     1.5 &     0.6 &     2.2 &    18.5 &     1.0 &    19.3 &    -1.8 &     0.7 &    -2.5\\
Cen A                &     3.0 &     1.1 &     2.9 &     1.1 &     0.5 &     0.8 &    18.2 &     0.6 &    18.1 &    -0.5 &     0.4 &    -0.5\\
\hline
\multicolumn{13}{l}{\bf{Warm Component}} \\
Mrk 231              &     3.7 &     0.1 &     3.7 &     3.3 &     0.1 &     3.4 &    16.8 &     0.9 &    15.9 &    -1.5 &     0.9 &    -0.6\\
IRAS F17207-0014     &     4.9 &     0.5 &     4.7 &     2.4 &     0.2 &     2.5 &    16.7 &     0.9 &    17.4 &    -1.5 &     0.8 &    -2.2\\
IRAS 09022-3615      &     4.5 &     0.3 &     4.5 &     2.8 &     0.2 &     2.8 &    16.5 &     0.9 &    17.4 &    -1.5 &     0.9 &    -2.5\\
Arp 220              &     4.1 &     0.6 &     4.9 &     2.8 &     0.3 &     2.4 &    17.4 &     0.9 &    16.5 &    -1.5 &     0.8 &    -0.9\\
Mrk 273              &     3.8 &     0.2 &     3.7 &     3.3 &     0.1 &     3.4 &    16.7 &     0.9 &    15.2 &    -1.5 &     0.8 &    -0.0\\
UGC 05101            &     3.3 &     0.4 &     3.6 &     3.2 &     0.2 &     3.4 &    16.9 &     0.9 &    15.9 &    -1.5 &     0.9 &    -0.9\\
NGC 6240             &     4.1 &     0.3 &     4.1 &     3.0 &     0.2 &     3.1 &    17.4 &     0.9 &    16.3 &    -1.5 &     0.9 &    -0.4\\
Arp 299-C            &     2.3 &     0.2 &     2.1 &     3.3 &     0.1 &     3.4 &    18.7 &     1.0 &    19.7 &    -2.1 &     0.9 &    -3.0\\
Arp 299-B            &     2.5 &     0.3 &     2.7 &     3.4 &     0.1 &     3.5 &    17.9 &     0.9 &    18.2 &    -1.5 &     0.7 &    -2.0\\
Arp 299-A            &     3.3 &     0.2 &     3.1 &     3.2 &     0.1 &     3.2 &    17.5 &     0.8 &    18.0 &    -1.3 &     0.8 &    -1.7\\
NGC 1068             &     4.8 &     0.5 &     4.9 &     2.9 &     0.3 &     2.8 &    17.2 &     0.9 &    17.4 &    -1.5 &     0.9 &    -1.8\\
NGC 1365-SW          &     3.2 &     0.6 &     3.7 &     2.8 &     0.2 &     2.6 &    18.3 &     0.9 &    17.9 &    -1.5 &     0.7 &    -1.4\\
NGC 1365-NE          &     3.1 &     0.7 &     2.1 &     2.7 &     0.1 &     2.8 &    18.9 &     1.1 &    20.4 &    -2.1 &     0.8 &    -2.9\\
NGC 4038             &     4.6 &     1.2 &     5.5 &     3.0 &     0.3 &     3.0 &    16.3 &     1.2 &    15.0 &    -1.5 &     0.8 &    -0.9\\
NGC 4038 (Overlap)   &     3.3 &     0.9 &     3.1 &     2.9 &     0.2 &     2.9 &    17.4 &     1.1 &    18.4 &    -1.4 &     0.7 &    -2.1\\
M82                  &     3.9 &     0.4 &     4.1 &     2.9 &     0.2 &     2.8 &    18.2 &     0.8 &    18.2 &    -1.1 &     0.6 &    -1.3\\
NGC 1222             &     4.2 &     0.5 &     4.3 &     2.4 &     0.2 &     2.4 &    16.7 &     0.9 &    15.7 &    -1.5 &     0.9 &    -0.7\\
M83                  &     2.9 &     0.3 &     2.0 &     2.9 &     0.2 &     2.9 &    18.8 &     0.5 &    19.7 &    -1.9 &     0.5 &    -2.0\\
NGC 253              &     3.1 &     0.3 &     3.5 &     3.2 &     0.2 &     3.0 &    19.2 &     0.8 &    19.5 &    -1.7 &     0.8 &    -2.3\\
NGC 1266             &     3.6 &     0.3 &     3.7 &     3.2 &     0.1 &     3.2 &    17.1 &     0.8 &    16.0 &    -1.4 &     0.8 &    -0.5\\
Cen A                &     2.6 &     0.4 &     2.2 &     3.1 &     0.2 &     3.3 &    18.5 &     0.8 &    19.0 &    -1.8 &     0.6 &    -2.1
\enddata
\end{deluxetable*}
\begin{deluxetable*}{l rrr | rrr | rrr | rrr }
\tabletypesize{\scriptsize}
\tablecaption{CO Fitting Results: Derived Parameters \label{tbl:co2}}
\tablehead{
\colhead{FTS Name}  & \multicolumn{3}{c}{Log $L_{\rm CO}$ [erg s$^{-1}$]} & \multicolumn{3}{c}{Log $P$ [K cm$^{-3}$]} & \multicolumn{3}{c}{Log $<N_{\rm CO}>$ [cm$^{-2}$]}   & \multicolumn{3}{c}{Log \mhtwo\ [\ms]}\\
\colhead{}    & \colhead{Mean} & \colhead{$\sigma$} & \colhead{Best}  & \colhead{Mean} & \colhead{$\sigma$} & \colhead{Best}  & \colhead{Mean} & \colhead{$\sigma$} & \colhead{Best}  & \colhead{Mean} & \colhead{$\sigma$} & \colhead{Best} 
}
\startdata
\multicolumn{13}{l}{\bf{Cool Component}} \\
Mrk 231              &    40.6 &     0.2 &    40.5 &     5.0 &     0.8 &     6.1 &    16.1 &     0.2 &    16.1 &     9.3 &     0.2 &     9.3\\
IRAS F17207-0014     &    41.2 &     0.2 &    41.2 &     5.6 &     1.0 &     5.2 &    16.9 &     0.4 &    17.3 &    10.1 &     0.4 &    10.5\\

Arp 220              &    41.0 &     0.4 &    41.3 &     5.0 &     0.8 &     4.4 &    17.1 &     0.4 &    17.5 &     9.6 &     0.4 &    10.0\\
Mrk 273              &    40.7 &     0.3 &    40.5 &     5.0 &     0.9 &     6.2 &    16.4 &     0.4 &    16.6 &     9.5 &     0.4 &     9.7\\
UGC 05101            &    40.5 &     0.4 &    41.0 &     5.0 &     1.0 &     6.6 &    16.2 &     0.4 &    16.5 &     9.3 &     0.4 &     9.6\\
NGC 6240             &    41.6 &     0.2 &    41.6 &     6.0 &     0.7 &     7.0 &    16.6 &     0.1 &    16.6 &     9.3 &     0.1 &     9.3\\
Arp 299-C            &    40.2 &     0.6 &    40.7 &     5.1 &     1.1 &     7.0 &    17.0 &     0.3 &    17.2 &     9.0 &     0.3 &     9.2\\
Arp 299-B            &    39.6 &     0.6 &    40.3 &     4.7 &     1.2 &     6.8 &    17.0 &     0.4 &    17.3 &     9.0 &     0.4 &     9.3\\
Arp 299-A            &    39.9 &     0.3 &    39.7 &     4.3 &     1.0 &     4.4 &    17.2 &     0.4 &    17.3 &     9.2 &     0.4 &     9.3\\
NGC 1068             &    40.8 &     0.1 &    41.0 &     5.0 &     0.4 &     5.4 &    17.6 &     0.3 &    17.4 &     8.8 &     0.3 &     8.5\\
NGC 1365-SW          &    40.0 &     0.4 &    40.2 &     4.6 &     1.0 &     4.3 &    17.6 &     0.4 &    17.3 &     9.1 &     0.4 &     8.8\\
NGC 1365-NE          &    40.4 &     0.2 &    40.5 &     3.8 &     0.2 &     4.0 &    18.4 &     0.1 &    18.3 &     9.9 &     0.1 &     9.8\\
NGC 4038             &    40.3 &     0.3 &    40.5 &     4.8 &     0.6 &     4.8 &    17.2 &     0.3 &    17.2 &     8.7 &     0.3 &     8.7\\
NGC 4038 (Overlap)   &    40.0 &     0.6 &    39.8 &     5.1 &     0.9 &     4.7 &    17.3 &     0.4 &    18.0 &     8.7 &     0.4 &     9.5\\
M82                  &    39.6 &     0.5 &    39.9 &     5.2 &     0.7 &     5.3 &    18.1 &     0.3 &    17.9 &     7.9 &     0.3 &     7.7\\
NGC 1222             &    39.2 &     0.3 &    39.7 &     4.7 &     1.0 &     6.5 &    16.5 &     0.4 &    17.4 &     8.3 &     0.4 &     9.2\\
M83                  &    38.6 &     0.9 &    38.7 &     5.1 &     1.1 &     3.7 &    17.6 &     0.5 &    18.6 &     8.0 &     0.5 &     9.1\\
NGC 253              &    39.3 &     1.0 &    40.0 &     5.0 &     0.9 &     4.6 &    18.5 &     0.4 &    18.6 &     8.3 &     0.4 &     8.4\\
NGC 1266             &    39.6 &     0.4 &    39.9 &     4.9 &     0.9 &     4.4 &    16.7 &     0.4 &    16.8 &     8.4 &     0.4 &     8.5\\
Cen A                &    38.7 &     0.3 &    38.5 &     4.1 &     0.9 &     3.7 &    17.7 &     0.5 &    17.6 &     8.2 &     0.5 &     8.2\\
\hline
\multicolumn{13}{l}{\bf{Warm Component}} \\
Mrk 231              &   42.59 &    0.06 &   42.66 &     7.0 &     0.1 &     7.1 &    15.3 &     0.1 &    15.3 &     8.4 &     0.1 &     8.5\\
IRAS F17207-0014     &   42.37 &    0.03 &   42.37 &     7.3 &     0.3 &     7.1 &    15.2 &     0.1 &    15.2 &     8.4 &     0.1 &     8.4\\
IRAS 09022-3615      &   42.63 &    0.04 &   42.62 &     7.3 &     0.1 &     7.3 &    14.9 &     0.1 &    14.9 &     8.4 &     0.1 &     8.4\\
Arp 220              &   42.23 &    0.07 &   42.16 &     6.9 &     0.3 &     7.3 &    15.9 &     0.2 &    15.6 &     8.4 &     0.2 &     8.1\\
Mrk 273              &   42.41 &    0.07 &   42.48 &     7.1 &     0.1 &     7.1 &    15.1 &     0.1 &    15.2 &     8.2 &     0.1 &     8.3\\
UGC 05101            &   42.10 &    0.06 &   42.23 &     6.5 &     0.3 &     7.0 &    15.4 &     0.3 &    15.0 &     8.5 &     0.3 &     8.2\\
NGC 6240             &   42.86 &    0.06 &   42.88 &     7.2 &     0.1 &     7.2 &    15.9 &     0.1 &    15.9 &     8.6 &     0.1 &     8.6\\
Arp 299-C            &   41.41 &    0.08 &   41.36 &     5.6 &     0.2 &     5.4 &    16.6 &     0.2 &    16.8 &     8.6 &     0.2 &     8.8\\
Arp 299-B            &   41.49 &    0.04 &   41.51 &     5.9 &     0.3 &     6.1 &    16.4 &     0.2 &    16.2 &     8.4 &     0.2 &     8.2\\
Arp 299-A            &   41.86 &    0.04 &   41.87 &     6.5 &     0.2 &     6.4 &    16.2 &     0.1 &    16.3 &     8.2 &     0.1 &     8.4\\
NGC 1068             &   41.38 &    0.08 &   41.33 &     7.7 &     0.2 &     7.8 &    15.7 &     0.1 &    15.6 &     6.8 &     0.1 &     6.7\\
NGC 1365-SW          &   41.18 &    0.07 &   41.19 &     5.9 &     0.4 &     6.3 &    16.7 &     0.4 &    16.4 &     8.2 &     0.4 &     7.9\\
NGC 1365-NE          &   41.10 &    0.07 &   40.97 &     5.8 &     0.6 &     4.9 &    16.8 &     0.5 &    17.5 &     8.3 &     0.5 &     9.0\\
NGC 4038             &   40.62 &    0.37 &   40.79 &     7.5 &     1.1 &     8.5 &    14.8 &     0.8 &    14.1 &     6.2 &     0.8 &     5.6\\
NGC 4038 (Overlap)   &   40.69 &    0.44 &   40.79 &     6.2 &     0.9 &     6.0 &    15.9 &     1.0 &    16.3 &     7.4 &     1.0 &     7.8\\
M82                  &   40.63 &    0.05 &   40.61 &     6.8 &     0.3 &     7.0 &    17.1 &     0.3 &    16.8 &     6.9 &     0.3 &     6.6\\
NGC 1222             &   40.44 &    0.11 &   40.43 &     6.6 &     0.3 &     6.8 &    15.1 &     0.3 &    14.9 &     7.0 &     0.3 &     6.8\\
M83                  &   40.19 &    0.09 &   40.17 &     5.8 &     0.3 &     5.0 &    16.9 &     0.3 &    17.7 &     7.3 &     0.3 &     8.1\\
NGC 253              &   40.78 &    0.12 &   40.73 &     6.3 &     0.2 &     6.6 &    17.5 &     0.2 &    17.3 &     7.4 &     0.2 &     7.1\\
NGC 1266             &   41.25 &    0.05 &   41.22 &     6.8 &     0.2 &     6.9 &    15.6 &     0.2 &    15.5 &     7.4 &     0.2 &     7.2\\
Cen A                &   40.02 &    0.05 &   40.08 &     5.7 &     0.3 &     5.5 &    16.7 &     0.3 &    16.9 &     7.2 &     0.3 &     7.4
\enddata
\tablecomments{Luminosities and masses may be lower limits for significantly extended galaxies; all luminosities and masses are those contained in a 43\farcs{5} beam.}
\end{deluxetable*}

\begin{figure*}
\centering
\includegraphics[width=0.7\textwidth]{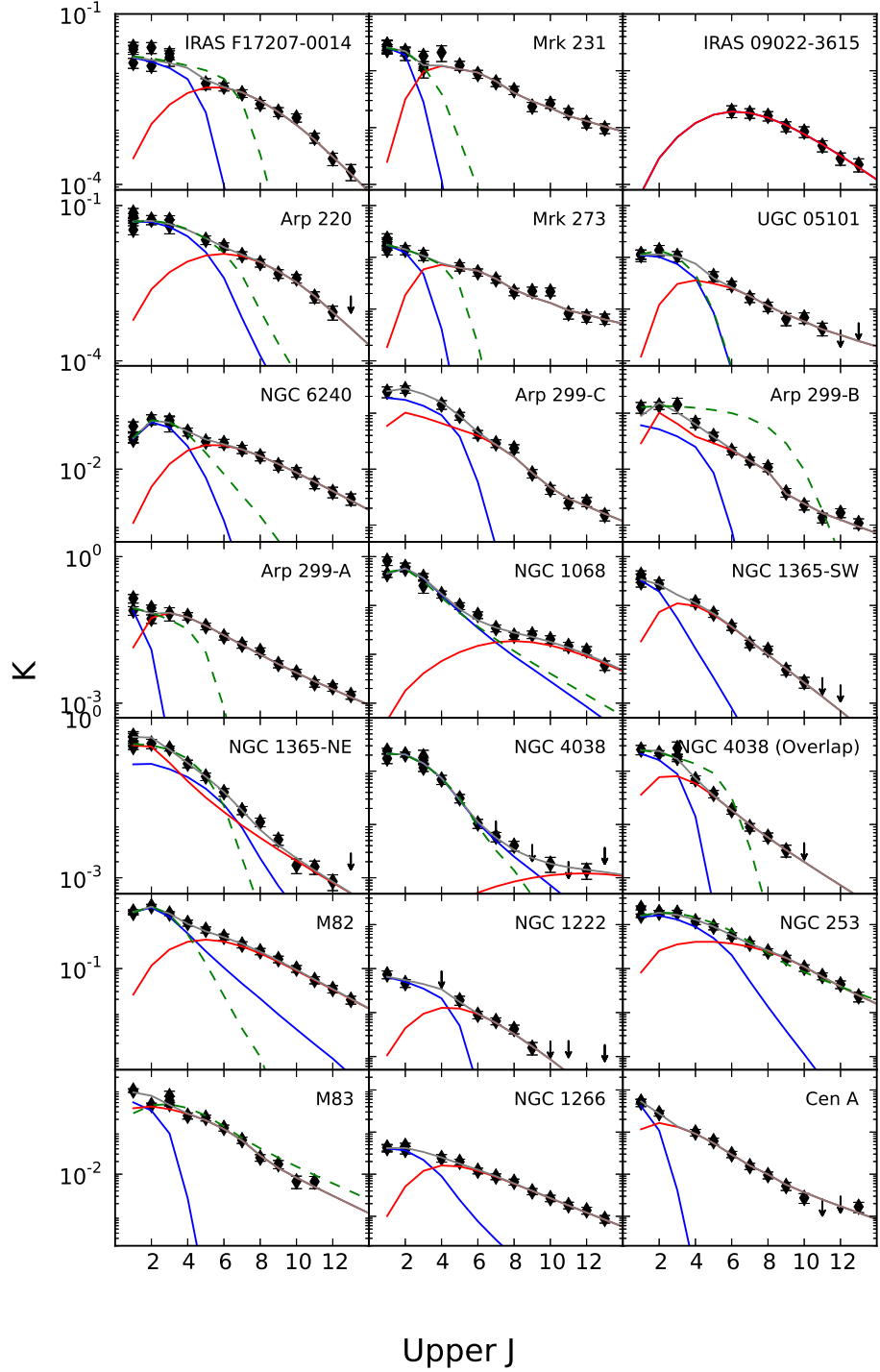} 
\caption[CO Modeling Spectral Line Energy Distributions]{CO modeling spectral line energy distributions.  The blue is the best-fit solution for the cold component, 
the red the best-fit solution for the warm, with their total in grey.  Those galaxies that contain 
all three of the lowest-J transitions also have a green line (dashed), which is the fit to only those three lines (see Section \ref{sec:disc:twocomp1}).
\label{fig:cosleds}}
\end{figure*}

\begin{figure*}
\centering
\includegraphics[width=0.7\textwidth]{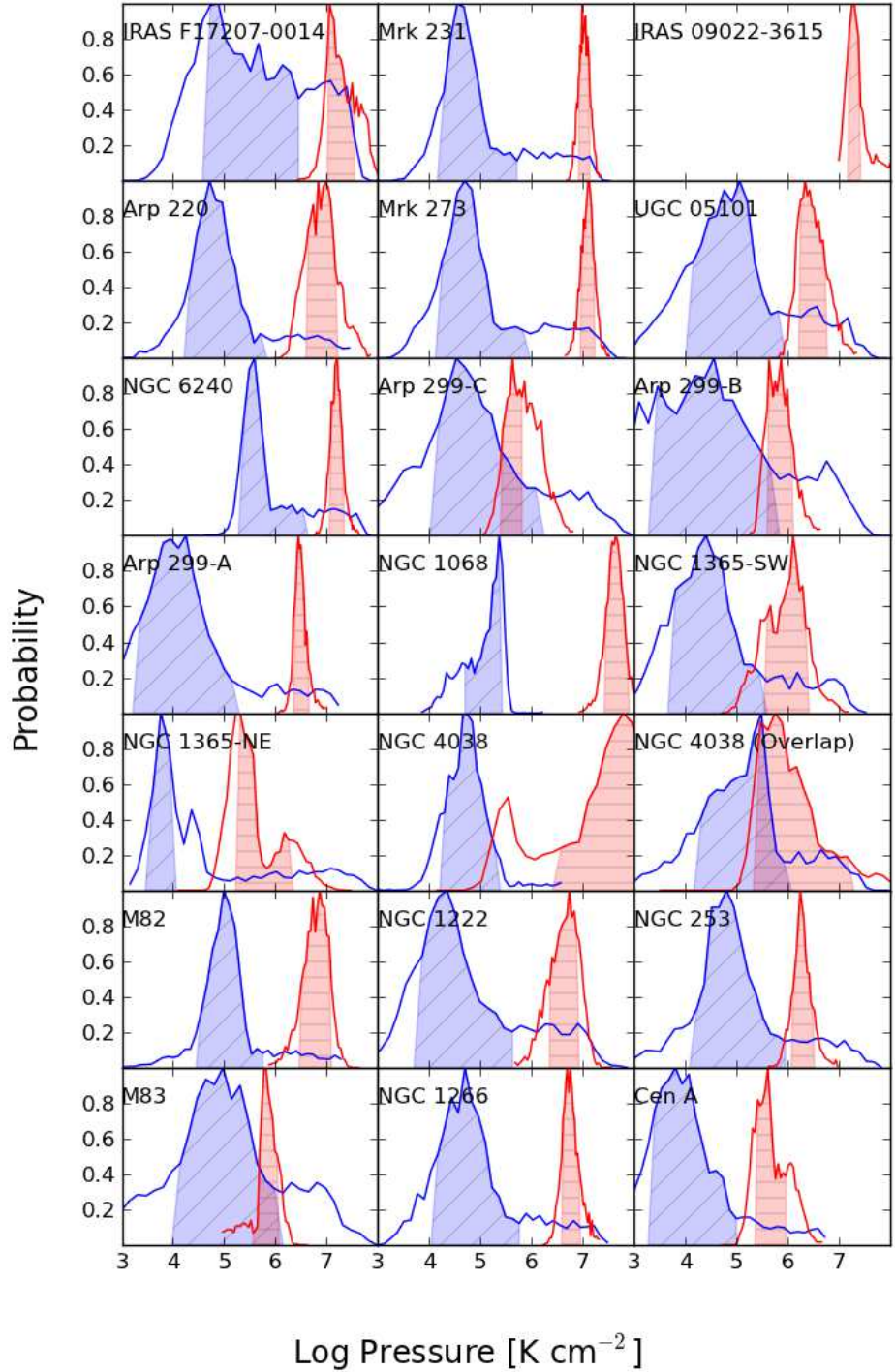}
\caption[CO modeling likelihood results: pressure]{CO modeling likelihood results: pressure.  Blue/red (diagonal/horizontal hatches) represent cool/warm components.  Shaded areas indicate the 1$\sigma$ uncertainty region, defined here as the mode median +/- the error (symmetric) on the parameter.  Vertical lines represent the best-fit model parameter.  IRAS~F17207-0014 is an example of a galaxy with best-fit parameter values close to the median, whereas UGC~05101 is an example where the best-fit and median values do not align.\label{fig:copress}}
\end{figure*}

\begin{figure*}
\centering
\includegraphics[width=0.7\textwidth]{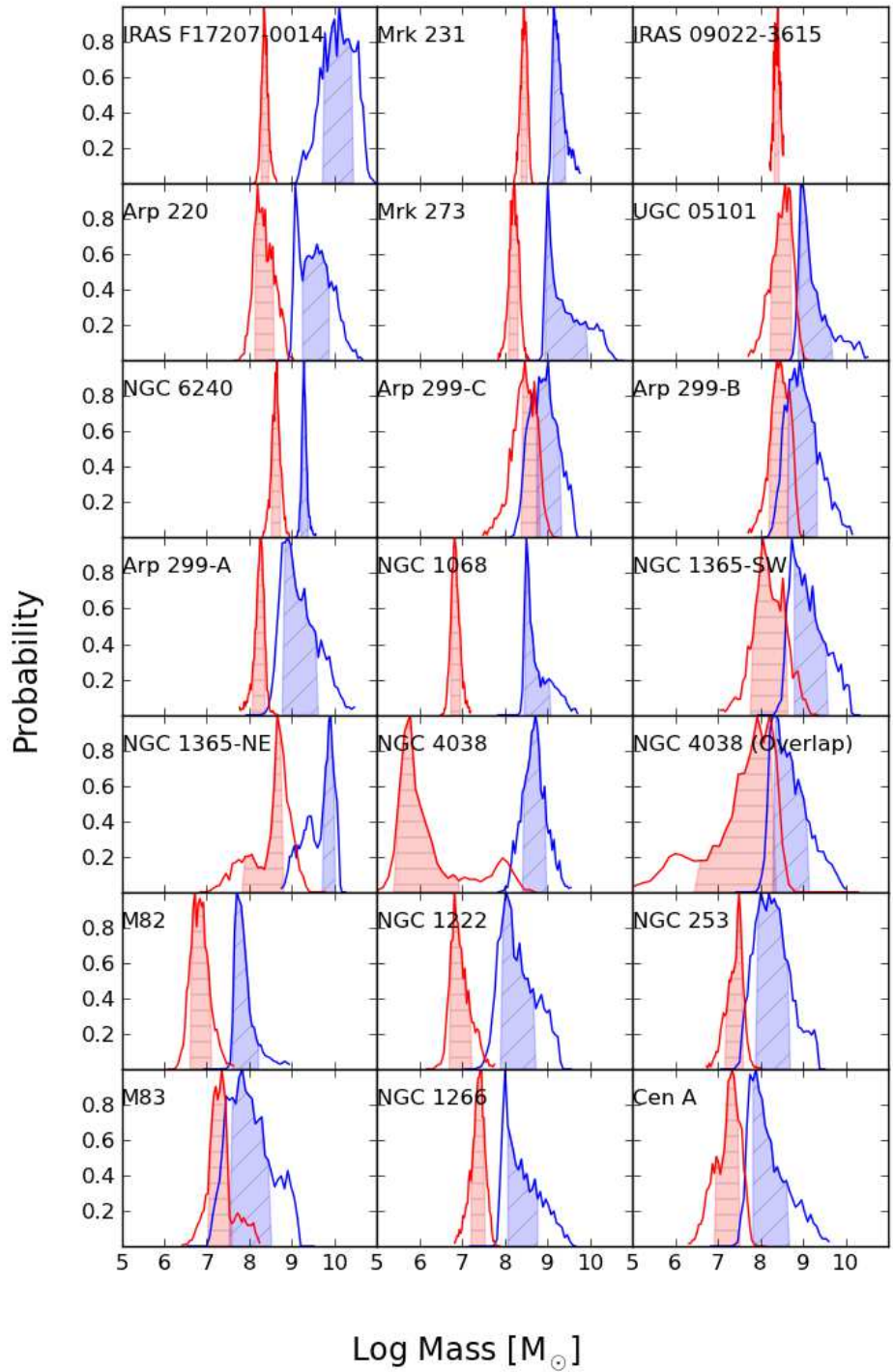}
\caption[CO modeling likelihood results: mass]{CO modeling likelihood results: mass.  See Figure \ref{fig:copress} for more information.\label{fig:comass}}
\end{figure*}

\begin{figure*}
\centering
\includegraphics[width=0.7\textwidth]{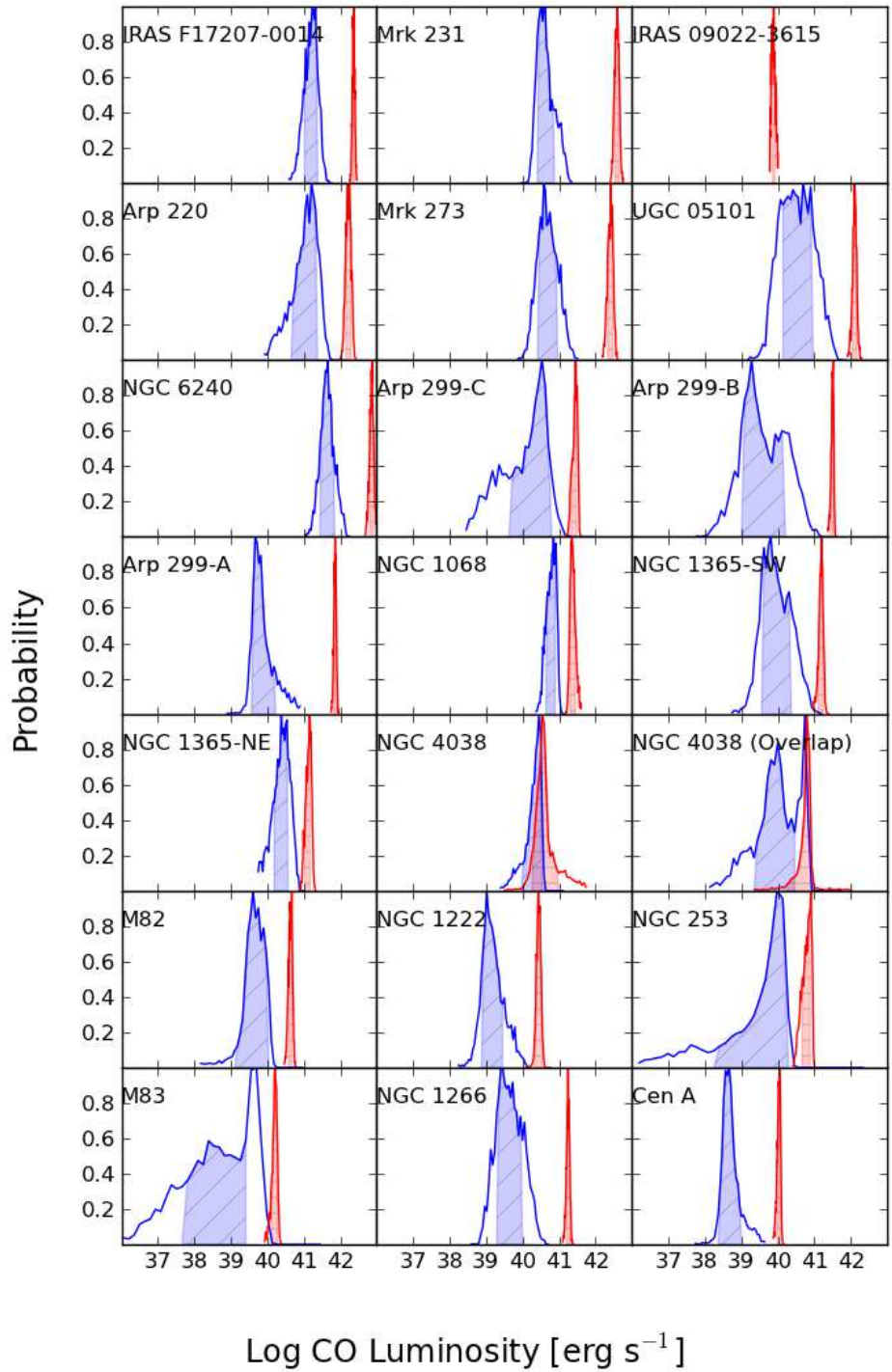}
\caption[CO modeling likelihood results: luminosity]{CO modeling likelihood results: luminosity.  See Figure \ref{fig:copress} for more information.\label{fig:colum}}
\end{figure*}

\begin{figure} 
\includegraphics[height=\columnwidth,angle=270]{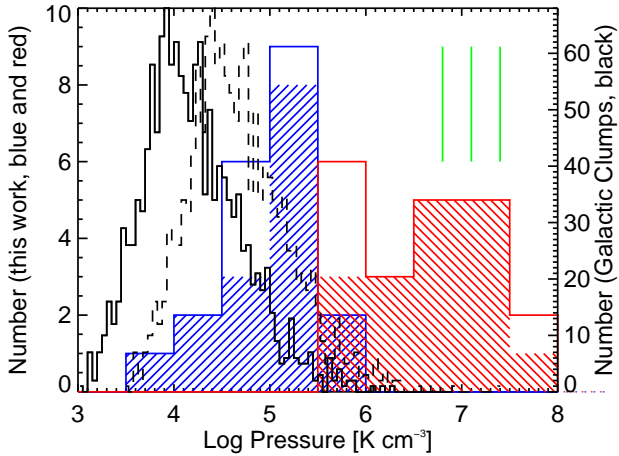}
\caption[CO modeling histograms for pressure]{CO modeling histograms for pressure.  Cold component is in blue (upward slant), warm component in red 
(downward slant).  Duplicate pointings of galaxies are not filled in by diagonal lines.  
The black solid (dashed) histogram is the pressure of galactic molecular clumps at a temperature of 10 (30) K, 
based on densities determined by the Bolocam Galactic Plane Survey (BGPS, Ellsworth-Bowers et al., in prep). 
The temperature range is chosen from the 20 K mean gas kinetic temperature found for BGPS sources in \citet{Dunham2010}.
The y-axis on the left is the number for the distributions in this work, and the right axis is the number for the Galactic clumps. 
The vertical green lines indicate the pressures, from left to right, for the Sgr B2(N), Sgr A*, and Sgr B2(M) warm extended emission, 
discussed further in Section \ref{sec:disc:galactic} \citep{Etxaluze2013,Goicoechea2013}.  
\label{fig:cohist1}}
\end{figure}

\begin{figure*}  
\plotone{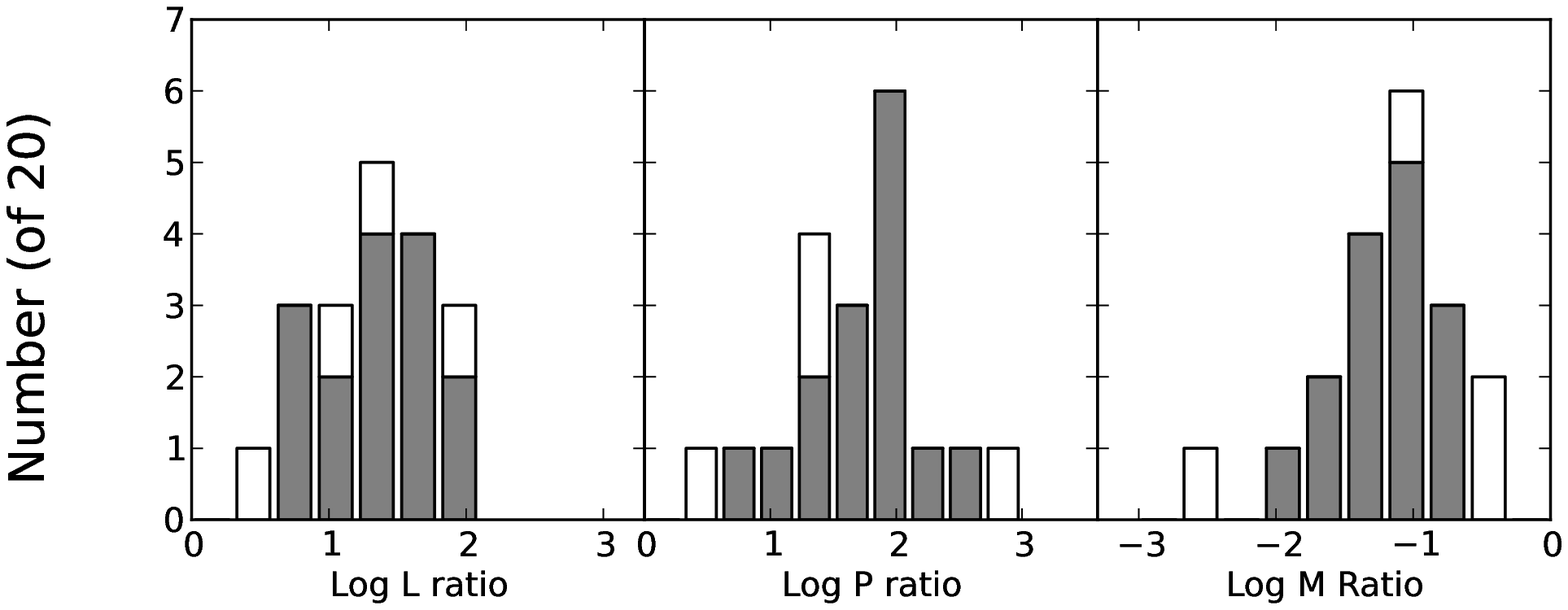} 
\caption[CO modeling histograms for derived parameter ratios]{CO modeling histograms for derived warm/cold component parameter ratios.  Duplicate pointings of galaxies are shown in white (one pointing per galaxy remains gray) and not used for fitting average values. 
\label{fig:cohist2}}
\end{figure*}

The results are presented in Figures \ref{fig:cosleds}, \ref{fig:copress}, \ref{fig:comass}, \ref{fig:colum}, 
\ref{fig:cohist1}, \ref{fig:cohist2} and Tables \ref{tbl:co1}, \ref{tbl:co2}, \ref{tbl:co3}.  
As with the dust modeling results in the previous section, we present the mode mean, mode sigma, and best-fit results 
(recall Section \ref{sec:analysis} for these terms).  However, the likelihood distributions themselves (in the accompanying figures) are not as simply described as in the dust modeling case.  In many instances, the best fit result (and the associated SLED illustrated in \ref{fig:cosleds}) does not correspond to the mean or mode of a marginalized parameter distribution.  Additionally, the results for some galaxies included contributions from multiple modes, which can be thought of as separate islands in parameter space.  The mode mean and sigma presented here are for that of the mode containing the highest integrated likelihood; this mode had an obvious distinction as the far more likely mode than any other ones found.

\subsubsection{Effects of Extinction Correction and Modeling Priors}\label{sec:disc:priors}

In most cases, our use of extinction correction and mass and length priors did not significantly impact the results of the 
CO modeling, especially in the determination of the molecular mass, which will become important in the discussion of the results.

The extinction correction affected the high-J CO lines the most, increasing 
their fluxes more than low-J lines and thus affecting the shapes of the SLEDs.  
When we use non-extinction corrected fluxes in modeling, we found that the distribution of parameters were not statistically 
significantly different given the errors bars.  The most 
noticeable difference was the warm pressure, which was only 0.1 dex higher when dust correction is included 
(this is expected because overall shape of the SLED is determined by the pressure).
The CO luminosity generally peaks at mid-J lines not as strongly affected by extinction (e.g. \jsix).

Of the maximum mass and length priors, the length turned out to limit only a very small section of the allowed parameter space.
The dynamical mass was the more restrictive prior.
When modeled without the maximum mass and length priors, the marginalized parameter likelihoods for most of our galaxies were  
indistinguishable from modeling with the priors.  This means the priors did not have a significant effect (i.e. the combinations of 
parameters violating these priors did not contribute significantly to the likelihood anyway due to poor match to the data).
Only a few showed an appreciable difference (M82, Arp~220, and NGC~253), generally by allowing an unconstrained high-mass shoulder, 
but not changing the location of the main likelihood peak.

\begin{deluxetable*}{l rrr | rrr | rrr}
\tabletypesize{\scriptsize}
\tablecaption{CO Fitting Results: Derived Parameter Ratios\label{tbl:co3}}
\tablehead{
\colhead{FTS Name}  & \multicolumn{3}{c}{Log $L_{warm}/L_{cool}$} & \multicolumn{3}{c}{Log $P_{warm}/P_{cool}$} & \multicolumn{3}{c}{Log $M_{warm}/M_{cool}$}\\
\colhead{}    & \colhead{Mean} & \colhead{$\sigma$} & \colhead{Best}  & \colhead{Mean} & \colhead{$\sigma$} & \colhead{Best}  & \colhead{Mean} & \colhead{$\sigma$} & \colhead{Best} 
}
\startdata
Mrk 231              &     1.9 &     0.2 &     2.1 &     2.0 &     0.9 &     1.0 &    -0.8 &     0.2 &    -0.8\\
IRAS F17207-0014     &     1.2 &     0.2 &     1.1 &     1.7 &     1.0 &     2.0 &    -1.7 &     0.4 &    -2.1\\

Arp 220              &     1.2 &     0.4 &     0.9 &     1.9 &     0.9 &     2.9 &    -1.2 &     0.5 &    -1.9\\
Mrk 273              &     1.7 &     0.3 &     2.0 &     2.0 &     0.9 &     0.9 &    -1.2 &     0.5 &    -1.4\\
UGC 05101            &     1.6 &     0.5 &     1.2 &     1.5 &     0.9 &     0.4 &    -0.8 &     0.5 &    -1.5\\
NGC 6240             &     1.2 &     0.2 &     1.3 &     1.2 &     0.7 &     0.1 &    -0.7 &     0.2 &    -0.7\\
Arp 299-C            &     1.2 &     0.6 &     0.7 &     0.5 &     1.1 &    -1.6 &    -0.4 &     0.3 &    -0.5\\
Arp 299-B            &     1.9 &     0.6 &     1.2 &     1.2 &     1.2 &    -0.7 &    -0.6 &     0.4 &    -1.1\\
Arp 299-A            &     2.0 &     0.3 &     2.1 &     2.2 &     0.9 &     2.0 &    -1.0 &     0.5 &    -1.0\\
NGC 1068             &     0.6 &     0.2 &     0.4 &     2.6 &     0.5 &     2.4 &    -1.9 &     0.3 &    -1.9\\
NGC 1365-SW          &     1.2 &     0.5 &     1.0 &     1.3 &     1.0 &     1.9 &    -0.9 &     0.5 &    -0.9\\
NGC 1365-NE          &     0.7 &     0.3 &     0.4 &     2.0 &     0.6 &     0.9 &    -1.5 &     0.6 &    -0.8\\
NGC 4038             &     0.4 &     0.4 &     0.3 &     2.7 &     1.3 &     3.7 &    -2.5 &     0.8 &    -3.1\\
NGC 4038 (Overlap)   &     0.7 &     0.9 &     1.0 &     1.1 &     1.2 &     1.2 &    -1.3 &     0.9 &    -1.7\\
M82                  &     1.1 &     0.5 &     0.7 &     1.6 &     0.7 &     1.7 &    -1.1 &     0.4 &    -1.1\\
NGC 1222             &     1.3 &     0.4 &     0.7 &     1.9 &     1.0 &     0.3 &    -1.4 &     0.5 &    -2.5\\
M83                  &     1.6 &     1.0 &     1.5 &     0.7 &     1.1 &     1.3 &    -0.7 &     0.4 &    -0.9\\
NGC 253              &     1.5 &     1.1 &     0.7 &     1.3 &     1.0 &     1.9 &    -0.9 &     0.5 &    -1.4\\
NGC 1266             &     1.6 &     0.4 &     1.3 &     1.9 &     0.9 &     2.5 &    -1.1 &     0.5 &    -1.3\\
Cen A                &     1.3 &     0.3 &     1.6 &     1.5 &     0.9 &     1.8 &    -1.0 &     0.5 &    -0.7\\
Weighted Average  &     1.2 &     0.1 &  &     1.8 &     0.2 &  &    -0.9 &     0.1 & 
\enddata
\tablecomments{IRAS~09022-3615 is not included because only one component was modeled.}
\end{deluxetable*}

\subsection{LTE Analysis of {\rm \ci}}\label{sec:CI}

\begin{deluxetable}{lrrrr}
\tabletypesize{\scriptsize}
\tablecaption{\ci\ LTE Temperatures\label{tbl:ci}}
\tablehead{
\colhead{FTS Name} & \colhead{T$_{ex}$          } & \colhead{$\sigma$        } & \colhead{M$_{\rm \ci}$} & \colhead{$\sigma$} \\
\colhead{        } & \colhead{[K]}               & \colhead{[K]}               & \colhead{[$10^4$ \ms]}    & \colhead{[$10^4$ \ms]}
}
\startdata
IRAS F17207-0014     & 24\phantom{.} & 5 & 350\phantom{.00} & 30\phantom{.00}\\
IRAS 09022-3615      & 20. & 3 & 800\phantom{.00} & 70\phantom{.00}\\
Arp 220              & 24\phantom{.} & 5 & 190\phantom{.00} & 20\phantom{.00}\\
Mrk 273              & 23\phantom{.} & 3 & 230\phantom{.00} & 10\phantom{.00}\\
NGC 6240             & 44\phantom{.} & 6 & 300\phantom{.00} & 20\phantom{.00}\\
Arp 299-C            & 28\phantom{.} & 4 & 41\phantom{.00} & 1\phantom{.00}\\
Arp 299-B            & 27\phantom{.} & 4 & 41\phantom{.00} & 2\phantom{.00}\\
Arp 299-A            & 35\phantom{.} & 6 & 52\phantom{.00} & 2\phantom{.00}\\
NGC 1068             & 28\phantom{.} & 1 & 31\phantom{.00} & 0.9\phantom{0}\\
NGC 1365-SW          & 25\phantom{.} & 1 & 27\phantom{.00} & 1\phantom{.00}\\
NGC 1365-NE          & 24\phantom{.} & 1 & 32\phantom{.00} & 1\phantom{.00}\\
NGC 4038             & 35\phantom{.} & 10 & 4.4\phantom{0} & 0.4\phantom{0}\\
NGC 4038 (Overlap)   & 23\phantom{.} & 1 & 13\phantom{.00} & 0.5\phantom{0}\\
M82                  & 36\phantom{.} & 3 & 3.9\phantom{0} & 0.1\phantom{0}\\
M83                  & 39\phantom{.} & 4 & 1.6\phantom{0} & 0.07\\
NGC 253              & 33\phantom{.} & 2 & 4.5\phantom{0} & 0.1\phantom{0}\\
NGC 1266             & 30. & 4 & 11\phantom{.00} & 0.4\phantom{0}\\
Cen A                & 50. & 6 & 0.86 & 0.05\\
Average &    26\phantom{.} &     8 & & 
\enddata
\end{deluxetable}

Two forbidden lines of neutral carbon (\ci) were present in our spectra.  The ratios of the line intensities in Jy km s$^{-1}$ 
(\jtwo/\jone) 
were used to derive excitation temperatures, equivalent to the kinetic temperatures of the \ci\ emitting gas
under the assumption of local thermodynamic equilibrium:

\begin{equation}\label{eqn:Tlte}
T_{ex} = E_{2-1} \bigg[ {\rm ln} \bigg( \frac{g_2}{g_1} \frac{A_{2-1} S_{1-0}}{A_{1-0} S_{2-1}} \bigg) \bigg] ^{-1} . 
\end{equation}

\noindent In the above equation, $A$ is the Einstein A coefficient for each line, $g$ the statistical weight of a level, $E_{2-1}$ 
the difference in energy levels in K (as we will use through the remainder of the section), 
and $S$ the flux in Jy km s$^{-1}$.  (If using the ratio in K km s$^{-1}$ or W m$^{-2}$, one must 
also include $\lambda_{1-0}/\lambda_{2-1}$ in the natural log term above.)

We corrected each flux to account for the absorption by dust as described in Section \ref{sec:coground}.  
Each integrated flux in LTE is proportional to the population level of the upper state of the line, where the total 
number of atoms or molecules in the state, $N_j$ equals $L_{j \too i}/(A_{j \too i} h \nu_{j \too i})$.  The total number of atoms or molecules can be found 
by dividing by the fraction of those in that state, where $f_i = g_i e^{-E_i / T_{ex}} / Z(T_{ex})$ and $Z(T_{ex})$ 
is the partition function, $Z(T_{ex}) = \sum_J g_J e^{-E_J/T_{ex}}$.  The total mass is therefore $m N_i/f_i$, where $m$ 
is the mass of the atom or molecule.  Temperatures and masses (calculated from the J=2 energy level, as these lines have the lowest error in flux) are 
in Table \ref{tbl:ci}.  We further discuss these results in Section \ref{sec:disc:c}.

\subsection{LTE Analysis of \htwo}\label{sec:H2}

\begin{deluxetable*}{l rr rr rr rr rr rr r r}
\tabletypesize{\scriptsize}
\tablecaption{\htwo\ Line Measurements: $10^{-16}$ W m$^{-2}$ \label{tbl:h2lines}}
\tablehead{
\colhead{FTS Name}  & \colhead{S(0)} & \colhead{$\sigma$} & \colhead{S(1)} & \colhead{$\sigma$} & \colhead{S(2)} & \colhead{$\sigma$} & \colhead{S(3)} & \colhead{$\sigma$} & \colhead{S(5)} & \colhead{$\sigma$} & \colhead{S(7)} & \colhead{$\sigma$} & \colhead{$\tau_{9.7}$} & \colhead{Ref}
}
\startdata
Mrk 231              & $<$ 67.5 & \ldots & 1.14 & 0.28 & 0.39 & 0.14 & 0.43 & 0.10 & \ldots & \ldots & $<$ 122\phantom{.00} & \ldots & 0.8 & 1,2 \\
IRAS F17207-0014     & $<$ 23.9 & \ldots & 0.88 & 0.01 & 0.50 & 0.09 & 0.57 & 0.11 & \ldots & \ldots & $<$ 8.50 & \ldots & 0.0 & 2 \\
IRAS 09022-3615      & $<$ 3.30 & \ldots & 0.12 & 0.01 & $<$ 0.20 & \ldots & $<$ 0.20 & \ldots & \ldots & \ldots & $<$ 0.04 & \ldots & 0.0 & 2 \\
Arp 220              & $<$ 97.0 & \ldots & 1.86 & 0.17 & 0.98 & 0.13 & 0.73 & 0.02 & \ldots & \ldots & 1.29 & 0.38 & 3.3 & 2 \\
Mrk 273              & $<$ 26.3 & \ldots & 1.02 & 0.01 & 0.56 & 0.07 & 1.04 & 0.09 & \ldots & \ldots & $<$ 15.0\phantom{0} & \ldots & 2.0 & 2 \\
UGC 05101            & $<$ 10.0 & \ldots & 0.50 & 0.05 & 0.27 & 0.05 & 0.28 & 0.03 & \ldots & \ldots & $<$ 19.5\phantom{0} & \ldots & 1.6 & 2 \\
NGC 6240             & 0.5 & 0.2 & 4.27 & 0.34 & 3.55 & 0.26 & 5.83 & 0.43 & \ldots & \ldots & 5.54 & 0.50 & 0.0 & 2 \\
Arp 299-C            & $<$ 0.7 & \ldots & 4.11 & 1.0\phantom{0} & 1.92 & 0.48 & \ldots & \ldots & 1.46 & 0.36 & \ldots & \ldots & 1.2 & 3 \\
Arp 299-B            & $<$ 0.7 & \ldots & 4.11 & 1.0\phantom{0} & 1.92 & 0.48 & \ldots & \ldots & 1.46 & 0.36 & \ldots & \ldots & 1.2 & 3 \\
Arp 299-A            & $<$ 1.21 & \ldots & 3.70 & 0.90 & 1.80 & 0.45 & \ldots & \ldots & 1.97 & 0.50 & \ldots & \ldots & 1.2 & 3 \\
NGC 1068             & $<$ 1.87 & \ldots & 6.50 & 1.6\phantom{0} & $<$ 8.00 & \ldots & 5.76 & 1.4\phantom{0} & 6.40 & 1.6\phantom{0} & 3.50 & 0.87 & 0.5 & 3,4 \\
NGC 1365-SW          & 1.59 & 0.1 & 5.69 & 1.4\phantom{0} & $<$ 3.13 & \ldots & \ldots & \ldots & 2.01 & 0.50 & \ldots & \ldots & 0.2 & 3,5 \\
NGC 1365-NE          & 1.59 & 0.1 & 5.69 & 1.4\phantom{0} & $<$ 3.13 & \ldots & \ldots & \ldots & 2.01 & 0.50 & \ldots & \ldots & 0.2 & 3,5 \\
NGC 4038             & 0.7 & 0.2 & 3.95 & 0.99 & 1.60 & 0.40 & \ldots & \ldots & \ldots & \ldots & \ldots & \ldots & 4.7 & 3,6 \\
NGC 4038 (Overlap)   & 3.67 & 0.3 & 7.02 & 0.37 & 2.38 & 0.47 & $<$ 2.15 & \ldots & \ldots & \ldots & \ldots & \ldots & 0.0 & 6 \\
M82                  & 7.80 & 1.9 & 15.0\phantom{0} & 3.8\phantom{0} & 12.0\phantom{0} & 3.0\phantom{0} & \ldots & \ldots & 11.5\phantom{0} & 2.9\phantom{0} & 4.80 & 1.2\phantom{0} & 1.8 & 3 \\
NGC 1222             & 0.3 & 0.0 & 0.95 & 0.03 & 0.29 & 0.02 & \ldots & \ldots & \ldots & \ldots & \ldots & \ldots & 0.0 & 5 \\
M83                  & $<$ 1.04 & \ldots & 7.29 & 1.8\phantom{0} & \ldots & \ldots & \ldots & \ldots & 2.84 & 0.70 & \ldots & \ldots & 0.3 & 3 \\
NGC 253              & 2.13 & 0.5 & 19.6\phantom{0} & 4.9\phantom{0} & 12.0\phantom{0} & 3.0\phantom{0} & \ldots & \ldots & 11.5\phantom{0} & 2.9\phantom{0} & 8.40 & 2.1\phantom{0} & 1.8 & 3 \\
NGC 1266             & 0.2 & 0.1 & 1.49 & 0.07 & 1.22 & 0.07 & 1.90 & 0.12 & 2.42 & 0.20 & 1.92 & 0.29 & 0.2 & 7 \\
Cen A                & 2.51 & 0.6 & 8.64 & 2.2\phantom{0} & 5.40 & 1.4\phantom{0} & 5.81 & 1.5\phantom{0} & 4.54 & 1.1\phantom{0} & 3.26 & 0.80 & 1.8 & 3
\enddata
\tablecomments{See Appendix \ref{appendix:individual} for information about extended sources. {\bf References.} (1) \citet{Armus2007}; (2) \citet{Higdon2006}; (3) \citet{Rigopoulou2002}; (4) \citet{Lutz2000}; (5) \citet{Bernard-Salas2009}; (6) \citet{Brandl2009}; (7) \citet{Roussel2007}}
\end{deluxetable*}

\begin{figure*}  
\includegraphics[height=\textwidth,angle=270]{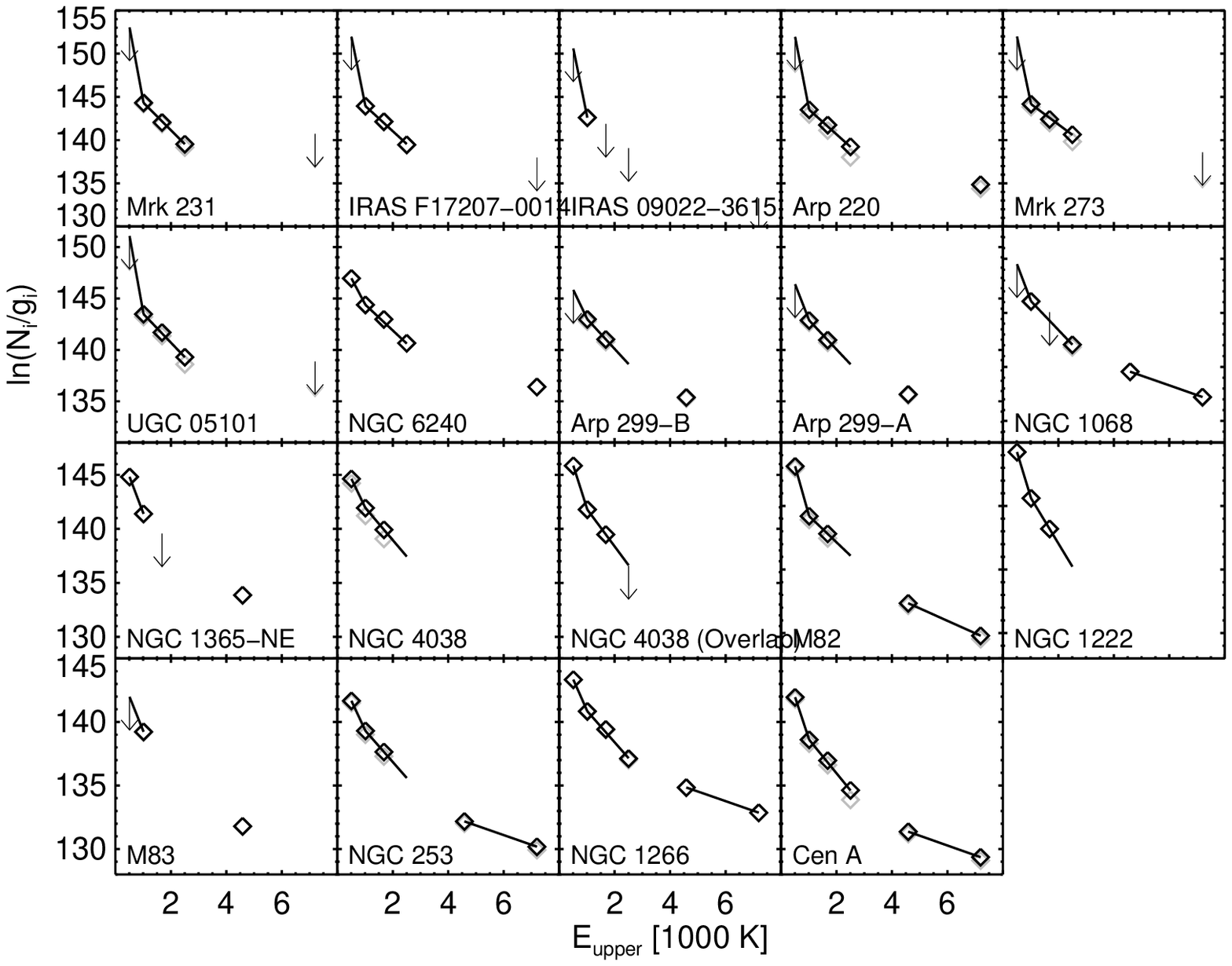} 
\caption[\htwo\ Excitation Diagrams]{\htwo\ Excitation Diagrams.  For each line flux available, the total number 
of molecules is calculated by the equations in \ref{sec:CI}.  
The inverse of the solid line slopes are the excitation temperatures; the S(3), S(2), and S(1) generally lie on a fairly constant line, but the higher and 
lower energy fluxes demonstrate a gradient in excitation temperatures.  
The non-extinction-corrected lines are shown in gray.
\label{fig:h2}}
\end{figure*}

CO is used as a proxy for total molecular gas, most of which is molecular hydrogen.  
The electric quadrupole transitions of \htwo\ are difficult to observe, 
but were available for many of these bright galaxies.  
Here we consider the S(0), S(1), S(2), S(3), S(5), and S(7) transitions of \htwo.
The hydrogen fluxes and optical depths are in Table \ref{tbl:h2lines};
 see Appendix \ref{appendix:individual} for detailed information on extended galaxies.
The derived temperatures and masses are in Table \ref{tbl:h2}, with excitation diagrams in Figure \ref{fig:h2}.
The ground state transition (at 510 K) is very insensitive to cold (tens of K) gas, so the hydrogen-derived masses in
 Table \ref{tbl:h2} necessarily miss much of the molecular mass.

With Equation \ref{eqn:Tlte}, we calculated excitation temperatures for the following combination of lines (when available): S(1) and S(0); 
S(3), S(2), and S(1); S(7) and S(5).
We also corrected these lines for dust extinction, but in this portion of the spectrum the notable feature is the 9.7 $\mu$m silicate absorption 
feature.  
Based on the extinction models of \citet{Draine1989}, we used $\tau_\lambda/\tau_{9.7} = 0.19, 0.35, 0.43, 0.99, 0.20, 0.30$ for 
S(0), S(1), S(2), S(3), S(5), and S(7), and $A_\lambda = 1.086 \tau$, $A_V = 17 \tau_{9.7}$.  
The values that we use for $\tau_{9.7}$ are listed in the last column of Table \ref{tbl:h2}.
In many cases, only an upper limit was available for S(0), meaning the S(1)-S(0) excitation temperature is a lower limit, 
and the S(0)-derived mass an upper limit.

One could calculate an excitation temperature from any pair of lines.  The excitation temperature is the inverse of the slope 
of the lines presented in the excitation diagrams (Figure \ref{fig:h2}), which can be used to fit an excitation temperature 
to multiple lines, as we did with S(3), S(2), and S(1).  
One excitation temperature generally could not be fit for all lines from S(0) through S(7), which is why we present three.
We did not calculate, for example, an excitation temperature from S(7)-S(3) when the S(5) line was unavailable, for consistency across the sample.

The extinction correction most dramatically increased the flux for the S(3) line, because of its alignment in wavelength with the 
silicate absorption feature.  Cen A is a good example of a galaxy with high extinction and many lines measured.  
The extinction correction raised the three temperatures by 6, 39, and 50 K.  
If we calculated separate excitation temperatures S(2)-S(1) and S(3)-S(2), with no extinction correction we found 
$392 \pm 82$ and $300 \pm 39$ K.  With extinction correction, the temperatures were $406 \pm 88$ and $351 \pm 54$ K.
The correction had the effect of increasing the two temperatures, but more importantly, bringing these three mid-excitation lines into 
better alignment, giving us a better estimate of one middle excitation temperature.  (Notice that the S(2)-S(1) and S(3)-S(2) temperatures 
quoted above are not distinguishable within the error bars.)

For the conditions studied here, the first 5 \htwo\ levels are likely in LTE, 
based on the collision rate coefficients in \citet{LeBourlot1999}.
This assumption may become increasingly poor for the S(5) and S(7) lines, 
and the temperatures and masses from these lines should be treated as approximations.

\begin{deluxetable*}{l rrr rr rr rr rr rr r}
\tabletypesize{\scriptsize}
\tablecaption{\htwo\ LTE Results \label{tbl:h2}}
\tablehead{
\colhead{FTS Name}  & \colhead{} & \colhead{T$_1$} & \colhead{$\sigma$} & \colhead{T$_2$} & \colhead{$\sigma$} & \colhead{T$_3$} & \colhead{$\sigma$} & \colhead{M$_1$} & \colhead{$\sigma$} & \colhead{M$_2$} & \colhead{$\sigma$} & \colhead{M$_3$} & \colhead{$\sigma$} & \colhead{$\tau_{9.7}$} \\
\colhead{}  & \colhead{} & \colhead{[K]} & \colhead{[K]} & \colhead{[K]}  & \colhead{[K]}  & \colhead{[K]}  & \colhead{[K]}  & \multicolumn{2}{c}{[$10^6$ \ms]}  & \multicolumn{2}{c}{[$10^6$ \ms]} & \multicolumn{2}{c}{[$10^3$ \ms]} & \colhead{}
}
\startdata
Mrk 231              &  $>$ & 57 & \ldots & 310 & 20 & \ldots & \ldots & \ldots & \ldots & 159\phantom{.000} & 36\phantom{.000} & \ldots & \ldots & 0.80\\
IRAS F17207-0014     &  $>$ & 63 & \ldots & 340 & 10 & \ldots & \ldots & \ldots & \ldots & 94.6\phantom{00} & 7.1\phantom{00} & \ldots & \ldots & 0.00\\
IRAS 09022-3615      &  $>$ & 63 & \ldots & \ldots & \ldots & \ldots & \ldots & \ldots & \ldots & \ldots & \ldots & \ldots & \ldots & 0.00\\
Arp 220              &  $>$ & 59 & \ldots & 350 & 7 & \ldots & \ldots & \ldots & \ldots & 61.8\phantom{00} & 4.4\phantom{00} & \ldots & \ldots & 3.30\\
Mrk 273              &  $>$ & 64 & \ldots & 420 & 9 & \ldots & \ldots & \ldots & \ldots & 81.3\phantom{00} & 2.8\phantom{00} & \ldots & \ldots & 2.00\\
UGC 05101            &  $>$ & 66 & \ldots & 360 & 10 & \ldots & \ldots & \ldots & \ldots & 55.4\phantom{00} & 5.2\phantom{00} & \ldots & \ldots & 1.60\\
NGC 6240             &     & 200 & 40 & 400 & 10 & \ldots & \ldots & 786\phantom{.00} & 470\phantom{.0} & 119\phantom{.000} & 8.2\phantom{00} & \ldots & \ldots & 0.00\\
Arp 299-C            &  $>$ & 180 & \ldots & 340 & 60 & \ldots & \ldots & \ldots & \ldots & 35.6\phantom{00} & 14\phantom{.000} & \ldots & \ldots & 1.18\\
Arp 299-B            &  $>$ & 180 & \ldots & 340 & 60 & \ldots & \ldots & \ldots & \ldots & 35.6\phantom{00} & 14\phantom{.000} & \ldots & \ldots & 1.18\\
Arp 299-A            &  $>$ & 140 & \ldots & 350 & 60 & \ldots & \ldots & \ldots & \ldots & 30.7\phantom{00} & 12\phantom{.000} & \ldots & \ldots & 1.18\\
NGC 1068             &  $>$ & 150 & \ldots & 370 & 30 & 1110 & 170 & \ldots & \ldots & 4.56\phantom{0} & 1.2\phantom{00} & 77.7\phantom{0} & 35\phantom{.00} & 0.47\\
NGC 1365-SW          &     & 150 & 10 & \ldots & \ldots & \ldots & \ldots & 168\phantom{.00} & 36\phantom{.0} & \ldots & \ldots & \ldots & \ldots & 0.15\\
NGC 1365-NE          &     & 150 & 10 & \ldots & \ldots & \ldots & \ldots & 168\phantom{.00} & 36\phantom{.0} & \ldots & \ldots & \ldots & \ldots & 0.15\\
NGC 4038             &     & 190 & 30 & 330 & 60 & \ldots & \ldots & 83.6\phantom{0} & 32\phantom{.0} & 13.3\phantom{00} & 5.2\phantom{00} & \ldots & \ldots & 4.70\\
NGC 4038 (Overlap)   &     & 120 & 2 & 290 & 30 & \ldots & \ldots & 758\phantom{.00} & 78\phantom{.0} & 16.1\phantom{00} & 3.4\phantom{00} & \ldots & \ldots & 0.00\\
M82                  &     & 130 & 10 & 480 & 100 & 1020 & 140 & 44.7\phantom{0} & 16\phantom{.0} & 0.481 & 0.16\phantom{0} & 11.3\phantom{0} & 5.2\phantom{0} & 1.76\\
NGC 1222             &     & 140 & 4 & 280 & 7 & \ldots & \ldots & 111\phantom{.00} & 15\phantom{.0} & 5.53\phantom{0} & 0.39\phantom{0} & \ldots & \ldots & 0.00\\
M83                  &  $>$ & 180 & \ldots & \ldots & \ldots & \ldots & \ldots & \ldots & \ldots & \ldots & \ldots & \ldots & \ldots & 0.29\\
NGC 253              &     & 210 & 30 & 400 & 90 & 1310 & 230 & 3.47 & 1.2 & 0.676 & 0.25\phantom{0} & 4.69 & 2.0\phantom{0} & 1.76\\
NGC 1266             &     & 200 & 40 & 400 & 8 & 1320 & 110 & 20.1\phantom{0} & 12\phantom{.0} & 3.25\phantom{0} & 0.14\phantom{0} & 66.0\phantom{0} & 13\phantom{.00} & 0.24\\
Cen A                &     & 150 & 20 & 370 & 30 & 1300 & 220 & 8.97 & 3.2 & 0.396 & 0.089 & 2.12 & 0.87 & 1.76
\enddata
\tablecomments{T$_1$ is derived from the S(1)-S(0) lines, T$_2$ from the S(1), S(2), and S(3) lines, and T$_3$ from the S(7)-S(5) lines. 
Some values of T$_1$ are lower limits because the S(0) lines were upper limits.  M$_1$ is derived from the S(0) line at T$_1$, 
M$_2$ from the S(1), S(2), or S(3) line using T$_2$, and M$_3$ from the S(5) or S(7) line and T$_3$.  
Note that the numbers in M$_3$ are presented as 3 orders of magnitude smaller than those in the other mass columns. 
See Section \ref{sec:H2} for more explanation, and Table \ref{tbl:h2lines} for references for $\tau_{9.7}$.}
\end{deluxetable*}

\subsection{{\rm \cii}\ and {\rm \nii}\ Line Ratios}\label{sec:nii}

Finally, we collected \cii\ 158 $\mu$m fluxes for our sample of galaxies to compare to our FTS measured \nii\ 205 $\mu$m line.
The ratio of these two lines (\cii/\nii) from ionized gas is fairly constant, about 2.5 to 4.3, independent of ionized 
gas density \citep{Oberst2006}.  This implies that ratios higher than this value indicate excess \cii\ emission from other sources,
especially PDRs.
For example, a ratio of 30 indicates 8 to 14\% of \cii\ emission from ionized gas, hence 92 to 86\% from other sources
(for an assumed ionized ratio range of 2.5 to 4.3, respectively).

We present our results in Table \ref{tbl:ciilines}, which includes the \cii\ line fluxes used and the percentage of emission 
from ionized gas.
In most of our galaxies, the majority of \cii\ emission is from PDRs and other non-ionized sources, ranging from 48\% to 96\%.
The median is 75-85\%.  We discuss the meaning of this in Section \ref{sec:disc:c}.  

\begin{deluxetable*}{l rr rr rr}
\tabletypesize{\scriptsize}
\tablecaption{\cii\ Line Measurements: $10^{-16}$ W m$^{-2}$ \label{tbl:ciilines}}
\tablehead{
\colhead{FTS Name}  & \colhead{\cii} & \colhead{$\sigma$} & \colhead{\% Ionized\tablenotemark{a}} & \colhead{$\sigma$} & \colhead{\% Ionized\tablenotemark{b}} & \colhead{$\sigma$}
}
\startdata
Mrk 231              & 3.70 & 0.20 & 20.\phantom{0} & 3\phantom{.00} & 35\phantom{.0} & 5\phantom{.00} \\
IRAS F17207-0014     & 6.70 & 0.80 & 12\phantom{.0} & 2\phantom{.00} & 20.\phantom{0} & 3\phantom{.00} \\
IRAS 09022-3615      & 6.69 & 0.22 & 12\phantom{.0} & 1\phantom{.00} & 21\phantom{.0} & 2\phantom{.00} \\
Arp 220              & 9.40 & 0.40 & 30.\phantom{0} & 7\phantom{.00} & 52\phantom{.0} & 10\phantom{.00} \\
Mrk 273              & 5.50 & 0.30 & 17\phantom{.0} & 2\phantom{.00} & 29\phantom{.0} & 3\phantom{.00} \\
UGC 05101            & 5.59 & 0.28 & 26\phantom{.0} & 2\phantom{.00} & 44\phantom{.0} & 4\phantom{.00} \\
NGC 6240             & 27.2\phantom{0} & 0.60 & 15\phantom{.0} & 0.5\phantom{0} & 25\phantom{.0} & 0.9\phantom{0} \\
Arp 299-C            & \ldots & \ldots & \ldots & \ldots & \ldots & \ldots \\
Arp 299-B            & \ldots & \ldots & \ldots & \ldots & \ldots & \ldots \\
Arp 299-A            & 82.2\phantom{0} & 1.2\phantom{0} & 5.3 & 0.3\phantom{0} & 9.1 & 0.5\phantom{0} \\
NGC 1068             & 214\phantom{.00} & 1.7\phantom{0} & 20.\phantom{0} & 0.5\phantom{0} & 35\phantom{.0} & 0.8\phantom{0} \\
NGC 1365-SW          & 104\phantom{.00} & 2.5\phantom{0} & 22\phantom{.0} & 0.6\phantom{0} & 38\phantom{.0} & 1\phantom{.00} \\
NGC 1365-NE          & 104\phantom{.00} & 2.5\phantom{0} & 18\phantom{.0} & 0.5\phantom{0} & 31\phantom{.0} & 0.8\phantom{0} \\
NGC 4038             & 50.2\phantom{0} & 1.3\phantom{0} & 6.7 & 0.3\phantom{0} & 11\phantom{.0} & 0.5\phantom{0} \\
NGC 4038 (Overlap)   & \ldots & \ldots & \ldots & \ldots & \ldots & \ldots \\
M82                  & 1320\phantom{.00} & 4.3\phantom{0} & 8.0 & 0.06 & 14\phantom{.0} & 0.1\phantom{0} \\
NGC 1222             & 24.2\phantom{0} & 0.90 & 4.2 & 0.2\phantom{0} & 7.2 & 0.4\phantom{0} \\
M83                  & 181\phantom{.00} & 3.5\phantom{0} & 15\phantom{.0} & 0.3\phantom{0} & 25\phantom{.0} & 0.6\phantom{0} \\
NGC 253              & 499\phantom{.00} & 5.2\phantom{0} & 8.7 & 0.5\phantom{0} & 15\phantom{.0} & 0.8\phantom{0} \\
NGC 1266             & 5.00 & 1.0\phantom{0} & 18\phantom{.0} & 4\phantom{.00} & 31\phantom{.0} & 8\phantom{.00} \\
Cen A                & 295\phantom{.00} & 2.1\phantom{0} & 4.2 & 0.05 & 7.2 & 0.09 \\
Median &  &  &    14\phantom{.0}   &  &    25\phantom{.0}   & 
\enddata
\tablecomments{All are from \citet{Brauher2008} except UGC~05101 and IRAS~09022-3615 from \citet{Diaz-Santos2013}. 
\nii\ line fluxes were given in Table \ref{tbl:flux3}.}
\tablenotetext{a}{Calculated using \cii/\nii ratio = 2.5 in ionized medium.}
\tablenotetext{b}{Calculated using \cii/\nii ratio = 4.3 in ionized medium.}
\end{deluxetable*}

\section{Discussion}\label{sec:disc}

This survey sought to answer a variety of questions related to the molecular ISM in galaxies, specifically 
those detailed in Section \ref{sec:intro}.
The first subsection here addresses systematic effects of our modeling procedure, including aspects of two-component modeling that 
ought to be considered in future CO SLED modeling.  
The next two focus on two useful quantities derived from our data: the luminosity-to-mass 
conversion factor (Section \ref{sec:disc:lummass}) and gas-to-dust mass ratios (Section \ref{sec:disc:dust}).
Section \ref{sec:disc:compare} discusses trends in the molecular gas properties within our sample, and 
Section \ref{sec:disc:c} specifically examines the 
difference in properties derived from C, C$^+$, and CO.
Finally, Section \ref{sec:disc:galactic} compares the molecular gas 
properties to those of the Galactic Center and Sgr B2.

In the plots that follow, most parameters are plotted in 
log-log space, with bottom panels showing the ratio of the parameters.
This is a small sample of diverse galaxies; the lines are meant to illustrate the presence or absence of general trends, but should be interpreted with caution.
In some cases, our primary interest is whether two parameters have a linear relationship or not; to robustly determine 
the uncertainties, we used a case re-sampling nonparametric bootstrap method.
For $n$ galaxies included in the fit, we drew $n$ samples (allowing re-selection of the same sample) and fit a line 1000 times; 
the error was then the 68\% interval of the probability density function of the resulting parameter fits.

\subsection{Discussion of Systematic Effects of Simultaneous Two-Component Modeling}\label{sec:disc:systematic}

Most previous studies of CO molecular gas properties were based on the few lines available 
in atmospheric windows.  With three or four lines, only one component of cool molecular gas could be 
described with molecular excitation models such as RADEX, relying on four parameters.  
The first detections of CO \jsix\ by \citet{Harris1991} indicated higher-excitation gas than 
explained by CO \jone. 
Now with 10 additional lines available with  {\it Herschel}, by visual inspection alone (Figure \ref{fig:slednorm}) 
one can see that additional excitation is required to explain high-J lines.
This point has already become noticeable in the past few years 
\citep[e.g.][]{Panuzzo2010,Kamenetzky2012,Rangwala2011,Spinoglio2012,Rigopoulou2013,Hailey-Dunsheath2012,Pereira-Santaella2013} 
and was a major motivation for this study.
For the past few years, FTS SLEDs have often been modeled using two or three distinct components, each 
described by its own set of physical parameters.  This section will analyze the statistical validity and astrophysical meaning of such models.

\subsubsection{Distinct Components vs. a Continuous Distribution}\label{sec:disc:continuous}

We wish to emphasize that {\it it is unlikely that all molecular gas is described either by our ``cool" or ``warm" component conditions.}
Rather, we recognize that it is likely that we are really measuring the sum of the emission from a wide distribution of molecular cloud properties.
There may also be multiple components comprised of different ranges of conditions or distributions.

For example, the \htwo\ excitation ladders (Figure \ref{fig:h2}) clearly indicate a gradient in excitation temperatures (though 
we note that \htwo\ rotational lines are not sensitive to the coldest molecular gas).  Higher lines of \htwo\ indicate higher 
temperatures and lower masses.  In general, Table \ref{tbl:h2} shows that the mass in the few hundred K component (T$_2$) is 1-20\% 
of that in the colder gas traced by S(0).  The S(5) and S(7) lines, which trace $> 1000$ K gas, come from orders of magnitude smaller gas masses.  
In our two-component SLED modeling, we found typically 10\% of the mass in very warm gas 
and 90\% in cooler gas, but these two components are presumably sums over a range of gas conditions.

Increased effort has been made to describe the SLEDs that would result from, for example, a continuous distribution of 
temperatures.  \citet{Neufeld2012} created models with a power-law distribution in gas temperatures ($dN/dT \propto T^{-\beta}$, 
using a constant density) and found such models can describe CO rotational diagrams in the {\it Herschel} PACS range 
that cannot be described by an isothermal medium. 
\citet{Pereira-Santaella2014} applied the same concept to a sample of Seyferts observed by the FTS (including one from this work, UGC~05101); 
to fit both CO and \htwo, they required a broken power law, and found that $\beta$ and $n_{H_2}$ were degenerate, for the same reasons 
we find $T$ and $n_{H_2}$ degenerate.  
Such models can influence the further discussion of excitation mechanisms, e.g. the extent to 
which PDRs or shocks may explain the CO emission.  

However, a yet unresolved problem is the extent to which even high S/N CO SLEDs can {\it distinguish} a power-law distribution 
from distinct components.  The resulting SLED from a power-law distribution can exactly imitate the emission 
from gas with a uniform temperature and different density, due to the degeneracy between $n_{H_2}$ and $T$.  
For example, \citet{Goicoechea2013} found the CO emission from Sgr A* can be described by either a single, hot, 
low-density component of gas or multiple, cooler components at higher density.  The CO SLEDs alone do not provide the 
information necessary to distinguish the {\it uniqueness} of either type of model.
Numerical simulations of galaxies, coupled with radiative transfer, could potentially 
deepen our understanding of how a gradient of excitation conditions adds up to produce the total emission we can measure \citep[e.g.][]{Narayanan2014}.

Therefore, differences in ``cool" and ``warm" component properties among galaxies can thus be indicating different {\it underlying distributions} 
of cloud physical conditions.  In the rest of the discussion we continue to refer to these distinct components, 
but we caution the reader to interpret our conclusions with this important section in mind.  
The following subsections refer only to distinct component modeling, in order to compare with the majority of the {\it Herschel} FTS literature.  

\subsubsection{Adding Another Distinct Component}\label{sec:disc:twocomp2}

In the context of distinct components, we must address 
the question of why we modeled 2 components, and not 3 or more.
The simplest answer is that the SLEDs were statistically well described 
by 8 parameters and hence there is no justification for adding a more complicated model where the number of 
free parameters approaches or exceeds the number of data points.  
One could cite ``Occam's razor" and stop here.
However, a more in-depth investigation 
is still informative, especially given our understanding of the more complicated physical situation we are attempting to 
describe.
We emphasize that this subsection concerns our {\it ability to detect} the presence of three distinct components from the thirteen lines of data we have; 
as already discussed, there is likely a continuous distribution of gas conditions, and we do not mean to imply that 
there are {\it not actually} three or more components physically present in the gas.

We sought to investigate if we could discern the temperature/mass components we see in \htwo\ (now referred to as Components I, II, and III) 
with the CO SLEDs.  We used the same method 
as described in \ref{sec:comodel} (now requiring 12 free parameters) for two well-defined galaxy SLEDs (Arp~220 and M82).  
We applied some additional constraints:  $T_{\rm I} < T_{\rm II} < T_{\rm III}$, $M_{\rm I} > M_{\rm II} > M_{\rm III}$, and 
we restricted the temperature ranges for each component, such that $0.5/2/2.3 < {\rm Log}(T_{\rm I}/T_{\rm II}/T_{\rm III}) < 2.3/3/3.5$.
These ranges were meant to encompass the ranges seen in \htwo, but we found qualitatively similar results without such restrictions (see Appendix {\ref{appendix:individual} for Arp~220).
The same priors as previously used were applied for optical depths, maximum length of any one component, and maximum total mass (Section \ref{sec:copriors}).

The resultant best fits were nearly identical with the 2-component fits (no improvement in total $\chi^2$), with 
Component I acting as the cool component and Component II as the warm.
By this we mean that the pressure, luminosity, and mass distributions were significantly overlapping with our previous results (though discernible).  
We illustrate this in Appendix \ref{appendix:individual} for the case of Arp~220.
Component III had negligible mass, negligible contribution to the total luminosity, and was poorly constrained otherwise.  
In other words, the CO SLEDs are not sensitive to the extremely low mass component derived from S(7)-S(5) hydrogen lines, 
{\it absent additional strict prior constraints on the parameter space}.
Though one could exactly fix the mass and temperatures of each component to match that of hydrogen, we have no reason to believe 
that there is a one-to-one correspondence between the CO and \htwo\ components, in addition to doubts about the LTE approximation 
for the S(7) and S(5) lines, and 
the knowledge that the \htwo\ cannot probe the coolest gas anyway.  More importantly, fixing the mass of Component III orders of 
magnitude lower than the other components (to match \htwo) would produce the same result: a SLED described by the other two components, 
where the temperature of Component II would simply move up or down (allowable due to the $n-T$ degeneracy) based on the imposed 
constraints to preserve the same pressure.  This was our best-fit result {\it without} requiring that the $M_{III}$ be orders of 
magnitude lower than $M_{II}$, only that it be lower.  The best-fit SLED was statistically indistinguishable from the best-fit 
two-component SLED (same total $\chi^2$), because the third component contributed negligible flux to each line.  
We ran the 3-component fit again, this time {\it not} requiring that $M_{\rm III}$ be lower than the other masses (only that 
the sum be less than the dynamical mass prior, as before).  We still found the equivalent of a two-component fit, with the 
remaining component contributing negligible mass and luminosity.  

One could undoubtedly find a good fit to the SLEDs using any number ($\ge 2$) of distinct components.  
Though the addition of more components perhaps becomes closer to the real situation of a continuous distribution of conditions, 
the uniqueness of the parameter solutions significantly decreases, and thus the physical parameters themselves become less well constrained.
We point out the 3-component models end up indistinguishable from the 2-component models not because the 2-components are the ``right" 
answers, for the reasons discussed in Section \ref{sec:disc:continuous}, but because the 3-component models do not improve the quality of the 
fit and cannot be constrained (a wide range of combinations of parameters could reproduce the SLEDs).

\subsubsection{Two- vs. One-Component (Low-J only) Modeling}\label{sec:disc:twocomp1}

How does simultaneously modeling two components {\it change} 
our understanding of the cold component?  To answer this, we also ran single-component likelihoods using only 
the \jone, \jtwo, and \jthree\ lines, if all three 
were available in the literature (for 17 of our 21 pointings, those with green dashed SLEDs shown in Figure \ref{fig:cosleds}), 
allowing for multiple measurements for each line.

We found that the luminosity of cold CO and thermal gas pressure in the cold component was overestimated when modeled alone (using only \jthree\ and below)  
by an average of 0.5 and 0.6 dex, respectively.  
No effect on the temperature and density could be determined, because these two parameters were not as well constrained 
as the pressure (this is a consequence of their degeneracy in excitation modeling and not specific to the methods compared here).
In all cases, modeling only the cold CO will severely underestimate the total CO luminosity, which is dominated by warm gas.
The parameter that was usually correctly determined is the CO mass, because it is most highly dependent on the 
J = 0, 1, and 2 population levels.  This point will become very important when discussing luminosity-to-mass conversion 
factors in Section \ref{sec:disc:lummass}. 

Though low-z galaxies are often modeled with low-J lines, high-redshift SMGs are often detected in only a few mid-J lines.
\citet{Carilli2013} reviewed studies of cool molecular 
gas in high-$z$ galaxies and found that quasars require a high-excitation component, related to the AGN, to explain mid-J line flux.
We showed that even galaxies with no AGN require a high-excitation component.  Furthermore, they note that SMGs (in contrast with quasars) 
demonstrate less excited molecular gas and excess emission in the CO \jone\ transition.  
We have demonstrated that the 
problem is perhaps best approached from another direction: the ``excess" emission is not in the CO \jone\ transition.
Instead, the CO \jone\ should be considered as entirely emitted by the coldest gas, and the real excess is in the mid-J lines, requiring 
the higher-excitation component.
Just such a component was found using the \jthree\ to \jnine\ lines of the $z = 2.56$ Cloverleaf quasar \citep[thermal pressure $> 10^6$ K cm$^{-3}$; ][]{Bradford2009}.
One cannot disentangle this question without 
a more complete SLED, as we show here for a range of low-$z$ galaxies (with and without AGN, with varying degrees of star formation).
By more complete, we mean a good distribution of lines from \jone\ to \jnine\ or higher.
Moreover, the mass estimated from mid-J lines alone will be an underestimate of the total molecular mass if CO \jone\ is unavailable.
We tested the extent of this effect by modeling only the \jthree\ to \jsix\ lines as one component of molecular gas.
On average, the log of the ratio of cold component mass (from our two-component models) to these mid-J masses was $0.56 \pm 0.34$.
This means masses using mid-J lines will be underestimated by a factor of 1.7 - 7.9, or 3.6 on average.
For example, with ALMA, the \jone\ (\jtwo) line is unavailable above $z$ = 0.4 (1.7), so it will be difficult to accurately 
estimate molecular mass at high-redshift.  The SLEDs shown here could also be used as analogues for missing lines in future high-$z$ molecular gas modeling.

\subsubsection{Iterative vs. Simultaneous Two-Component Modeling}\label{sec:disc:twocomp_it}

Having compared the difference between one and two components, it is also useful to examine 
the effect that simultaneous modeling of an 8-dimensional parameter space has compared to 
an iterative modeling approach, which has been used in the past \citep[e.g.][]{Kamenetzky2012,Schirm2014}.  
This iterative modeling involves first modeling high-J lines alone, above some 
cutoff such as the \jsix\ line.  The low-J line fluxes predicted by the best-fit model are then subtracted from the measured 
low-J line fluxes, and the remainder is modeled as the cold component.  One can continue alternating between warm 
and cool components until the two solutions converge.  
Specific comparisons of the results from iterative vs. simultaneous modeling are given in Appendix \ref{appendix:individual} for 
M82, Arp~220, and NGC~4038.  

This approach has three major issues.  First, as discussed near the beginning of 
Section \ref{sec:analysis}, most likelihood analysis codes suitable for a large number of parameters are not well 
designed to refine the best-fit solution, which can differ significantly from the median parameters.  This means 
that the selection of the model flux to subtract from the measured fluxes for modeling the second component is not 
representative of the probability distribution function.  Second, the choice of where to break the SLED apart for the two 
component modeling is somewhat arbitrary, introducing an additional prior, or restriction of parameter space.  
\citet{Schirm2014} illustrated the differences that can be caused 
by choosing different lines to break the spectrum; they found the statistical ranges of the parameters found for three different choices 
do not vary significantly in the case of the Antennae galaxy.  Third, as a consequence of the previous two drawbacks, 
iterative modeling uncouples the uncertainties and covariances of the warm 
and cool components, which falsely lifts some degeneracies in the analysis.  
We believe that our parameter estimates and uncertainties are a more accurate description 
of a two-component model, and that future two-component modeling should allow for covariance between the components' parameters, as we do here.

\subsection{Luminosity to Mass Conversion Factors}\label{sec:disc:lummass}

It is common to use an empirically measured value to convert from CO \jone\ luminosity to total molecular mass.  
For Milky Way clouds, this is the ``X-factor" in units of cm$^{-2}$ (K km s$^{-1}$): 
$X({\rm CO}) = N({\rm H_2})/ I({\rm CO})$, where $I({\rm CO}) = \int T dv$.
For extragalactic work, one cannot resolve individual clouds, but an ensemble of emitting clouds would have a similar 
proportionality known as $\alpha_{\rm CO} = M / L^\prime$ [\ms (K km s$^{-1}$ pc$^{2}$)$^{-1}$] \citep{Dickman1986}.  
For simplicity, we do not include the units of $\alpha_{\rm CO}$ and $X({\rm CO})$ from now on.
$L^\prime$ is the area-integrated source brightness temperature, $I({\rm CO})$ times the area on the sky in pc$^2$.
If neither (or both) factors take helium into account, then $X({\rm CO}) = 6.6 \times 10^{19} \alpha$ (in the respective units given above).
However, for an X-factor in terms of $N_{\rm H_2}$ (standard) and masses in total molecular gas (including He), the conversion is
  $X({\rm CO}) = 4.5 \times 10^{19} \alpha$ due to the factor of 1.4 in our Equation \ref{eqn:mass}.

\citet{Bolatto2013} discussed the theoretical basis for the CO-to-\htwo\ conversion and presented a comprehensive summary of the techniques and results of its measurement for 
Galactic and extragalactic molecular clouds.  A typical $X({\rm CO})$ for the disk of the Milky Way is 
$2 \times 10^{20}$ ($\alpha_{\rm CO} \approx 4$), but for the Galactic center, $X({\rm CO})$ is 3-10 times lower.  Normal spiral galaxies 
have values close to that of the Milky Way disk, whereas starburst galaxies and (U)LIRGS have values 
$X({\rm CO}) < 1 \times 10^{20}$ ($\alpha_{\rm CO} < 2$).  For these highly excited galaxies, the lower $X({\rm CO})$ values are often attributed 
to higher gas temperatures and large velocity dispersions \citep[in excess of self-gravity, see simulations by][]{Narayanan2011}.  
Additionally, if the CO emission is extended and not confined in self-gravitating molecular clouds, 
$X({\rm CO})$ would also be lower, a condition that is likely present for ULIRGS \citep{Bolatto2013}.
Other factors may influence the CO-to-\htwo\ conversion factor, 
such as metallicity and gas-to-dust mass ratio.  Lower metallicity or gas/dust ratios will decrease the depth of the CO-emitting layer in 
molecular clouds, decreasing the CO intensity and increasing $X({\rm CO})$.  
Additionally, $X({\rm CO})$ is only sensitive to CO-bright gas; we may be missing CO-faint \htwo\, reservoirs of \htwo\ where C$^{+}$ or C 
is the dominant form of carbon, instead of CO (see further discussion in Section \ref{sec:disc:c}).
We first present our derived values of $\alpha_{\rm CO}$ and then discuss the systematic 
effects that could affect these values \citep[see also][]{Maloney1988}.
  
The masses and luminosities used to derive $\alpha_{\rm CO}$ are plotted in Figure \ref{fig:alpha}, and resulting values in Table \ref{tbl:alpha}.
When we collected multiple measurements of the \jone\ line, we used a weighted average (after converting all to a 43.5\as\ beam following Section \ref{sec:sourcebeam}).  
We used the total CO mass (cool and warm components), but point out that 
$\alpha_{\rm CO}$ is only reduced by about 10\% if only using cold CO mass, though the errors remain roughly the same because they are dominated by the errors of the cold CO mass.  As discussed in the previous section, the mass estimate using only the first three lines of CO should 
be the similar to our estimate derived from the entire SLED.  This means we do not differ significantly from previous studies also using 
radiative transfer models to determine mass for this conversion factor.

\begin{deluxetable}{l rrrr}
\tabletypesize{\scriptsize}
\tablecaption{Derived Values of $\alpha_{\rm CO}$ [\ms (K km s$^{-1}$ pc$^{2}$)$^{-1}$] \label{tbl:alpha}}
\tablehead{
\colhead{FTS Name}  & \colhead{Log($\alpha_{\rm CO}$)} &  \colhead{$\sigma_{\rm Log}$} & \colhead{$\alpha_{\rm CO}$} & \colhead{$\sigma$}
}
\startdata
Mrk 231              & -0.6\phantom{0} & 0.1 & 0.3 & 0.09 \\
IRAS F17207-0014     & 0.05 & 0.4 & 1\phantom{.0} & 1\phantom{.00} \\
IRAS 09022-3615      & \ldots & \ldots & \ldots & \ldots \\
Arp 220              & -0.3\phantom{0} & 0.4 & 0.6 & 0.5\phantom{0} \\
Mrk 273              & -0.3\phantom{0} & 0.4 & 0.5 & 0.5\phantom{0} \\
UGC 05101            & -0.4\phantom{0} & 0.4 & 0.4 & 0.3\phantom{0} \\
NGC 6240             & -0.5\phantom{0} & 0.2 & 0.3 & 0.1\phantom{0} \\
Arp 299-C            & -0.2\phantom{0} & 0.2 & 0.6 & 0.3\phantom{0} \\
Arp 299-B            & -0.3\phantom{0} & 0.3 & 0.5 & 0.4\phantom{0} \\
Arp 299-A            & -0.3\phantom{0} & 0.4 & 0.5 & 0.5\phantom{0} \\
NGC 1068             & -0.5\phantom{0} & 0.4 & 0.4 & 0.3\phantom{0} \\
NGC 1365-SW          & -0.1\phantom{0} & 0.4 & 0.8 & 0.7\phantom{0} \\
NGC 1365-NE          & 0.5\phantom{0} & 0.2 & 3\phantom{.0} & 1\phantom{.00} \\
NGC 4038             & -0.2\phantom{0} & 0.3 & 0.6 & 0.4\phantom{0} \\
NGC 4038 (Overlap)   & -0.3\phantom{0} & 0.4 & 0.5 & 0.5\phantom{0} \\
M82                  & -0.4\phantom{0} & 0.3 & 0.4 & 0.3\phantom{0} \\
NGC 1222             & -0.2\phantom{0} & 0.4 & 0.6 & 0.6\phantom{0} \\
M83                  & -0.2\phantom{0} & 0.4 & 0.6 & 0.6\phantom{0} \\
NGC 253              & -0.05 & 0.4 & 0.9 & 0.8\phantom{0} \\
NGC 1266             & -0.2\phantom{0} & 0.4 & 0.6 & 0.5\phantom{0} \\
Cen A                & 0.5\phantom{0} & 0.4 & 3\phantom{.0} & 3\phantom{.00} \\
Average &    -0.2 &     0.3 &     0.3 &     1.1
\enddata
\tablecomments{See Section \ref{sec:disc:lummass}.  IRAS~09022-3615 is not included because the cold gas mass was not modeled. 
Averages do not include the following duplicate pointings: Arp~299-B, Arp~299-C, NGC~4038, NGC~1365-SW}
\end{deluxetable}

\begin{figure}
\includegraphics[height=\columnwidth,angle=270]{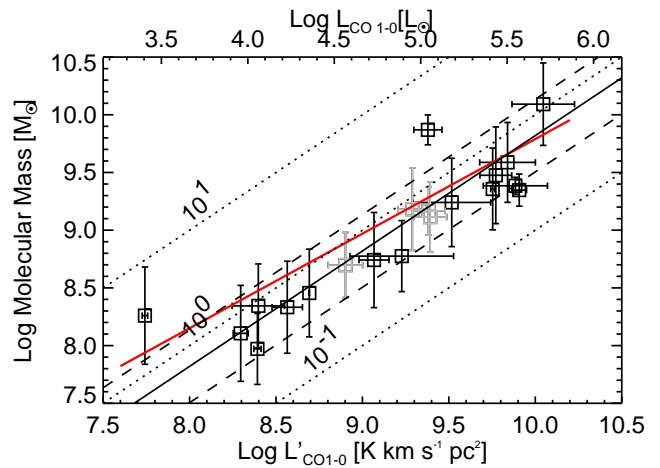} 
\caption[CO-to-\htwo\ Conversion Factor, $\alpha_{\rm CO}$]{CO-to-\htwo\ Conversion Factor, $\alpha_{\rm CO}$.  See Section \ref{sec:disc:lummass}.  The x-axis is luminosity of CO \jone, averaged if multiple lines were available.  The y-axis is the total molecular mass from the CO modeling.  The dotted diagonal lines represent 
the slopes corresponding to $\alpha_{\rm CO}$ =10, 1, and 0.1.  The solid line is our best-fit ratio, excluding the gray duplicate galaxy points, Log($\alpha_{\rm CO}$) = -0.2 $\pm 0.3$.  Dashed lines indicate the $\pm 1 \sigma$ lines.  Linearly this corresponds to $\alpha_{\rm CO}$ = 0.7 $\pm$ 0.5.
The solid red line is out best-fit when allowing for a nonlinear slope, $M = 55 (L^\prime)^{0.80}$, excluding NGC~6240.
Gray points are 
duplicate galaxy pointings excluded from line fitting.\label{fig:alpha}}
\end{figure}

The dotted lines in Figure \ref{fig:alpha} clearly demonstrate that we find $\alpha_{\rm CO} < 10$ in all cases, and in fact $\alpha_{\rm CO} \le 1$ 
in all but two cases (both with $\alpha_{\rm CO} \approx 3$), Cen A and NGC~1365-NE.  NGC~1365-SW, however, shows $\alpha_{\rm CO} = 0.8$, and we would not expect the two to be dramatically different 
(or particularly independent, given their close separation relative to the FTS beam size, see Appendix \ref{appendix:individual}). 
The discrepancy is due to the higher mass found for NGC~1365-NE.  Cen A is often an outlier in this sample; its lower gas excitation and lower mass would in fact 
lead us to expect a higher $\alpha_{\rm CO}$ for reasons described below.  Our derived values match well with those in the literature as summarized by \citet{Bolatto2013}, 
though generally on the lower end of ranges given, e.g. NGC~1068 \citep{Papadopoulos1999}, NGC~4038 \citep{Zhu2003}, Arp~299 \citep{Sliwa2013}, 
Mrk~231, NGC~6240 \citep{Bryant1999}.
We are somewhat lower for M82 \citep[$0.4 \pm 0.3$ vs. $1.2-2.4$, ][]{Wild1992}, NGC~4038 Overlap \citep[$0.5 \pm 0.5$ vs. $1.2-2.4$, ][]{Zhu2003}, 
and Arp~220 \citep[$0.6 \pm 0.5$ vs. 2.4, ][]{Scoville1997}.

Even with the two outliers, our best-fit $\alpha_{\rm CO}$ is approximately $0.66 \pm 0.48$ ($X({\rm CO}) = 0.26 \pm 0.21 \times 10^{20}$). 
This is on the low, but overlapping end of the range of estimates by \citet{Yao2003} for starburst galaxies 
($X({\rm CO}) = 0.3 \pm 0.8 \times 10^{20}$) and 
\citet{Papadopoulos2012} for LIRGS ($X({\rm CO}) = 0.3 \times 10^{20}$).  
If we instead fit a line to Figure \ref{fig:alpha}, allow for slope not equal to one, we find $M = 1312 (L^\prime)^{0.65}$, which corresponds to 
$\alpha_{\rm CO}$ from 1.9 to 0.4 over the approximate range of $L^\prime = 10^8 - 10^{10}$ (K km s$^{-1}$ pc$^{2}$).  
The bootstrapped estimate of the slope is $0.68 \pm 0.18$.
This fit is largely fixed by the very low mass error bars on NGC~6240; excluding this point, the linear fit is $M = 55 (L^\prime)^{0.80}$, $\alpha_{\rm CO}$ from $1.5 - 0.6$.  Here the bootstrapped slope is $0.79 \pm 0.18$, so we cannot rule out a linear relationship.

A variety of systematic effects could change our derived $\alpha_{\rm CO}$.  
First, using a factor of 1.36 (correcting for helium only, as is often done), rather than 1.4 would lower $\alpha_{\rm CO}$ by only a few percent.
A different value of the relative abundance of CO to \htwo\ ({\it not} the same as $X({\rm CO})$ in this section) would 
also modify our mass calculation.  We use  $3 \times 10^{-4}$; another commonly used value, $1 \times 10^{-4}$, would lower $\alpha_{\rm CO}$
by a factor of 3.  

We confirm that a conversion factor $\alpha_{\rm CO} < 1$ is appropriate for CO \jone\ emission from LIRGs and other submillimeter-bright local galaxies.
This factor may not scale linearly with CO \jone\ luminosity.
We attempted to discern a similar relationship between the CO \jsix\ emission and the warm molecular mass, but did not find one.
This is not particularly surprising; the theoretical basis for $\alpha_{\rm CO}$ relies on CO \jone\ being thermalized 
and $T_{\rm kinetic}$ being the same in all sources \citep{Maloney1988}.
We find very subthermal excitation in the \jsix\ line ($T_{\rm excitation} \ll T_{\rm kinetic}$).  The best-fit excitation temperatures for the 
CO \jsix\ line range from 3-30 K (with one, IRAS~09022-3615 at 140 K), while the kinetic temperatures are 200-3000 K, and the two are not correlated.
Even if the \jsix\ line were thermalized, the warm gas temperature varies in a non-systematic way from galaxy to galaxy.  
Thus, CO \jsix\ is a poor tracer of mass (but a good tracer of warm component luminosity, as we show in Section \ref{sec:disc:compare}).

\subsection{Gas-to-Dust Mass Ratios}\label{sec:disc:dust}

In Figure \ref{fig:summarydmass}, we show the dust mass vs. 
the molecular gas mass (for both warm and cold components).  For comparison, we also show the 1:1, 10:1, and 100:1 gas:dust mass 
ratios with dotted lines.  Our mean results (for the cold component of CO) lie between the 10:1 and 100:1 lines; the blue best-fit 
line varies from a ratio of 76 at the lowest end to 42 at the higher end, but the uncertainty in the slope is high 
enough that we cannot rule out a constant ratio (a linear dust-to-gas relationship).
(Recall that the warm mass is generally only 10\% of the contribution to the total molecular gas, but that our conclusions utilizing 
the cold component below are uncertain by a larger factor.)  

\begin{figure}
\includegraphics[height=\columnwidth,angle=270]{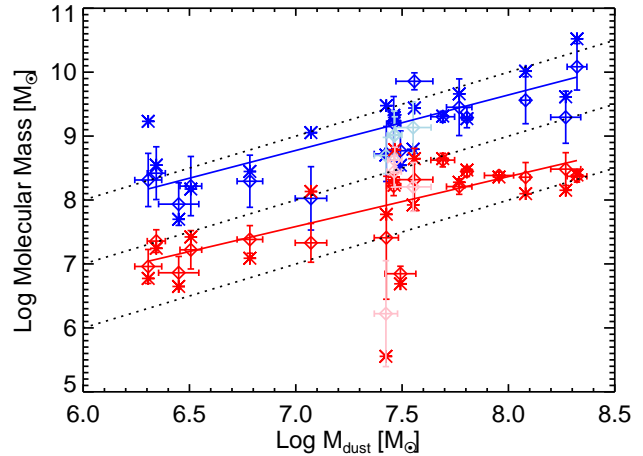} 
\caption[Gas vs. Dust Mass]{Gas vs. Dust Mass.
Log($M_{H_2, cool}) = (0.87 \pm 0.1)$Log($M_d) + (2.7 \pm 1)$, 
Log($M_{H_2, warm}) = (0.78 \pm 0.07)$Log($M_d) + (2.1 \pm 0.5)$.  The dotted lines correspond, from bottom to top, to the 1:1, 10:1, and 100:1 ratios. 
Diamonds are medians with one sigma error bars, asterisks are best-fit, red/blue represents warm/cool components, and lighter-colored points are 
duplicate galaxy pointings excluded from line fitting.\label{fig:summarydmass}}
\end{figure}

In a survey of 14 galaxies with the Submillimeter Array (half of which are in this sample), 
the average gas/dust ratio in the center of galaxies was found to be about 120 $\pm$ 28, though with large  variation \citep[][W08]{Wilson2008}.  
The gas/dust mass ratios derived from total galaxy luminosities (comparable to this work) showed even greater variation, with ratios from 29 to 725.  
Our ratios vary from about 10 to 120, with considerable error bars, but we do not find ratios in the hundreds.  
Our estimates for the dust and gas mass are both calculated differently than in W08, where the gas mass was calculated from the CO \jthree\ flux, assuming 
an area-integrated luminosity ratio of (\jthree/\jone) of 0.5 and $\alpha$ = 0.8 (recall Section \ref{sec:disc:lummass}).  
For all our available (\jthree/\jone) line ratios, we find a median of 0.7 and a standard deviation of 0.4.
The dust mass in W08 was 
calculated by scaling the 850 or 800 $\mu$m fluxes to 880 $\mu$m using $\beta = 1.5 \pm 0.5$ and $\kappa = 0.9$ cm$^2$ g$^{-1}$.
When our ratios of gas/dust are different, half the time it is due to the dust mass being very different, and the other half of the time, the gas mass.
Our molecular masses are determined by non-LTE modeling, 
and our dust masses from full SED modeling; we believe these methods to be more robust, and thus rule out gas/dust masses much greater than 120, 
our largest ratio.  
This is lower than that of the local region of the Milky Way \citep[140, ][]{Draine2007}.

\citet{Remy-Ruyer2014} found that metallicity was the most important parameter in determining gas/dust mass ratios.  
According to their broken power law model, 
gas/dust ratios of 1, 10, and 100 correspond to metallicities of 2.2, 1.2, and 0.2 dex above solar.  Gas/dust ratios 
above $\sim 160$ indicate sub-solar metallicity.  
Our ratios, ranging from 76 to 42, would correspond to metallicities of 0.3 to 0.6 dex above solar, increasing with far-infrared luminosity 
(and proportionally, star formation rate).  Their relationship (which they note is derived from data with considerable scatter), 
and our data points, are not sufficient to determine the metallicity of individual galaxies with 
precision.

\subsection{Molecular Gas Properties in Context: Comparisons Among Galaxies}\label{sec:disc:compare}

An examination of the SLED shapes in Figure \ref{fig:slednorm} shows the variety of excitation conditions present in our sample.
Some have bright mid-J excitation but turn over at high-J (e.g., M82, NGC~253).  Others remain somewhat flat at high-J (e.g., Mrk~273, Mrk~231).
On one extreme end, the CO \jten\ luminosity of NGC~6240 is over 240 times that of \jone, while for Cen A, the ratio is less than 6.

Having derived a variety of molecular gas properties for the warm and cool components (most reliably luminosity, pressure, and mass), 
we now examine which of those properties are shared among the sample and which vary with, for example, galaxy \lfir.  
To examine these relationships, 
we compared each likelihood parameter against the \lfir\ derived from the dust modeling.  No discernible relationships 
were found for temperature or density, but pressure is discussed below.  

The warm and cold component CO luminosities and masses, perhaps not surprisingly, 
are proportional to \lfir\ (Figures \ref{fig:summarylum} and \ref{fig:summarymass}); massive/luminous galaxies tend to be more luminous across 
the electromagnetic spectrum.
To determine if this were the only relation we were observing,
 we also examined the slope of the CO luminosity vs. mass for the cold (warm) component, finding a slope of 
0.8 $\pm$ 0.3 (1.3 $\pm$ 0.2), consistent with unity. 
We also looked at the CO luminosity per unit mass for each component vs. \lfir, and
found a slope of 0.4 $\pm$ 0.2 for the cold component, consistent with zero.
For the warm component, we find 0.3 $\pm$ 0.1, which may imply a non-zero relationship; that is, that the ratio 
of CO luminosity per mass in the warm component may increase slightly with increasing \lfir.
Note that in the two aforementioned figures, 
the cold and warm components are plotted separately; the {\it total} mass/luminosity is dominated by the cool/warm components, 
respectively.  
Given the large uncertainties in the slopes, we cannot discern a different relationship between mass (or luminosity) and \lfir\ 
for the warm and cold components.
The total CO luminosity is about 4 $\times 10^{-4}$ \lfir\ (Figure \ref{fig:summarylum}).  The luminosity 
in only the cold component is (2.3 - 5) $\times 10^{-5}$ \lfir.
The total CO luminosity is also well correlated with the CO \jsix\ line luminosity (Figure \ref{fig:summarylum6}).

A consequence of the aforementioned relationships is that the luminosity-to-mass ratio in the warm component is two orders of 
magnitude higher than that of the cold component.  Additionally, we do not detect any obvious outliers from the luminosity relationships 
in the AGN-dominated galaxies, though this is a small sample.  The two notable outliers in $L_{CO}$/\lfir\ in Figure \ref{fig:summarylum}, that 
lie above the trend, are NGC~1266 and NGC~6240, both discussed in Appendix \ref{appendix:individual}.

\begin{figure}
\includegraphics[width=\columnwidth]{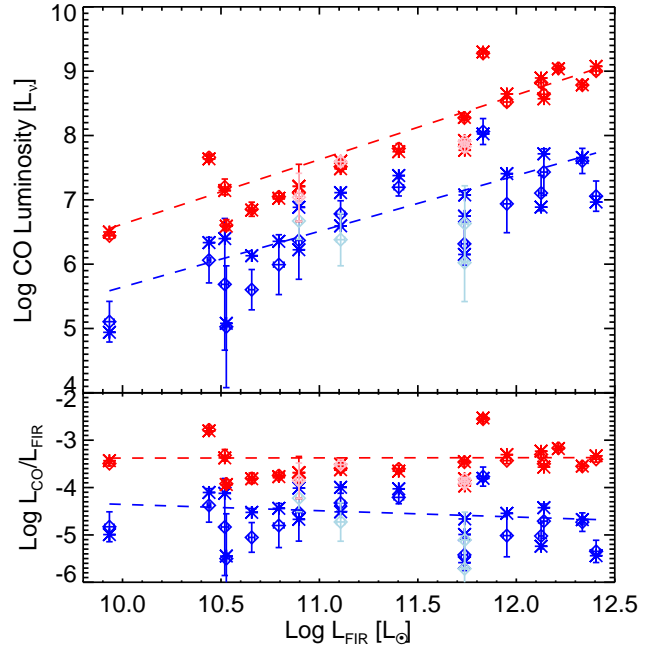} 
\caption[CO modeling likelihood results vs. \lfir: Luminosity]{CO modeling likelihood results vs. \lfir: total CO luminosity. 
Log($L_{CO, cool}) = (0.87 \pm 0.2)$Log($L_{FIR}) + (-3.0 \pm 2)$, 
Log($L_{CO, warm}) = (1.0 \pm 0.12)$Log($L_{FIR}) + (-3.4 \pm 2)$. 
Diamonds are medians with one sigma error bars, asterisks are best-fit, red/blue represents warm/cool components, and lighter colored points are 
duplicate galaxy pointings excluded from line fitting.
\label{fig:summarylum}}
\end{figure}

\begin{figure}
\includegraphics[width=\columnwidth]{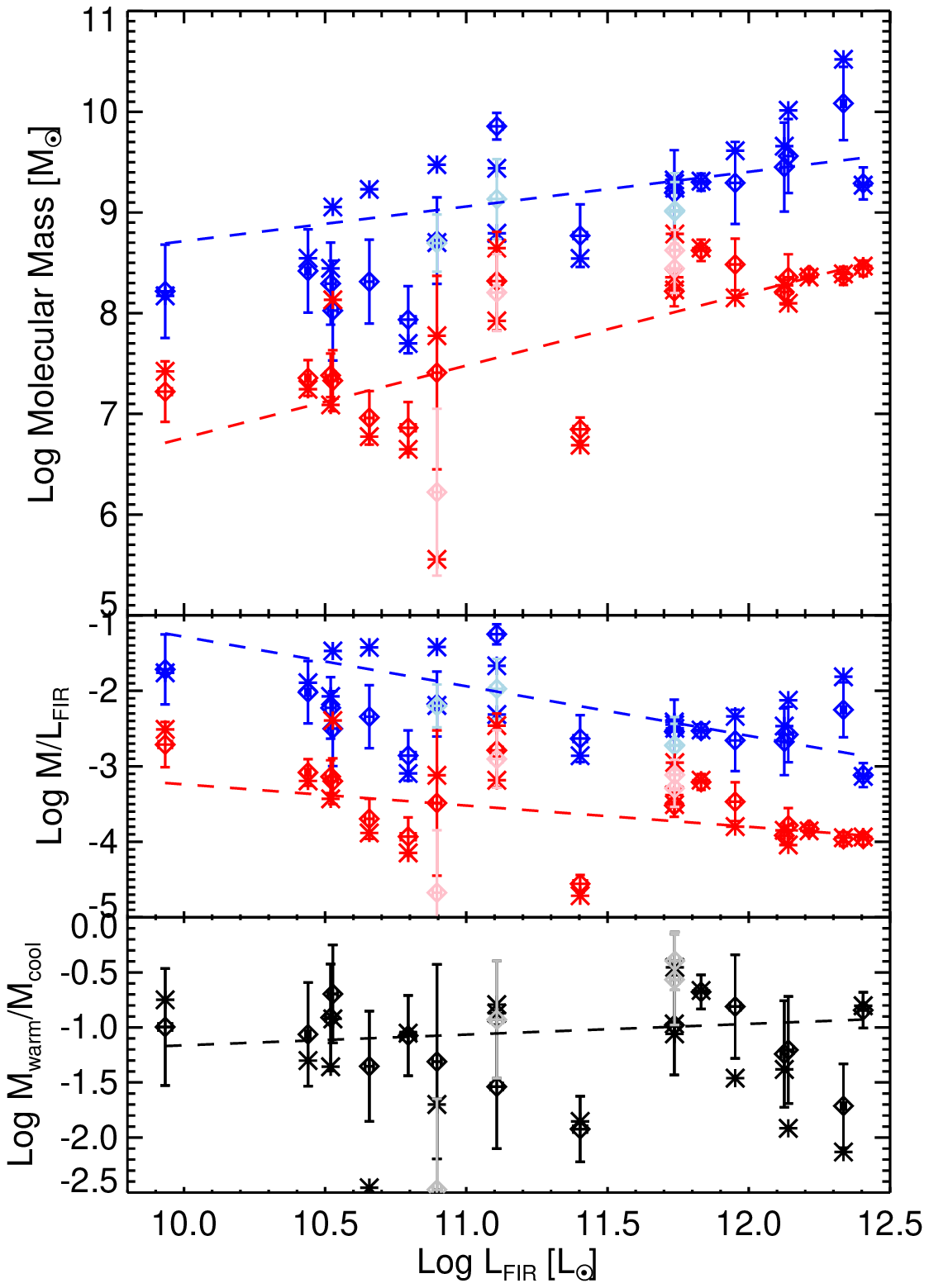}
\caption[CO modeling likelihood results vs. \lfir: Mass]{CO modeling likelihood results vs. \lfir: mass.
Log($M_{H_2, cool}) = (0.34 \pm 0.3)$Log($L_{FIR}) + (5.3 \pm 4)$, 
Log($M_{H_2, warm}) = (0.70 \pm 0.1)$Log($L_{FIR}) + (-0.4 \pm 1.6)$.
Diamonds are medians with one sigma error bars, asterisks are best-fit, red/blue represents warm/cool components, and lighter colored points are 
duplicate galaxy pointings excluded from line fitting.\label{fig:summarymass}}
\end{figure}

\begin{figure}
\includegraphics[width=\columnwidth]{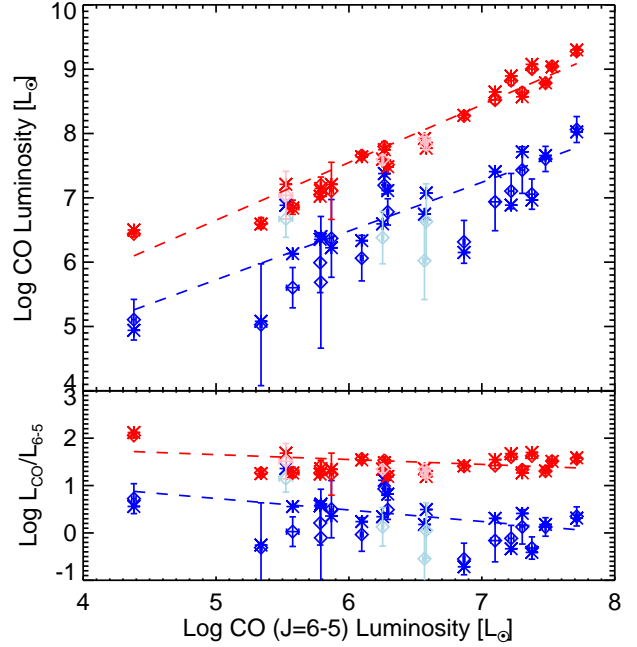}
\caption[CO modeling likelihood results vs. $L_{CO_{6-5}}$: Luminosity]{CO modeling likelihood results vs. $L_{CO_{6-5}}$: total CO luminosity.
Log($L_{CO, cool}) = (0.76 \pm 0.13)$Log($L_{CO_{6-5}}) + (1.9 \pm 1)$, 
Log($L_{CO, warm}) = (0.90 \pm 0.10)$Log($L_{CO_{6-5}}) + (2.2 \pm 1)$.
Diamonds are medians with one sigma error bars, asterisks are best-fit, red/blue represents warm/cool components, and lighter colored points are 
duplicate galaxy pointings excluded from line fitting.
\label{fig:summarylum6}}
\end{figure}

\begin{figure}
\includegraphics[width=\columnwidth]{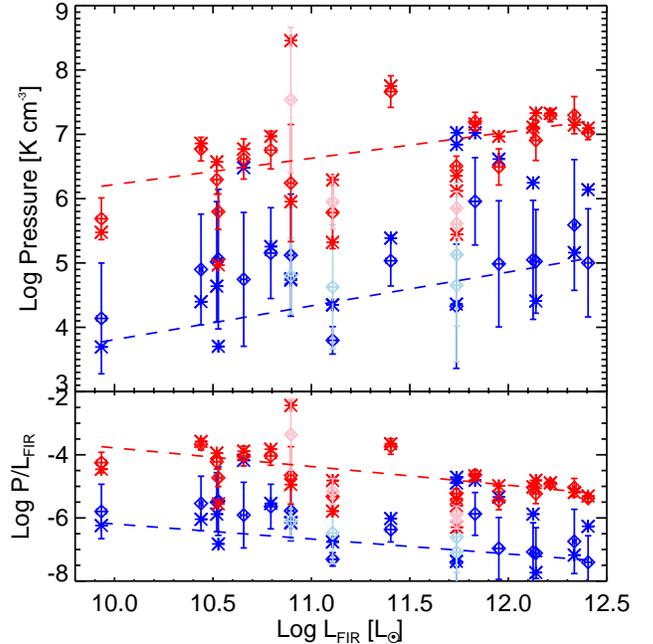} 
\caption[CO modeling likelihood results vs. \lfir: Pressure]{CO modeling likelihood results vs. \lfir: pressure.
Log($P_{cool}) = (0.5 \pm 0.3)$Log($L_{FIR}) + (-1.4 \pm 3.4)$, 
Log($P_{warm}) = (0.4 \pm 0.1)$Log($L_{FIR}) + (2.1 \pm 1.4)$.
Diamonds are medians with one sigma error bars, asterisks are best-fit, red/blue represents warm/cool components, and lighter colored points are 
duplicate galaxy pointings excluded from line fitting.\label{fig:summarypress}}
\end{figure}

In Figure \ref{fig:cohist1}, we show histograms of the warm and cool component pressures in red and blue, respectively.
We also plot the pressure histogram for molecular cloud clumps derived from the densities measured 
from the Bolocam Galactic Plane Survey (Ellsworth-Bowers et al., in prep), assuming a temperature of 10 K (solid line) 
or 30 K (dashed line).  
In addition to the temperature dependence, the BGPS distributions in Figure \ref{fig:cohist1} also depend (linearly) on the 
dust opacity used to calculate the clump mass and density \citep{Ossenkopf1990}. 
The warm component clearly has a higher pressure, independent of those assumptions, meaning it is not like most Galactic molecular clumps (though 
see the comparisons to Sgr B2 and Sgr A* in Section \ref{sec:disc:galactic})

The bulk of the molecular mass in the Galaxy is in lower pressure ($\sim 10^{4}$ K cm$^{-3}$) clouds, 
not clumps.  
Were we to be simply summing or ``counting" Galactic-type giant molecular clouds in these galaxies, we would 
expect the cold CO pressures to be similar; instead, ours are higher.
(Recall from Section \ref{sec:disc:twocomp1}, our cold pressure is 0.5 dex {\it lower} than if we modeled this 
component alone; simultaneous modeling is {\it not} the reason our pressure is {\it higher} than Galactic.)
This means that the bulk of the molecular gas in this sample of high-SFR galaxies is more energetic (higher thermal pressure, 
and hence greater thermal energy per unit volume) than the bulk of 
molecular gas in our Galaxy.  Additionally, the bulk excitation may be similar to that of denser Galactic clumps, 
but this additional interpretation relies on the aforementioned assumptions.

One explanation could be the high cosmic-ray energy densities caused by the higher star 
formation rates in these galaxies \citep{Abdo2010}.
Cosmic rays can volumetrically heat the gas, including the dense UV-shielded cores that set the initial 
conditions for star formation; cosmic-ray-dominated regions (CRDRs) heat the gas to 80-240 K in compact extreme starbursts, 
closer to the cold component temperatures we find here \citep{Papadopoulos2010a}.
Even if cosmic rays do not dominate the heating, their influence will still heat the gas more  
than PDRs alone and will increase the Jeans mass, and hence the stellar initial mass function mass scale \citep{Papadopoulos2010a}.
The higher temperatures in our cool gas component may be a direct feedback mechanism of star formation; not from the UV light of O and B stars, but from 
cosmic rays.

It could still be that we are summing/``counting" molecular clouds that typically have higher pressures than Galactic clouds; in that case, 
we would expect the mass and luminosity to increase with increasing galaxy mass or luminosity, but the average
pressure to remain the same.  Though it is hard to discern a relationship between pressure and luminosity (Figure \ref{fig:summarypress}), we find 
best-fit slopes of 0.53 and 0.41 for the cool and warm components, respectively.  The bootstrap method yields errors on these parameters 
of $\pm$ 0.27 and 0.12.  For the cold component, we cannot exclude a zero slope at the 2$\sigma$ level, but for the 
warm component, the bulk average pressure appears slightly correlated with \lfir\ (to a significance of 3.4$\sigma$)."
This implies that the energetics of the warm component in these galaxies are different, not that we are just viewing ``more" of the 
same molecular gas components with increased \lfir.  The slight correlation between the luminosity/mass ratio and \lfir\ for the warm component may also be 
indicative of this.

Other properties we sought to investigate were the relative {\it ratios} between the warm and cool component pressure, mass, and luminosity, 
shown in Figure \ref{fig:cohist2}.
We did not detect any trend with \lfir\ (or SFR), the presence of an AGN, $L_{CO_{6-5}}$, or dust mass.  On average, the log ratios of the warm/cold CO pressure, mass, and luminosity
were $1.8 \pm 0.2$, $-1.0 \pm 0.08$, and $1.2 \pm 0.08$.  Linearly, these correspond to ratios of $60 \pm 30$, $0.11 \pm 0.02$, and $15.6 \pm 2.7$.  
The pressure is the least well determined ratio.  It is dependent upon the relative shapes of the SLEDs of the two components; they can ``trade off" 
a significant amount in the mid-J lines and still fit the overall shape.  We find that the two components are not in pressure equilibrium; once equilibrium is not enforced, we have no expectation for what the ratio should be.  
Aside from the broad constraints that the gas be both dense enough and cool enough to be molecular, but not so dense and cold that 
the timescale for gravitational collapse is short compared to a dynamical time, there is no obvious limitations on the allowed ranges 
of $n$ and $T$, so we might expect a broad distribution.  Furthermore, if we see such different excitation among galaxies, as described in the previous paragraph, 
we would also expect different distributions of excitation mechanisms within galaxies.  
The mass and luminosity are global properties (a sum), whereas the pressure is a local quantity (here, an average).  Mass and luminosity are anchored by the lowest-J (for cool) and highest-J (for warm) lines.  This reaffirms previous conclusions in the literature, from studying individual galaxies, that the low-J CO dominates the mass 
and the high-J CO dominates the luminosity and hence the cooling \citep[e.g.][]{Kamenetzky2012,Rangwala2011,Spinoglio2012,Rigopoulou2013}.

\subsection{Carbon in Various Forms: C, C$^{+}$, CO}\label{sec:disc:c}

Near newly formed, bright O and B stars, 
CO only exists where it is adequately shielded from dissociation by UV photons.  In the traditional model of a molecular cloud, 
this will be in the interior of the cloud, surrounded by a transition layer in which carbon is mostly neutral and atomic, but the hydrogen is 
still substantially molecular, and then another layer in which the carbon is mostly ionized and the hydrogen atomic \citep{Hollenbach1997}.
The molecular gas in the transition layer will not be traced by CO emission, so it is referred to as 
``CO-dark."  The PDR models of \citet{Wolfire2010} indicate CO-dark gas may account for 30\% of the molecular gas mass.
New observations indicate that a significant fraction of molecular gas in the Milky Way 
is CO-dark \citep{Pineda2013}.  Here we compare the view of galaxies studied with C, C$^{+}$, CO, and dust.

We first discuss the column density ratios (and overall relative abundances) of C, C$^{+}$, and CO.  
Our beam-averaged column densities for the cold and warm components of CO are presented in Table \ref{tbl:co2}, 
and the total mass from \ci\ in Table \ref{tbl:ci}.  Though we could not calculate excitation temperatures for 
\cii\ to use in the equations in Section \ref{sec:CI}, we used 150 K which corresponds to a fraction of atoms 
in the upper state of 0.52 (the fraction approaches 2/3 for $T \gg 92$ K).  
Even with this assumption, the uncertainties in the column density ratios are dominated by the 
uncertainties in the CO column densities.
The distributions are shown in Figure \ref{fig:ccolumns}; 
we found a median $N_{\rm C}/N_{\rm CO} \sim 0.5$ and a median $N_{\rm C^+}/N_{\rm C} \sim 0.5$.
Almost all $N_{\rm C}/N_{\rm CO}$ values are less than 1, in line with those reported in \citet{Wilson1997}, except for NGC~6240, 
which appears to have \ci\ emission even more abnormally luminous than its CO emission.  There is large dispersion in these ratios.  

\begin{figure}
\includegraphics[width=\columnwidth]{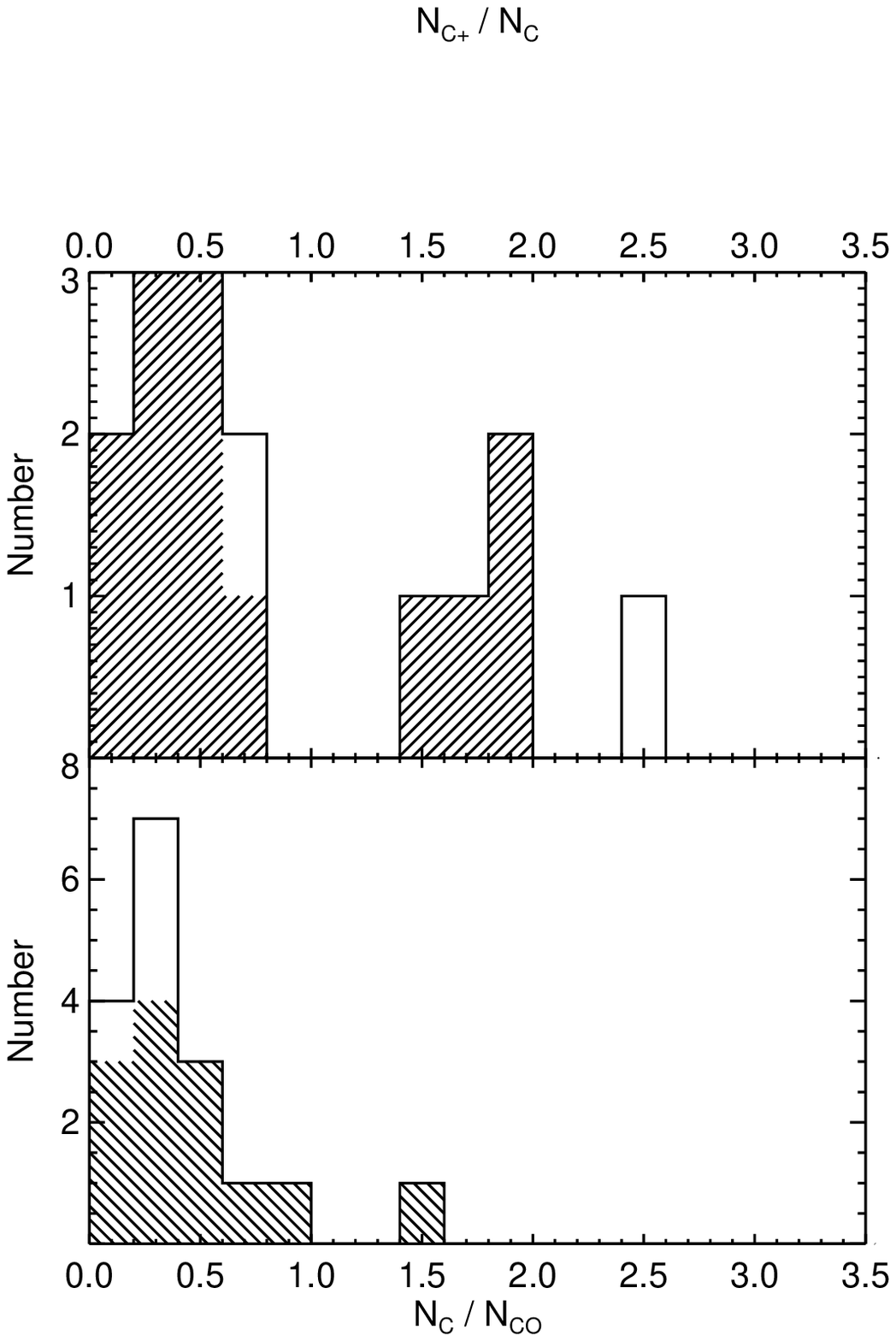} 
\caption[Column Density Ratios of C, C$^+$, CO]{Column Density Ratios of C, C$^+$, CO.  Top panel is the ratio of C$^+$ to C column density,
bottom panel is ratio of C to CO column density.  Duplicate pointings of galaxies are not filled in by diagonal lines.  The only galaxy not included in the bottom 
plot is NGC~6240, with an abnormally high ratio of 38.\label{fig:ccolumns}}
\end{figure}

We next turn our attention to the temperatures derived from the two neutral atomic carbon lines in our spectra.  
The excitation temperatures of \ci, 
shown in Table \ref{tbl:ci} appear to be clustered between 20-40 K, regardless of 
galaxy, and are not correlated to other measures, such as total infrared luminosity, cold CO temperature, or dust temperature (Figure \ref{fig:temperatures}).
Without correcting fluxes 
for the dust absorption, which affects the \jtwo\ line slightly more than the \jone, the derived temperatures would be about 0.5 to 5 K lower.
We found an average temperature of 26.3 $\pm$ 9.2 K, in agreement with the 29.1 $\pm$ 6.3 K cited for a sample of high-$z$ galaxies \citep{Walter2011}.
This indicates that neutral C is likely tracing the same cool component of gas across a range of galaxy luminosities and redshifts, and is therefore 
not a particularly good distinguisher of excitation conditions.  

\begin{figure}
\includegraphics[height=\columnwidth,angle=270]{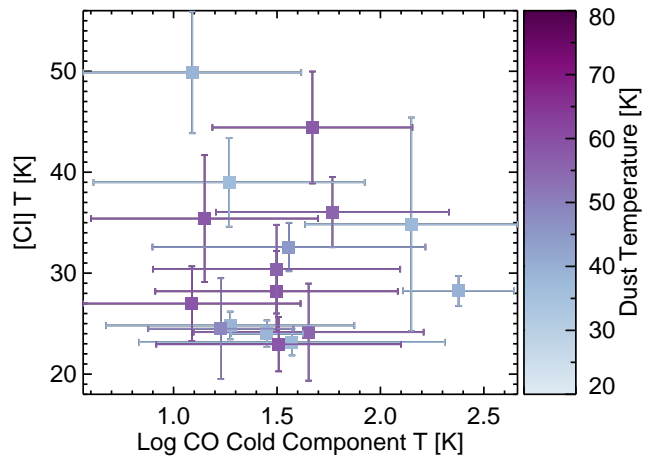} 
\caption[Carbon, CO, and Dust Temperatures]{Carbon, CO, and Dust Temperatures.  Note that the y-axis and color bar are linear.  There is no apparent correlation 
between temperatures.\label{fig:temperatures}}
\end{figure}

However, \citet{Papadopoulos2004b} proposed using the 
\ci\ \jone\ line to measure global molecular gas mass, finding good agreement with 
molecular mass measured by CO.   
We also found a correlation between neutral C mass and molecular mass measured from CO, consistent with a linear relationship (Figure \ref{fig:cimass}).
They adopted a relative abundance $X_{\rm [CI]/ H_2} = 3 \times 10^{-5}$.
The weighted average abundance of $X_{\rm [CI]/ H_2}$ = $M_C/(M_{gas}/1.4) \times m_{H_2}/m_C$; using the gas mass derived
from the cold component CO fitting (and therefore dependent upon $X_{CO}$), we found log($X_{\rm [CI]/ H_2}$) =  -4.0 $\pm$ 0.5.  
In a linear ratio, this is $10^{+20}_{-7} \times 10^{-5}$, a not particularly well-constrained value but consistent with 
the values presented in \citet{Papadopoulos2004b} for the Cloverleaf quasar, the Orion A and B clouds, and the nucleus of M82 (ranging from $1-5 \times 10^{-5}$), 
as well as with the mean of $8.4 \pm 3.5 \times 10^{-5}$ reported for high redshift galaxies in \citet{Carilli2013}.
We required a higher average abundance of \ci\ to match our CO-derived gas mass values than in \citet{Papadopoulos2004b}, and this discrepancy 
is almost entirely due to our higher measured \ci\ fluxes for the three galaxies that our samples share (NGC~6240, Arp~220, and Mrk~231).  
We differed in how we determined the populations of the J levels: we used both \ci\ lines, calculated an excitation temperature, and then used Boltzmann distributed populations, as opposed to using one line and an estimate based on the gas conditions as in \citet{Papadopoulos2004a}.  In the end the 
population levels were roughly the same, but the different line fluxes caused the difference in \ci\ mass.  We confirm the conclusions 
of \citet{Papadopoulos2004b}: \ci\ is as good of a tracer of total molecular mass as radiative transfer modeling of CO, 
though further refinements of the value of $X_{\rm [CI]/ H_2}$ will aid in its precision.

\begin{figure}
\includegraphics[height=\columnwidth,angle=270]{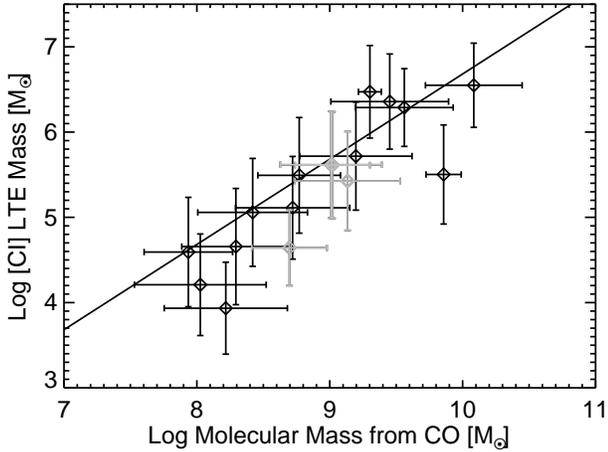} 
\caption[CI and CO mass]{CI and CO mass.  The horizontal line represents the average ($X_{\rm [CI]/ H_2}$) in Section \ref{sec:disc:c}.  \label{fig:cimass}}
\end{figure}

Our finding of differing temperatures between dust, \ci, and CO, is in line with the findings of others in the literature.
For example, \citet{Mangum2013} found the temperatures derived from ammonia (NH$_{3}$), a well-known kinetic temperature probe, 
differed from the dust temperatures for a sample of star-forming galaxies (seven of which are in our sample).  
We found higher dust temperatures by modeling the full SED instead of just using the 60 and 100 $\mu$m flux densities (6 to 11 K higher), but still confirm that $T_{\rm NH_3} > T_{\rm dust}$, and add $T_{\rm [CI]} < T_{\rm dust}$.
They concluded that dust temperature should not be used as a proxy for gas temperature, and that 
higher gas temperatures of NH$_3$ may be caused by turbulence and/or cosmic ray heating, not just radiative processes.  
\citet{Carilli2013} also note that ``the heating and cooling processes of the dust and molecular gas phases are quite different, 
and therefore thermal balance is not required."  
We confirm these findings, and add the additional caution that gas temperatures from different atomic or molecular species 
are likely tracing different conditions.  Specifically, \ci\ is not tracing the higher temperature gas (certainly not that of high-J CO, and possibly not that of low-J CO).
This means \ci\ is likely measuring neutral atomic gas unaffected by star formation \citep{Carilli2013}.

The picture becomes more complicated when considering the \cii\ 158 $\mu$m line, which has been found to be emitted by from a variety of sources, and 
may be tracing the CO-dark \htwo\ gas described above.
(Note that $\alpha_{\rm CO}$ is not sensitive to reservoirs of \htwo\ where C$^{+}$ or C 
is the dominant form of carbon.)
\citet{Pineda2013} found, via a study of the Galactic center, that \cii\ emission is produced by a combination 
of PDRs ($\sim$ 47\%), CO-dark \htwo\ gas ($\sim$ 28\%), cold atomic gas ($\sim$ 21\%), and ionized gas ($\sim$ 4\%).
\citet{Langer2014} also studied the column densities of CO-dark \htwo\ gas of individual clouds, a level of detail we do not have here.
We have already discussed some differences between our galaxies and the Milky Way; can these distributions be valid in 
starburst galaxies?
In Section \ref{sec:nii} and Table \ref{tbl:ciilines}, we presented the 
estimated percentage of C$^+$ emission from ionized gas using line ratios, and found that in most cases, the fractions are higher than 4\%, 
with a median of 14-25\%.  They are not correlated with \lfir\ or \cii/\lfir. 
This matches with the $27$\% (error range 19-46\%) found for the Carina Nebula \citep{Oberst2006}.
However, we cannot say anything about the distribution of the remaining source contributions, only that there is less (proportional) \cii\ line emission from 
the sum of PDRs, CO-dark \htwo\ gas, and cold atomic gas in these galaxies than in the Milky Way.  
\citet{Pineda2013} found that the fraction of mass from CO-dark \htwo\ increases with Galactocentric distance, from 20\% at 4 kpc to 80\% at 10 kpc .
Because the emission from our galaxies is more akin to that of the Galactic center, we expect 
lower fractions of CO-dark \htwo\ gas than in the Milky Way as a whole. 
A study similar to that of \citet{Pineda2013} 
and \citet{Langer2014} could be conducted for the nearest galaxies or Milky Way satellites comparing the distribution 
of HI, C$^+$, $^{12}$CO and $^{13}$CO, and possibly applied to this sample of galaxies.

Even absent formal modeling of PDRs, we see a picture that contradicts traditional PDR models, even with additional heating from mechanical 
turbulence or enhanced cosmic rays \citep[e.g.][]{Wolfire2010}.
Detailed studies of individual galaxies have consistently found that PDR models cannot explain the large luminosities in the mid- to high-J CO lines:
Arp~220 \citep{Rangwala2011}, M82 \citep{Kamenetzky2012}, M83 \citep{Wu2014}, NGC~6240 \citep{Meijerink2013}, Cen A \citep{Israel2014}, NGC~891 \citep{Nikola2011}, 
and the Galactic center \citep{Bradford2005}. 
In only a few instances have PDRs been found to be adequate, namely the Antennae \citep{Schirm2014}, IC324 \citep{Rigopoulou2013}, 
and the outer star-forming ring of NGC~1068 \citep{Spinoglio2012}.
Additionally, the low $\alpha_{\rm CO}$ we find requires some combination of higher temperatures 
(thereby raising the emissivity provided the line remains optically thick) or non-virialized molecular clouds.
Cosmic-ray dominated regions (CRDRs) could explain the elevated CO temperatures \citep{Papadopoulos2010a} or be combined with PDRs \citep{Meijerink2011}, which would imply a higher Jeans mass 
as a consequence.
The concurrence of evidence presented here (and in the cited literature) confirms that {\it high-J CO emission is generally powered by non-radiative processes}, 
a conclusion which future models must take into account.

\subsection{Comparison to the Galactic Center: Sgr A* and Sgr~B2}\label{sec:disc:galactic}

We have already compared our pressure distributions to those of molecular clumps in the Galactic plane (Section \ref{sec:disc:compare}).
Two specific regions in the Galaxy are more comparable to the galaxies in our sample: the warm gas and dust heated and ionized by the 
massive stars orbiting Sgr A*, and the giant molecular cloud Sgr~B2, approximately 120 pc away from Sgr A*.  The CO SLEDs of these 
observations were included in Figure \ref{fig:slednorm}.

\citet{Goicoechea2013} found the CO SLED from \jfour\ to \jtwentyfour\ in the warm gas within 1.5 pc of Sgr A* was consistent with either a single 
component of gas (T = $10^{3.1}$ K, n $\le 10^4$ cm$^{-3}$, pressure $\le 10^{7.1}$ K cm$^{-3}$) or multiple cooler components at higher density.  
In the single component case, this hot gas must fill a small fraction of the volume (not homogeneously distributed), and requires excitation in 
addition to PDRs.  Despite its proximity to our Galaxy's central black hole, the X-ray luminosity is too low to create an XDR, and cosmic rays would 
only heat the gas to a few tens of K.  The authors suggest low-density shocks contribute to the heating of this hot molecular gas, 
though it is unclear if they are from in-falling gas, clump-clump collisions, or outflows from stellar winds or protostars.

\citet{Etxaluze2013} resolved the three main compact cores, Sgr~B2(N), Sgr~B2(M), and Sgr~B2(S) from the extended envelope of the Sgr~B2 molecular cloud 
in both dust and molecular line emission and absorption.  In addition to determining the dust properties over a $\sim$ 58 arcmin$^2$ map, they mapped  
the line emission from the CO \jfour\ to \jeleven\ lines.  While the \jsix\ warm gas emission is spread over the molecular cloud, that of \jeleven\ 
is highly concentrated around the compact cores.  They conduct non-LTE modeling of the CO emission for the B2(N) and B2(M) cores and require two components 
(starting from the \jfour\ line, not \jone), which they denote as warm extended emission (60 and 100 K) and hot compact emission associated with the cores (560 and 320 K).
The log(pressure) for the warm components are 6.8 and 7.4, and for the hot components, 8.7 and 8.5, respectively, for B2(N) 
and B2(M).
For our galaxy-averaged spectra, {\it the pressures for our warm component are consistent with those of the Sgr~B2 extended molecular cloud emission}, 
and lower than that of the hot components.  

While very high molecular gas temperatures are not found in the Galactic plane as a whole, 
they are found in Sgr~B2 and Sgr A*.
Though we cannot resolve molecular clouds in nearby galaxies, it is clear that the high-J lines are emitted from
regions of highly excited gas.  
As one progresses from lower to higher J, the area filling factor of the emitted region becomes progressively smaller, 
as was demonstrated by \citet{Etxaluze2013} for Sgr~B2.  
The SLEDs of the Sgr~B2 cores, shown in Figure \ref{fig:slednorm}, peak at higher-J than the mid-J peak of the extended Sgr~B2 molecular 
cloud envelope (which is more similar to our galaxies).  This means that our warm component emission is likely dominated by 
regions resembling the 
warm extended molecular cloud envelopes (whose pressure matches those we measure here), not star-forming cores (of a higher pressure).
While such compact regions are undoubtedly present, it must be at a lower level, so the bulk of the emission we measure is from the extended molecular clouds, not cores.
We tested this by examining the total integrated flux of the Sgr~B2 SPIRE FTS map, as one would measure if it were a
distant point-source.  The resulting SLED is similar to that of the Sgr~B2 molecular cloud, not the cores, despite their brightness in high-J lines.
As discussed in Section \ref{sec:disc:twocomp2}, 
we know there are gradients in physical conditions in our SLEDs, as we saw from LTE analysis of \htwo\ lines; 
the emission from cores (the hottest material) contributes an undetectably small fraction of the total high-J CO emission over the whole galaxy.
The broad implications for future modeling are clear: the excitation conditions and the 
geometries (filling factors) of the mid- to high-J lines are different from e.g., \jone, but not 
completely independent from one another.  ALMA can achieve unprecedented spatial resolution 
in observations of the \jsix\ line; such information 
can be used to place further prior information on CO modeling and possibly disentangle the multiple components (e.g. 
the analogues to those seen in the Galactic plane, 
the Sgr~B2 molecular cloud, and the Sgr~B2 cores) within nearby galaxies.
The PACS instrument onboard {\it Herschel} was also able to observe even higher-J lines than SPIRE; \citet{Hailey-Dunsheath2012} was 
able to detect the even higher-excitation component of gas that lies beyond the SPIRE spectrum in NGC~1068.


\section{Conclusions}\label{sec:conclusions}

We presented spectra of 17 infrared-luminous galaxy systems at 21 different pointings observed 
with the {\it Herschel} FTS from 450-1550 GHz.
We have created a uniform, consistent pipeline which can perform analysis of such spectra, including source-beam coupling corrections, line fitting, and 
an 8 parameter likelihood analysis of the warm and cool CO gas for each source.  Such analysis for nearby galaxies can, at this time, only be performed with {\it Herschel} data, which contain enough CO lines to construct SLEDs up to \jthirteen.  

{\it Is the highly luminous warm component of molecular gas found in early {\it Herschel} SPIRE FTS studies generically present in high star-formation rate galaxies, 
and how does its presence modify what we already knew about cool molecular gas?  Accounting for the warm gas component luminosities, 
what $L_{\rm CO}$/\lfir\ are observed, and what other properties vary with e.g., \lfir?}
We found that high-excitation molecular gas is ubiquitous in this sample of 
galaxies with Log(\lfir) from $\sim$ 10 to 12.5.  
We clearly distinguish a low-pressure/high-mass component traced by low-J lines from a 
high-pressure/low-mass component in all systems from their CO SLEDs.
Most of the CO luminosity is emitted from the warmer, high-pressure component; 
the total CO luminosity is about $4 \times 10^{-4}$ \lfir, and is well-traced by the CO \jsix\ line.
The ratios of the warm/cold pressure, mass, and 
CO luminosity were $60 \pm 30$, 0.11 $\pm$ 0.02, and $15.6 \pm 2.7$.
Though the molecular gas mass and CO luminosity scale linearly with \lfir, the highly excited 
molecular gas pressure is proportional to \lfir$^{(0.4 \pm 0.1)}$, suggesting higher 
excitation per bulk mass of molecular material.  
The total mass of the low-pressure molecular gas is well traced by the CO \jone\ 
line (and \ci), but the pressure will be overestimated if not modeled simultaneously 
with the high-pressure component.
Likewise, future interpretation 
of high-$z$ CO emission, which often must be derived from just a few CO lines, 
should take the ratios derived here into account.

{\it Given the broad range of CO lines we now have with {\it Herschel}, can we reassess the CO luminosity-to-mass conversion factor, and dust-to-gas mass ratios?}
We found a luminosity-to-mass conversion factor of $\alpha_{\rm CO} \approx 0.7$, 
consistent with higher temperatures and/or non-virialized gas motions {\it in this low-pressure 
gas component}.  We measure gas-to-dust mass ratios of $\le 120$ (best-fit relationship from 42 to 76), though the CO gas and dust 
temperatures are not related.  

{\it What are the relative column densities of C$^+$, C, and CO in nearby starburst galaxies; in other words, where is the carbon not locked up in grains?}
We found $N_{\rm C}/N_{\rm CO}$ and $N_{\rm C^+}/N_{\rm C} \sim 0.5$, and that \ci\ can be used to measure total molecular mass.  \ci\ temperatures 
are clustered around 20-40 K.

{\it How do the molecular gas pressures of the galaxies in our sample compare to Galactic giant molecular clouds?}
Cosmic rays may be responsible for heating the CO 
gas above the very cool temperatures found in PDR models, and above that of Galactic molecular cloud clumps.
The high-pressure molecular gas excitation is consistent 
with the excitation in the extended molecular cloud 
emission of Sgr~B2 \citep{Etxaluze2013}.  The compact cores Sgr~B2(N) and Sgr~B2(M) 
are more highly excited than we measured in the extragalactic SLEDs; 
such emission is undoubtedly present, but being emitted from significantly 
smaller and lower mass regions, it cannot be resolved by non-LTE modeling 
of the CO SLEDs using discrete components.

{\it What are the systematic consequences of a discrete two-component model?}
The CO SLEDs available from {\it Herschel} are an enormous improvement upon previously 
available information; however, they still only offer $\le$ 13 data points for modeling 
what is likely a very complex distribution of gas conditions.  
We showed that 
while discrete two-component models may be an average over varying gas conditions 
(where different pressures may indicate different underlying distributions of pressures), 
they provide a sufficient fit to the data, and describe the SLEDs as well as adding an additional component would.
Likewise, powerlaw distributions of gas conditions could likely also 
fit the SLEDs well.  Further exploration of distributions of physical conditions is needed.  

Numerical simulations could place useful priors on models to 
help distinguish the physical parameters of such highly excited molecular gas; 
such information will be necessary to fully understand the excitation and feedback 
mechanisms taking place in star-forming galaxies. 
Additionally, ALMA is now offering opportunities to spatially map the 
distribution of this warm molecular gas and complement our galaxy-integrated observations.  
In most of these galaxies, non-radiative processes, such as shocks, turbulence, 
and stellar winds are required for the high-pressure molecular gas excitation.

\acknowledgments

SPIRE has been developed by a consortium of institutes led by Cardiff University (UK) and including Univ. Lethbridge (Canada); 
NAOC (China); CEA, LAM (France); IFSI, Univ. Padua (Italy); IAC (Spain); Stockholm Observatory (Sweden); Imperial College London, 
RAL, UCL-MSSL, UKATC, Univ. Sussex (UK); and Caltech, JPL, NHSC, Univ. Colorado (USA). This development has been supported by national 
funding agencies: CSA (Canada); NAOC (China); CEA, CNES, CNRS (France); ASI (Italy); MCINN (Spain); SNSB (Sweden); STFC (UK); and NASA (USA).
We would like to thank Rosalind Hopwood for HIPE reprocessing.
This paper is based upon work supported by NASA under award number NNX13AL16G.
John Bally provided excellent discussion which positively influenced the analysis of this work.

{\it Facilities:} \facility{Herschel (SPIRE)}.

\appendix

\section{Notes on Individual Galaxies}\label{appendix:individual}
In this appendix, we discuss both the treatment of extended galaxy data, when applicable, and 
compare to previous modeling results, which when done on an individual basis, 
often include more specifics than presented here.  To create a reasonable pipeline for application 
for a large number of galaxies with limited supplemental information, we did not include such information, 
but notable cases are discussed here.  Additionally, Figure \ref{fig:multiple} illustrates the 
position of the FTS beam for the three galaxies with multiple pointed observations.

\begin{figure*} 
\includegraphics[height=\textwidth,angle=270]{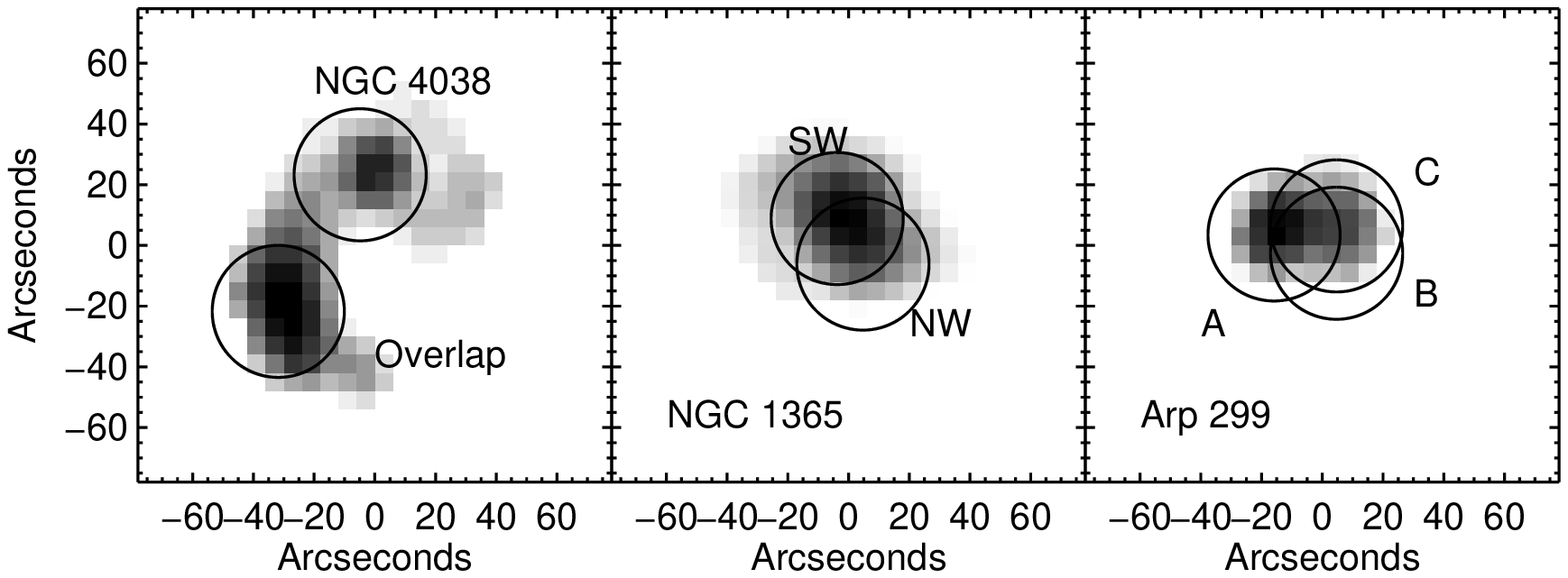} 
\caption[Locations of Multiple Pointings for 3 Galaxies.]{Locations of Multiple Pointings for 3 Galaxies.
Each circle represents the SPIRE-FTS beam with FWHM of 43\farcs5.\label{fig:multiple}}
\end{figure*}

\subsection{The Antennae: NGC~4038/Overlap}

This pair of merging galaxies is very extended on the sky (Figure \ref{fig:multiple}).  
Our \htwo\ lines come from \citep{Brandl2009}, which measured the emission at multiple locations in the overlap region.
We use a sum of Peaks 1, 2, 3, and 5 for this region; S(3) is an upper limit.  Only one pointing was 
measured for the NGC~4038 nucleus, which we only use for S(0), due to the small slit size of S(3).  
We instead use \citet{Rigopoulou2002} for S(1) and S(2).

The full FTS maps, from which our spectra are drawn, were modeled with RADEX by \citet{Schirm2014}.
Two components were also used, but using an iterative process to model the low-J lines, then high-J lines, and back and forth 
until convergence.  Additionally, they modeled multiple pixels in an FTS map, but we focus here 
on the two corresponding to the NGC~4038 nucleus and the Overlap region.  Their best-fit CO SLEDs 
indicate less warm component emission contribution to the mid-J lines than ours for NGC~4038, and more for 
the Overlap region.  This illustrates how an iterative modeling approach can restrict the allowed parameter space 
somewhat artificially, whereas our simultaneous modeling allows for more possibilities of ``trade-off" between 
the two components.  We recover the same total mass in the cold component (of both), with smaller uncertainties, though 
lower pressures (with overlapping 1$\sigma$ ranges) for NGC~4038.  For the warm component, we have higher uncertainties in the mass 
(though overlapping distributions), and lower pressures (only the NGC~4038 pressure distribution 1$\sigma$ range overlaps).
Recovering the same total mass, given the similarities of our methods, is very reassuring.  \citet{Schirm2014} also 
finds consistent \ci\ LTE temperatures of 10-30 K.

\subsection{M82}

For low-J CO lines, we use the measurements from \citet{Ward2003}, which are actually 
at two separate locations in M82, separated by $\sim 19$\as.  The correct value of $\eta$ 
for their 24\farcs{4} beam is 0.50, and we then used the photometer maps to determine the ratio 
of the flux when centered on the two separate lobes (Ward's measurements) 
to that when pointed at the center (our measurements).  We determined that we should sum the two fluxes 
and then divide by 0.29.  This is close to the number one would use if treating the two 
fluxes as uniformly extended (the ratio of the two beam sizes is 0.32).

A similar CO modeling analysis was done in both \citet{Panuzzo2010} and \citet{Kamenetzky2012}, using an 
iterative approach instead of simultaneous modeling.  We find similar results for the cold component pressure and mass, 
but slightly higher warm component pressure and mass.  Our Log($P_{\rm warm}$) is 6.8 $\pm$ 0.3 compared to $6.6^{+0.2}_{-0.5}$
in \citet{Kamenetzky2012}, and Log($M_{\rm warm}$) is 6.9 $\pm$ 0.3 compared to $6.2^{+0.5}_{-0.2}$.  As discussed 
for the Antennae, we would not expect exactly the same results when both components are allowed to vary against one another.

\subsection{NGC~1068}

NGC~1068's geometry poses a unique challenge for the FTS, because of the separate emission from the central circumnuclear disk (CND, 
$\sim 4$\as\ diameter) and the larger star-forming ring (SF-ring, $\sim 40$\as\ diameter), both of which are contained 
in the SPIRE beam.
\citet{Hailey-Dunsheath2012} found the emission from \jfourteen\ through \jthirty\ contained a clear inflection point, implying 
two components of medium (P = $10^{7.8}$ K cm$^{-3}$) and high (P = $10^{9.2}$ K cm$^{-3}$) excitation; 
this emission was coming from the central 10\as, 
with the high-excitation component blueshifted by 80 \kms.
\citet{Spinoglio2012} subtracted the medium-excitation component contribution from the lines in the FTS SSW detector
(\jnine\ to \jthirteen), which mainly originate from the CND.  The remainder was modeled with RADEX, and the contribution 
to the lower-J lines in the SLW were subtracted; that remainder was modeled again with RADEX to describe the SF-ring.  
Their 1$\sigma$ ranges for the log CND pressure were 6.5-6.8, and for the SF-ring, 4.3-5.2.
The pressure and mass for the SF ring overlaps with our cool component; our warm component is at a higher log pressure ($7.7 \pm 0.2$) 
because we model all of the emission through the \jthirteen\ lines; subtracting the ME component from \citet{Hailey-Dunsheath2012} 
drove their pressure lower.

In our two-component model, the warm component is likely dominated by emission from the CND, whereas the cool component 
may include significant contributions from both the CND and SF-ring.  We note that galaxy-integrated photometry fluxes, used to derive 
\lfir, dust mass, SFR, and stellar mass, will mask the underlying differences between the molecular gas in the CND and the SF-ring, 
influencing NGC~1068's place on e.g. the galaxy main sequence.

\subsection{NGC~1266}

NGC~1266 is unusual for a few reasons.  First, it contains a large concentration of \htwo\ in its nucleus, but 
shows no sign of an interaction or merger.  Second, \citet{Alatalo2011} found evidence 
for a large molecular outflow via high-resolution CO spectra; the wings of the lines require a low-amplitude, broad Gaussian 
to be fit properly.  We do not attempt to separate the relative contributions of the central velocity component and 
outflow in our line fits and modeling, though such work is in progress (Glenn et al., in preparation).  The possible consequence 
of our treatment is that the conditions we find may be an average between the conditions of the central and outflow components; they 
may be distinct from our average.

\subsection{Arp~299}

The FITS images presented in Figure 2 of \citet{Sliwa2013} were given to us by private 
communication with the author, such that we could convolve each map up to our 43\farcs5 beam 
and determine the integrated flux at each of the three pointings (A, B, and C) for CO \jtwo\ and \jthree.

\subsection{Arp~220}

\begin{figure}
\center
\plottwo{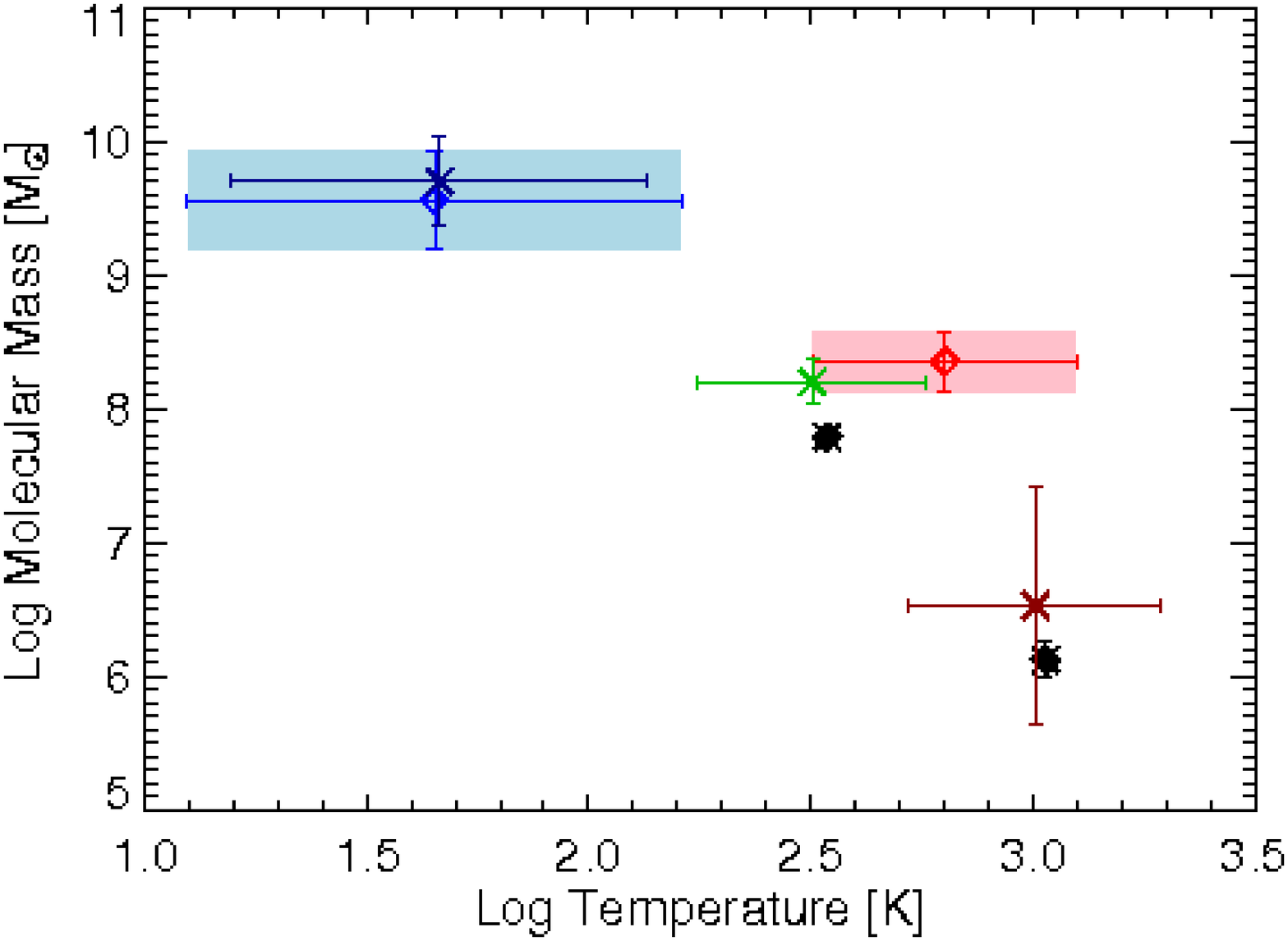}{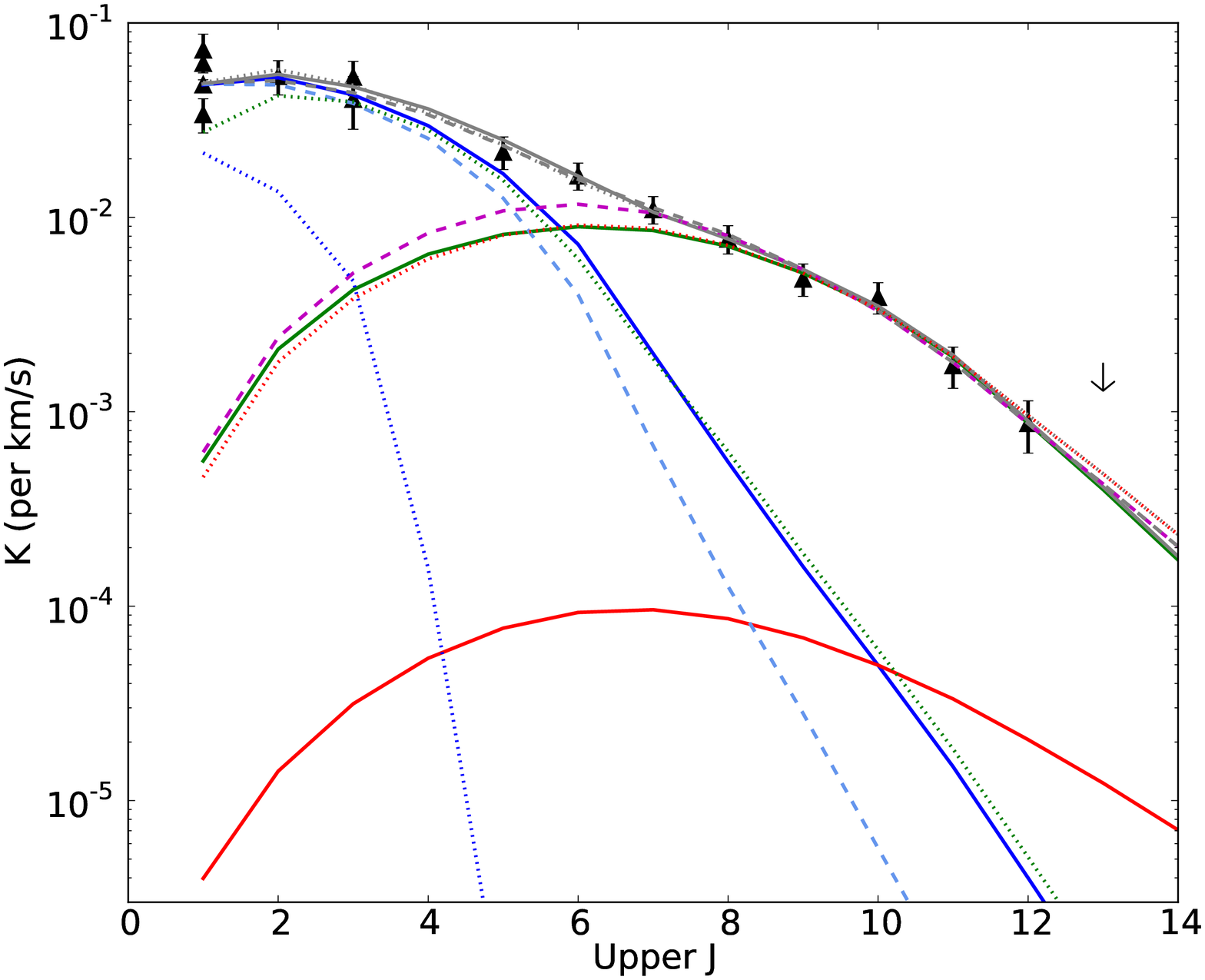}
\caption[Sample 3-Component Model Results for Arp~220: Temperature vs. Mass and SLED]{Sample 3-Component Model Results for Arp~220.  
Left:
Temperature vs. Mass.  
The red and blue shaded regions around the diamonds indicate the 
1$\sigma$ temperature and mass ranges from the two-component likelihood modeling.  The black asterisks are the temperatures and 
masses derived from molecular hydrogen lines (see Section \ref{sec:H2} and Table \ref{tbl:h2}).  
We attempted the three-component modeling described in Section \ref{sec:disc:twocomp2} to see if we could separate the warm CO gas (pink box)
into medium and higher temperature components similar to \htwo.  The dark blue, green, and dark red X's and 1$\sigma$ error bars 
denote the three-component modeling results for Components I, II, and III, respectively.  The distribution of the three 
components in temperature and mass now seem qualitatively more similar to that of \htwo.  However, Figures \ref{fig:3comp} 
and \ref{fig:3compparam} reveal that Component III is not constrained and negligible to the fit.  Component II is fulfilling the same 
role as the warm component, but was limited to a different temperature.  Right: SLED, the solid lines indicate the best-fit solution of the 
three-component model (Section \ref{sec:disc:twocomp2}) for Components I (blue), II (green), and III (red).  
The dashed lines are the best-fit two-component model for the cool (light blue) and warm (fuchsia) components.
The dotted lines are the best-fit three-component model when we remove the  $0.5/2/2.3 < {\rm Log}(T_{\rm I}/T_{\rm II}/T_{\rm III}) < 2.3/3/3.5$ constraint.  However, the best-fit as well as the median solutions still fall in those temperature ranges.  The total SLED (gray) is statistically indistinguishable in all three cases.  We emphasize that the best-fit model is only {\it one} of many models that can fit the data well (and 
is not necessarily representative of the likelihood distribution) and readers should not over-interpret the differences in the breakdown of individual components; this is why we examine the {\it full} parameter space.  
\label{fig:3comp}}
\end{figure}

\begin{figure}
\includegraphics[width=\columnwidth]{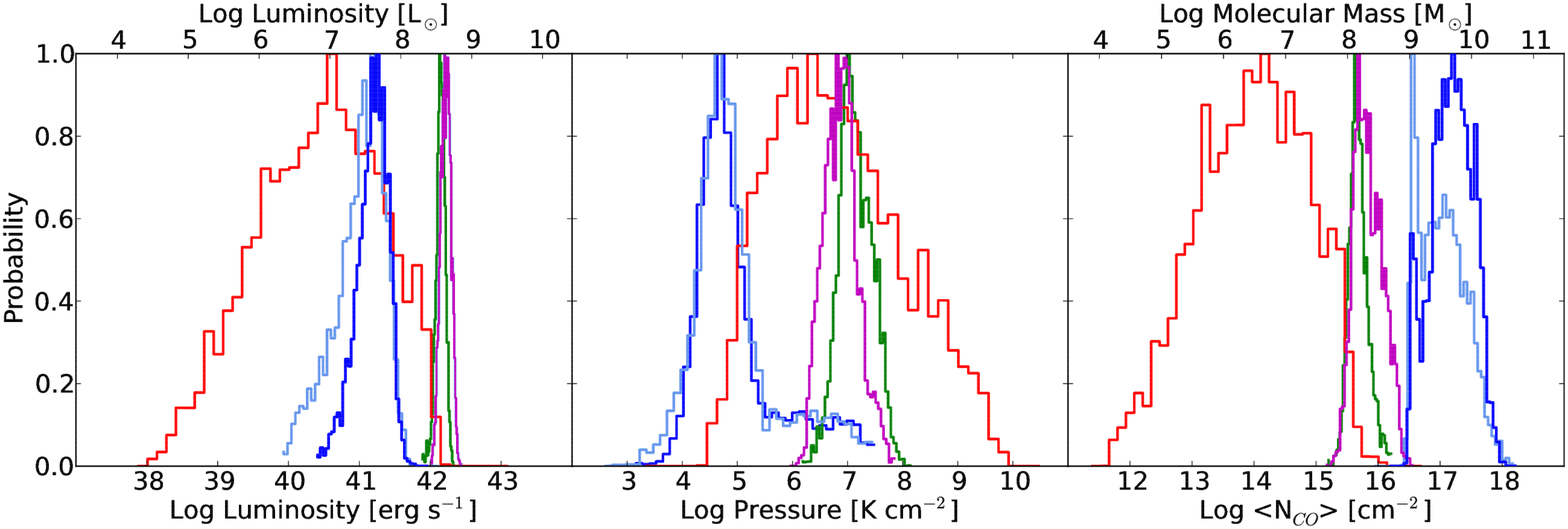}
\caption[Sample 3-Component Model Results for Arp~220: Derived Parameters]{
Sample 3-Component Model Results for Arp~220: Derived Parameters. 
From left to right: CO luminosity, pressure (product of temperature and density), and beam-averaged column density 
(product of column density and filling factor, proportional to mass, top axis).  
The dark blue, green, and red lines are the marginalized likelihoods of Components I, II, and III, respectively.
The light blue and fuchsia lines are for the cool and warm components of the two-component modeling.
Qualitatively, Component I and the cool component are the same, as are Component II and the warm component.
Component III is generally unconstrained so long as its mass is low enough that it does not modify the fit.\label{fig:3compparam}}
\end{figure}

\begin{figure}
\includegraphics[width=\columnwidth]{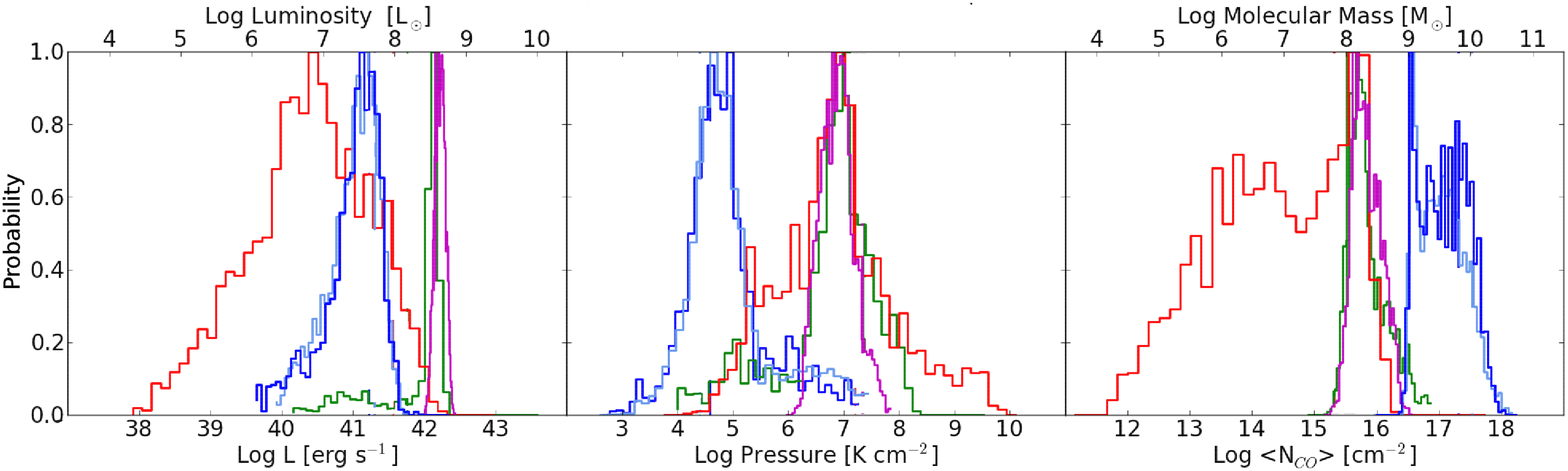}
\caption[Sample 3-Component Model Results for Arp~220: Derived Parameters, without Temperature Constraints]{
Sample 3-Component Model Results for Arp~220: Derived Parameters, without Temperature Constraints.
This figure is similar to Figure \ref{fig:3compparam} (see caption for explanation of colors), but we have 
removed the $0.5/2/2.3 < {\rm Log}(T_{\rm I}/T_{\rm II}/T_{\rm III}) < 2.3/3/3.5$ constraint.
The shapes of the likelihoods do shift slightly, one reason is because there is some likelihood 
in the higher temperatures for Component I and lower temperatures for Component II.
The conclusion is still much the same: the SLED is well 
fit by two components, with the third unconstrained.  
\label{fig:3compparamb}}
\end{figure}

Though not an extended galaxy, this merger was examined in a similar fashion by \citet{Rangwala2011}, 
who used additional interferometric information to constrain the source size of both the warm and cool components.
Additionally, the line fluxes of the different components were scaled by a different linewidth, and iterative, not simultaneous,
modeling was used.  Likely as a result of these changes, 
though we find overlapping distributions for the cold component mass and pressure, we find a higher warm component pressure 
(1$\sigma$ range Log(P) = 6.6-7.2 instead of 6.2-6.4), and lower mass (1$\sigma$ range Log(M) = 8.2-8.6 instead of 8.6-8.8).
In Section \ref{sec:disc:twocomp2}, we discussed three-component modeling, and present some sample results 
here for Arp~220 in Figures \ref{fig:3comp}, \ref{fig:3compparam}, and \ref{fig:3compparamb}, explained in the captions.

\subsection{Cen A} 

Centaurus A, the radio source in NGC~5128, is the nearest giant elliptical.  The aftermath of a merger, Cen A is notable for its bright, 
compact circumnuclear disk (CND) and extended thin disk (ETD).  The Herschel FTS beam is centered on the CND, and thus we are not probing the 
physical conditions in the ETD.  \citet{Israel2014} also examined the CO SLED, \ci, and \cii\ emission from the CND and multiple offset pointings 
of Cen A.  For their central pointing, they normalize the measured emission to that of a 22\as\ beam and find weak or negligible contribution from the 
ETD; our observations, with a larger beam, may have more contamination but are still dominated by the CND.  By a comparison to CO SLEDs of other well-known 
galaxies, they note that the falling CO SLED at high-J indicates Cen A has the ``coolest" CO ladder.  Our LVG modeling can quantify this statement: 
the warm component of CO in Cen A has one of the lowest pressures in our sample.  

\citep{Israel2014} models the $^{12}$CO SLEDs simultaneously with $^{13}$CO using two gas components, but does not present marginalized likelihood distributions, 
instead opting to present three solutions which match the observed SLED well.  Our results are not particularly comparable because we allow 
kinetic temperatures above 150 K.  Limiting the temperature will necessarily require a larger portion (of the mass) of the gas to appear much warmer.
They also use PDR/XDR models and note the potential importance of mechanical heating.  Their estimate of the mass of the CND, $8.4 \times 10^7$ \ms, 
is about a factor of 2 smaller than ours (which is their stated uncertainty); our larger beam area, and possible contamination from the ETD within it,
may be responsible for some of the difference.  We find the highest value of $\alpha_{\rm CO}$ in our sample for Cen A, but due to the 
difference in our mass estimate from \citet{Israel2014}, we find this in accordance with that of the Milky Way, not twice the value.
\citet{Parkin2014} also investigated atomic fine structure lines using PACS and SPIRE and compared to PDR models; they found that the matching 
PDR properties implies that Cen~A is similar to a normal disk galaxy, despite its unique morphology.

\subsection{NGC~6240}

Our results wholeheartedly agree with the conclusion of \citet{Meijerink2013}, that the CO line luminosity-to-continuum ratio is 
exceptionally high in this galaxy.  They argue, through shock modeling of CO and \htwo, that a high line-to-continuum ratio is a key diagnostic for shocks.
Their LVG models were in preparation at the time of this writing.

\subsection{Mrk~231}

The CO \jnine\ line of Mrk~231 seems abnormally low compared to the rest of the SLED.  For this spectrum, we are using a version reprocessed with HIPE 9 and off-axis background subtraction, but we find the same low flux with SPG v6.1.0 and SPG v11.1.0.
The v6.1.0 spectrum includes more frequency overlap between the SLW and SSW regions; fitting the CO \jnine\ line from the SLW band yields a higher flux, 652 $\pm 68$ Jy \kms.  In future versions, however, the \jnine\ line is not available in the more limited SLW frequency range.  We do not use the v6.1.0 spectrum because the background subtraction does not properly match the SLW to SSW, which should happen for a point source like Mrk~231.

\section{Tables of Photometry from the Literature}\label{appendix:lit}
\begin{deluxetable}{rrrrrrrrrr}
\tabletypesize{\scriptsize}
\tablecaption{Photometry Flux Densities: IRAS \label{tbl:photflux_IRAS}}
\tablehead{
\colhead{Galaxy} & \multicolumn{2}{c}{12 $\mu$m} &\multicolumn{2}{c}{25 $\mu$m} &\multicolumn{2}{c}{60 $\mu$m} &\multicolumn{2}{c}{100 $\mu$m} &  \colhead{Ref} \\
\colhead{} & \colhead{$F_\nu$} & \colhead{$\sigma_{tot}$}  & \colhead{$F_\nu$} & \colhead{$\sigma_{tot}$}  & \colhead{$F_\nu$} & \colhead{$\sigma_{tot}$}  & \colhead{$F_\nu$} & \colhead{$\sigma_{tot}$}  & \colhead{ } \\
\colhead{} & \colhead{[Jy]} & \colhead{[Jy]} & \colhead{[Jy]} & \colhead{[Jy]} & \colhead{[Jy]} & \colhead{[Jy]} & \colhead{[Jy]} & \colhead{[Jy]} & \colhead{ } 
}
\startdata
Mrk 231 & 1.83\phantom{0} & 0.184\phantom{0} & 8.84 & 0.884 & 30.8 & 3.08 & 29.7 & 2.98 & 1\\
IRAS F17207-0014 & 0.200 & 0.0320 & 1.61 & 0.164 & 32.1 & 3.21 & 36.1 & 3.65 & 1\\
IRAS 09022-3615 & 0.200 & 0.0377 & 1.19 & 0.121 & 11.6 & 1.17 & 11.1 & 1.16 & 1\\
Arp 220 & 0.610 & 0.0645 & 8.00 & 0.801 & 104\phantom{.0} & 10.4\phantom{0} & 115\phantom{.0} & 11.5\phantom{0} & 1\\
Mrk 273 & 0.240 & 0.0294 & 2.36 & 0.237 & 22.5 & 2.25 & 22.5 & 2.25 & 1\\
UGC 05101 & 0.250 & 0.0368 & 1.02 & 0.106 & 11.7 & 1.17 & 19.9 & 2.00 & 1\\
NGC 6240 & 0.590 & 0.0641 & 3.55 & 0.356 & 22.9 & 2.29 & 26.5 & 2.65 & 1\\
Arp 299-A & 3.97\phantom{0} & 0.398\phantom{0} & 24.5\phantom{0} & 2.45\phantom{0} & 113\phantom{.0} & 11.3\phantom{0} & 111\phantom{.0} & 11.1\phantom{0} & 1\\
NGC 1068 & 39.8\phantom{00} & 3.98\phantom{00} & 87.6\phantom{0} & 8.76\phantom{0} & 196\phantom{.0} & 19.6\phantom{0} & 257\phantom{.0} & 25.7\phantom{0} & 1\\
NGC 1365-NE & 5.12\phantom{0} & 0.513\phantom{0} & 14.3\phantom{0} & 1.43\phantom{0} & 94.3 & 9.43 & 166\phantom{.0} & 16.6\phantom{0} & 1\\
NGC 4038 (Overlap) & 1.94\phantom{0} & 0.199\phantom{0} & 6.54 & 0.655 & 45.2 & 4.52 & 87.1 & 8.71 & 1\\
M82 & 79.4\phantom{00} & 7.94\phantom{00} & 333\phantom{.00} & 33.3\phantom{00} & 1480\phantom{.0} & 148\phantom{.00} & 1370\phantom{.0} & 137\phantom{.00} & 1\\
NGC 1222 & 0.500 & 0.0550 & 2.28 & 0.231 & 13.1 & 1.31 & 15.4 & 1.54 & 1\\
M83 & 21.5\phantom{00} & 2.15\phantom{00} & 43.6\phantom{0} & 4.36\phantom{0} & 266\phantom{.0} & 26.6\phantom{0} & 524\phantom{.0} & 52.4\phantom{0} & 1\\
NGC 253 & 41.0\phantom{00} & 4.10\phantom{00} & 155\phantom{.00} & 15.5\phantom{00} & 968\phantom{.0} & 96.8\phantom{0} & 1290\phantom{.0} & 129\phantom{.00} & 1\\
NGC 1266 & 0.250 & 0.0391 & 1.20 & 0.124 & 13.1 & 1.31 & 16.9 & 1.70 & 1\\
Cen A & 22.2\phantom{00} & 2.22\phantom{00} & 28.3\phantom{0} & 2.83\phantom{0} & 213\phantom{.0} & 21.3\phantom{0} & 412\phantom{.0} & 41.2\phantom{0} & 1
\enddata
\tablecomments{{\bf References.} (1) \citet{Sanders2003}.}
\end{deluxetable}
\begin{deluxetable}{rrrrrrrr}
\tabletypesize{\scriptsize}
\tablecaption{Photometry Flux Densities: MIPS \label{tbl:photflux_MIPS}}
\tablehead{
\colhead{Galaxy} & \multicolumn{2}{c}{24 $\mu$m} &\multicolumn{2}{c}{70 $\mu$m} &\multicolumn{2}{c}{160 $\mu$m} &  \colhead{Ref} \\
\colhead{} & \colhead{$F_\nu$} & \colhead{$\sigma_{tot}$}  & \colhead{$F_\nu$} & \colhead{$\sigma_{tot}$}  & \colhead{$F_\nu$} & \colhead{$\sigma_{tot}$}  & \colhead{ } \\
\colhead{} & \colhead{[Jy]} & \colhead{[Jy]} & \colhead{[Jy]} & \colhead{[Jy]} & \colhead{[Jy]} & \colhead{[Jy]} & \colhead{ } 
}
\startdata
Mrk 231 & 4.34\phantom{0} & 0.485\phantom{0} & \ldots & \ldots & \ldots & \ldots & 1\\
Arp 220 & 4.01\phantom{0} & 0.449\phantom{0} & 80.8 & 14.6\phantom{0} & \ldots & \ldots & 1\\
Mrk 273 & 1.86\phantom{0} & 0.208\phantom{0} & 20.2 & 3.64 & 11.7 & 3.69 & 1\\
UGC 05101 & 0.808 & 0.0902 & 13.2 & 2.38 & 13.4 & 4.24 & 1\\
Arp 299-A & 8.66\phantom{0} & 0.0500 & \ldots & \ldots & \ldots & \ldots & 2\\
NGC 1068 & 80.0\phantom{00} & 8.00\phantom{00} & 180.\phantom{0} & 18.0\phantom{0} & \ldots & \ldots & 3\\
NGC 4038 (Overlap) & 6.13\phantom{0} & 0.0450 & \ldots & \ldots & \ldots & \ldots & 2\\
M83 & 42.0\phantom{00} & 4.20\phantom{00} & \ldots & \ldots & \ldots & \ldots & 3\\
NGC 253 & 140.\phantom{000} & 14.0\phantom{000} & \ldots & \ldots & \ldots & \ldots & 3\\
NGC 1266 & 0.880 & 0.0533 & 12.7 & 1.38 & 10.3 & 1.76 & 4,5
\enddata
\tablecomments{{\bf References.} (1) \citet{U2012}; (2) \citet{Lanz2013}; (3) NED; (4) \citet{Dale2007}; (5) \citet{Temi2009}.}
\end{deluxetable}
\begin{deluxetable}{rrrrrrrr}
\tabletypesize{\scriptsize}
\tablecaption{Photometry Flux Densities: PACS \label{tbl:photflux_PACS}}
\tablehead{
\colhead{Galaxy} & \multicolumn{2}{c}{75 $\mu$m} &\multicolumn{2}{c}{110 $\mu$m} &\multicolumn{2}{c}{170 $\mu$m} &  \colhead{Ref} \\
\colhead{} & \colhead{$F_\nu$} & \colhead{$\sigma_{tot}$}  & \colhead{$F_\nu$} & \colhead{$\sigma_{tot}$}  & \colhead{$F_\nu$} & \colhead{$\sigma_{tot}$}  & \colhead{ } \\
\colhead{} & \colhead{[Jy]} & \colhead{[Jy]} & \colhead{[Jy]} & \colhead{[Jy]} & \colhead{[Jy]} & \colhead{[Jy]} & \colhead{ } 
}
\startdata
Arp 299-A & 139\phantom{.0} & 13.9\phantom{0} & 127 & 12.7 & 74.2 & 7.42 & 1\\
NGC 4038 (Overlap) & 81.0 & 8.11 & 116 & 11.6 & 99.8 & 9.98 & 1\\
M82 & 1990\phantom{.0} & 198\phantom{.00} & \ldots & \ldots & 1290\phantom{.0} & 129\phantom{.00} & 1
\enddata
\tablecomments{{\bf References.} (1) \citet{Lanz2013}.}
\end{deluxetable}
\begin{deluxetable}{rrrrrrrrrrrr}
\tabletypesize{\scriptsize}
\tablecaption{Photometry Flux Densities: PLANCK \label{tbl:photflux_Planck}}
\tablehead{
\colhead{Galaxy} & \multicolumn{2}{c}{350 $\mu$m} &\multicolumn{2}{c}{550 $\mu$m} &\multicolumn{2}{c}{850 $\mu$m} &\multicolumn{2}{c}{1380 $\mu$m} &\multicolumn{2}{c}{2100 $\mu$m} &  \colhead{Ref} \\
\colhead{} & \colhead{$F_\nu$} & \colhead{$\sigma_{tot}$}  & \colhead{$F_\nu$} & \colhead{$\sigma_{tot}$}  & \colhead{$F_\nu$} & \colhead{$\sigma_{tot}$}  & \colhead{$F_\nu$} & \colhead{$\sigma_{tot}$}  & \colhead{$F_\nu$} & \colhead{$\sigma_{tot}$}  & \colhead{ } \\
\colhead{} & \colhead{[Jy]} & \colhead{[Jy]} & \colhead{[Jy]} & \colhead{[Jy]} & \colhead{[Jy]} & \colhead{[Jy]} & \colhead{[Jy]} & \colhead{[Jy]} & \colhead{[Jy]} & \colhead{[Jy]} & \colhead{ } 
}
\startdata
Mrk 231 & 1.87 & 0.154 & \ldots & \ldots & \ldots & \ldots & \ldots & \ldots & \ldots & \ldots & 1\\
Arp 220 & 13.9\phantom{0} & 0.997 & 3.64 & 0.281 & 0.943 & 0.103\phantom{0} & \ldots & \ldots & \ldots & \ldots & 1\\
Mrk 273 & 1.56 & 0.165 & \ldots & \ldots & \ldots & \ldots & \ldots & \ldots & \ldots & \ldots & 1\\
UGC 05101 & 3.23 & 0.258 & \ldots & \ldots & \ldots & \ldots & \ldots & \ldots & \ldots & \ldots & 1\\
NGC 6240 & 3.47 & 0.286 & \ldots & \ldots & \ldots & \ldots & \ldots & \ldots & \ldots & \ldots & 1\\
Arp 299-A & 9.62 & 0.695 & 2.56 & 0.200 & 0.645 & 0.0782 & \ldots & \ldots & \ldots & \ldots & 1\\
NGC 1068 & 48.9\phantom{0} & 3.64\phantom{0} & 12.8\phantom{0} & 0.937 & 2.48\phantom{0} & 0.213\phantom{0} & 0.720 & 0.0826 & \ldots & \ldots & 1\\
NGC 1365-NE & 43.2\phantom{0} & 3.17\phantom{0} & 11.9\phantom{0} & 0.886 & 2.50\phantom{0} & 0.196\phantom{0} & 0.577 & 0.0594 & \ldots & \ldots & 1\\
NGC 4038 (Overlap) & 17.3\phantom{0} & 1.25\phantom{0} & 4.91 & 0.375 & 0.862 & 0.114\phantom{0} & \ldots & \ldots & \ldots & \ldots & 1\\
M82 & 157\phantom{.00} & 11.4\phantom{00} & 37.2\phantom{0} & 2.69\phantom{0} & 8.18\phantom{0} & 0.598\phantom{0} & 2.69\phantom{0} & 0.196\phantom{0} & 0.958 & 0.0855 & 1\\
M83 & 118\phantom{.00} & 9.50\phantom{0} & 33.8\phantom{0} & 2.71\phantom{0} & 6.48\phantom{0} & 0.531\phantom{0} & 1.81\phantom{0} & 0.144\phantom{0} & \ldots & \ldots & 1\\
NGC 253 & 317\phantom{.00} & 25.1\phantom{00} & 91.7\phantom{0} & 7.54\phantom{0} & 17.4\phantom{00} & 1.54\phantom{00} & 4.57\phantom{0} & 0.388\phantom{0} & 1.28\phantom{0} & 0.110\phantom{0} & 1\\
Cen A & 115\phantom{.00} & 8.60\phantom{0} & 42.5\phantom{0} & 3.15\phantom{0} & 17.7\phantom{00} & 1.32\phantom{00} & \ldots & \ldots & \ldots & \ldots & 1
\enddata
\tablecomments{{\bf References.} (1) \citet{Ade2011}.}
\end{deluxetable}
\begin{deluxetable}{rrrrrr}
\tabletypesize{\scriptsize}
\tablecaption{Photometry Flux Densities: SCUBA \label{tbl:photflux_SCUBA}}
\tablehead{
\colhead{Galaxy} & \multicolumn{2}{c}{450 $\mu$m} &\multicolumn{2}{c}{850 $\mu$m} &  \colhead{Ref} \\
\colhead{} & \colhead{$F_\nu$} & \colhead{$\sigma_{tot}$}  & \colhead{$F_\nu$} & \colhead{$\sigma_{tot}$}  & \colhead{ } \\
\colhead{} & \colhead{[Jy]} & \colhead{[Jy]} & \colhead{[Jy]} & \colhead{[Jy]} & \colhead{ } 
}
\startdata
Mrk 231 & \ldots & \ldots & 0.0780 & 0.00780 & 1\\
IRAS F17207-0014 & 1.07 & 0.325 & 0.155\phantom{0} & 0.0471\phantom{0} & 2\\
Arp 220 & 6.29 & 0.629 & 0.832\phantom{0} & 0.0860\phantom{0} & 3,1\\
NGC 6240 & 1.00 & 0.304 & 0.150\phantom{0} & 0.0456\phantom{0} & 2\\
NGC 1222 & \ldots & \ldots & 0.0840 & 0.0160\phantom{0} & 3
\enddata
\tablecomments{{\bf References.} (1) NED; (2) \citet{Klaas2001}; (3) \citet{Dunne2000}.}
\end{deluxetable}
\begin{deluxetable*}{crrl|crrl|crrl}
\tabletypesize{\scriptsize}
\tablecaption{Photometry Flux Densities: ISO \label{tbl:photflux_ISO}}
\tablehead{
\colhead{$\lambda$} & \colhead{$F_\nu$} & \colhead{$\sigma_{tot}$}  & \colhead{Ref} & \colhead{$\lambda$} & \colhead{$F_\nu$} & \colhead{$\sigma_{tot}$}  & \colhead{Ref} & \colhead{$\lambda$} & \colhead{$F_\nu$} & \colhead{$\sigma_{tot}$}  & \colhead{Ref}\\
\colhead{[$\mu$m]}  & \colhead{[Jy]}    & \colhead{[Jy]}            & \colhead{ }  & \colhead{[$\mu$m]}  & \colhead{[Jy]}    & \colhead{[Jy]}            & \colhead{ }  & \colhead{[$\mu$m]}  & \colhead{[Jy]}    & \colhead{[Jy]}            & \colhead{ } 
}
\startdata
\multicolumn{4}{c}{\bf{Mrk 231}} & 
52 & 121\phantom{.000} & 10.8\phantom{00} & 2 & 
120 & 25.9\phantom{00} & 7.88\phantom{0} & 1\\
10 & 1.43\phantom{0} & 0.478 & 1 & 
57 & 134\phantom{.000} & 11.8\phantom{00} & 2 & 
122 & 20.8\phantom{00} & 3.70\phantom{0} & 2\\
12 & 2.40\phantom{0} & 0.805 & 1 & 
60 & 113\phantom{.000} & 38.0\phantom{00} & 1 & 
145 & 17.9\phantom{00} & 3.10\phantom{0} & 2\\
15 & 2.90\phantom{0} & 0.973 & 1 & 
63 & 148\phantom{.000} & 15.0\phantom{00} & 2 & 
150 & 18.9\phantom{00} & 5.75\phantom{0} & 1\\
25 & 8.66\phantom{0} & 2.90\phantom{0} & 1 & 
88 & 151\phantom{.000} & 18.7\phantom{00} & 2 & 
158 & 16.8\phantom{00} & 2.80\phantom{0} & 2\\
52 & 32.4\phantom{00} & 3.20\phantom{0} & 2 & 
90 & 112\phantom{.000} & 37.4\phantom{00} & 1 & 
170 & 11.5\phantom{00} & 2.10\phantom{0} & 2\\
57 & 37.2\phantom{00} & 3.60\phantom{0} & 2 & 
120 & 109\phantom{.000} & 33.2\phantom{00} & 1 & 
170 & 16.7\phantom{00} & 1.67\phantom{0} & 3\\
60 & 31.7\phantom{00} & 10.6\phantom{00} & 1 & 
122 & 118\phantom{.000} & 9.50\phantom{0} & 2 & 
180 & 12.7\phantom{00} & 3.87\phantom{0} & 1\\
63 & 42.9\phantom{00} & 4.00\phantom{0} & 2 & 
145 & 100.\phantom{000} & 10.1\phantom{00} & 2 & 
200 & 9.00\phantom{0} & 2.74\phantom{0} & 1\\
88 & 34.1\phantom{00} & 4.40\phantom{0} & 2 & 
150 & 87.9\phantom{00} & 26.7\phantom{00} & 1 & 
\multicolumn{4}{c}{\bf{Arp 299-A}}\\
90 & 27.3\phantom{00} & 9.17\phantom{0} & 1 & 
158 & 84.5\phantom{00} & 8.50\phantom{0} & 2 & 
52 & 129\phantom{.000} & 12.3\phantom{00} & 2\\
120 & 24.3\phantom{00} & 7.40\phantom{0} & 1 & 
170 & 77.1\phantom{00} & 6.70\phantom{0} & 2 & 
57 & 142\phantom{.000} & 13.1\phantom{00} & 2\\
122 & 19.5\phantom{00} & 1.70\phantom{0} & 2 & 
170 & 77.1\phantom{00} & 7.71\phantom{0} & 3 & 
63 & 151\phantom{.000} & 14.6\phantom{00} & 2\\
145 & 16.0\phantom{00} & 1.70\phantom{0} & 2 & 
180 & 64.0\phantom{00} & 19.4\phantom{00} & 1 & 
88 & 141\phantom{.000} & 19.9\phantom{00} & 2\\
150 & 14.7\phantom{00} & 4.48\phantom{0} & 1 & 
200 & 54.8\phantom{00} & 16.7\phantom{00} & 1 & 
122 & 86.4\phantom{00} & 7.90\phantom{0} & 2\\
158 & 16.1\phantom{00} & 2.20\phantom{0} & 2 & 
\multicolumn{4}{c}{\bf{Mrk 273}} & 
145 & 65.7\phantom{00} & 5.30\phantom{0} & 2\\
170 & 15.3\phantom{00} & 1.60\phantom{0} & 2 &  10 & 0.100 & 0.034 & 1 & 
158 & 58.6\phantom{00} & 7.50\phantom{0} & 2\\
170 & 15.3\phantom{00} & 1.53\phantom{0} & 3 &  12 & 0.250 & 0.084 & 1 & 
170 & 60.7\phantom{00} & 6.30\phantom{0} & 2\\
180 & 9.75\phantom{0} & 2.97\phantom{0} & 1 &  15 & 0.500 & 0.168 & 1 & 
\multicolumn{4}{c}{\bf{NGC 1365-NE}}\\
200 & 6.88\phantom{0} & 2.09\phantom{0} & 1 & 
25 & 2.07\phantom{0} & 0.694 & 1 &  15 & 4.44\phantom{0} & 0.444 & 3\\
\multicolumn{4}{c}{\bf{IRAS F17207-0014}} & 
57 & 22.3\phantom{00} & 2.00\phantom{0} & 2 & 
120 & 217\phantom{.000} & 21.7\phantom{00} & 3\\
10 & 0.080 & 0.027 & 1 &  60 & 27.5\phantom{00} & 9.21\phantom{0} & 1 & 
150 & 194\phantom{.000} & 19.4\phantom{00} & 3\\
12 & 0.200 & 0.067 & 1 &  63 & 24.9\phantom{00} & 2.50\phantom{0} & 2 & 
170 & 167\phantom{.000} & 16.7\phantom{00} & 3\\
15 & 0.250 & 0.084 & 1 &  88 & 22.6\phantom{00} & 2.30\phantom{0} & 2 & 
180 & 103\phantom{.000} & 10.3\phantom{00} & 3\\
25 & 1.32\phantom{0} & 0.443 & 1 & 
90 & 23.8\phantom{00} & 7.98\phantom{0} & 1 & 
200 & 85.2\phantom{00} & 8.52\phantom{0} & 3\\
52 & 40.7\phantom{00} & 4.90\phantom{0} & 2 & 
120 & 20.0\phantom{00} & 6.08\phantom{0} & 1 & 
\multicolumn{4}{c}{\bf{NGC 1222}}\\
57 & 31.6\phantom{00} & 2.70\phantom{0} & 2 & 
122 & 15.4\phantom{00} & 1.30\phantom{0} & 2 & 
57 & 20.2\phantom{00} & 1.80\phantom{0} & 2\\
60 & 32.2\phantom{00} & 10.8\phantom{00} & 1 & 
145 & 12.5\phantom{00} & 1.30\phantom{0} & 2 & 
63 & 20.4\phantom{00} & 2.00\phantom{0} & 2\\
63 & 41.9\phantom{00} & 3.90\phantom{0} & 2 & 
150 & 13.1\phantom{00} & 3.98\phantom{0} & 1 & 
122 & 12.9\phantom{00} & 1.10\phantom{0} & 2\\
88 & 48.5\phantom{00} & 4.90\phantom{0} & 2 & 
158 & 8.40\phantom{0} & 1.00\phantom{0} & 2 & 
145 & 11.4\phantom{00} & 1.10\phantom{0} & 2\\
90 & 31.9\phantom{00} & 10.7\phantom{00} & 1 & 
170 & 8.30\phantom{0} & 1.00\phantom{0} & 2 & 
158 & 10.8\phantom{00} & 2.20\phantom{0} & 2\\
120 & 30.0\phantom{00} & 9.12\phantom{0} & 1 & 
170 & 8.30\phantom{0} & 0.830 & 3 &  170 & 8.00\phantom{0} & 0.900 & 2\\
122 & 31.5\phantom{00} & 2.60\phantom{0} & 2 & 
180 & 8.69\phantom{0} & 2.64\phantom{0} & 1 & 
170 & 8.00\phantom{0} & 0.800 & 3\\
145 & 27.0\phantom{00} & 2.50\phantom{0} & 2 & 
200 & 7.40\phantom{0} & 2.25\phantom{0} & 1 &  \multicolumn{4}{c}{\bf{M83}}\\
150 & 23.0\phantom{00} & 7.00\phantom{0} & 1 & 
\multicolumn{4}{c}{\bf{NGC 6240}} & 
15 & 20.1\phantom{00} & 2.01\phantom{0} & 3\\
158 & 24.3\phantom{00} & 2.90\phantom{0} & 2 &  10 & 0.259 & 0.087 & 1 & 
\multicolumn{4}{c}{\bf{NGC 1266}}\\
170 & 26.4\phantom{00} & 3.20\phantom{0} & 2 &  12 & 0.750 & 0.252 & 1 & 
57 & 15.5\phantom{00} & 1.40\phantom{0} & 2\\
170 & 26.4\phantom{00} & 2.64\phantom{0} & 3 & 
15 & 1.00\phantom{0} & 0.335 & 1 & 
63 & 18.5\phantom{00} & 1.80\phantom{0} & 2\\
180 & 17.5\phantom{00} & 5.32\phantom{0} & 1 & 
25 & 3.31\phantom{0} & 1.11\phantom{0} & 1 & 
88 & 21.6\phantom{00} & 2.70\phantom{0} & 2\\
200 & 12.5\phantom{00} & 3.80\phantom{0} & 1 & 
52 & 16.5\phantom{00} & 2.50\phantom{0} & 2 & 
122 & 16.7\phantom{00} & 1.30\phantom{0} & 2\\
\multicolumn{4}{c}{\bf{Arp 220}} & 
57 & 27.5\phantom{00} & 4.00\phantom{0} & 2 & 
145 & 13.3\phantom{00} & 1.30\phantom{0} & 2\\
10 & 0.147 & 0.049 & 1 &  60 & 23.6\phantom{00} & 7.92\phantom{0} & 1 & 
158 & 14.8\phantom{00} & 4.80\phantom{0} & 2\\
12 & 0.600 & 0.201 & 1 &  63 & 23.5\phantom{00} & 3.40\phantom{0} & 2 & 
170 & 10.1\phantom{00} & 1.00\phantom{0} & 2\\
15 & 1.14\phantom{0} & 0.382 & 1 & 
88 & 25.8\phantom{00} & 4.00\phantom{0} & 2 & 
170 & 10.1\phantom{00} & 1.01\phantom{0} & 3\\
25 & 8.28\phantom{0} & 2.78\phantom{0} & 1 & 
90 & 26.7\phantom{00} & 8.96\phantom{0} & 1 &   & & & 
\enddata
\tablecomments{{\bf References.} ( 1) \citet{Klaas2001}; ( 2) \citet{Brauher2008}; ( 3) NED}
\end{deluxetable*}
\begin{deluxetable*}{crrl|crrl}
\tabletypesize{\scriptsize}
\tablecaption{Photometry Flux Densities: All Others \label{tbl:photflux_other}}
\tablehead{
\colhead{$\lambda$} & \colhead{$F_\nu$} & \colhead{$\sigma_{tot}$}  & \colhead{Ref} & \colhead{$\lambda$} & \colhead{$F_\nu$} & \colhead{$\sigma_{tot}$}  & \colhead{Ref}\\
\colhead{[$\mu$m]}  & \colhead{[Jy]}    & \colhead{[Jy]}            & \colhead{ }  & \colhead{[$\mu$m]}  & \colhead{[Jy]}    & \colhead{[Jy]}            & \colhead{ } 
}
\startdata
\multicolumn{4}{c}{\bf{Mrk 231}} &  880 & 0.056 & 0.006 & 1\\
350 & 1.73\phantom{0} & 0.173 & 1 &  \multicolumn{4}{c}{\bf{NGC 6240}}\\
880 & 0.080 & 0.008 & 1 &  350 & 2.48\phantom{0} & 0.248 & 1\\
\multicolumn{4}{c}{\bf{Arp 220}} &  \multicolumn{4}{c}{\bf{M82}}\\
350 & 9.74\phantom{0} & 0.974 & 1 & 
250 & 363\phantom{.000} & 25.4\phantom{00} & 2\\
880 & 0.490 & 0.049 & 1 &  350 & 122\phantom{.000} & 8.50\phantom{0} & 2\\
\multicolumn{4}{c}{\bf{Mrk 273}} & 
500 & 49.6\phantom{00} & 4.96\phantom{0} & 1\\
350 & 1.77\phantom{0} & 0.177 & 1 &   & & & 
\enddata
\tablecomments{{\bf References.} ( 1) NED; ( 2) \citet{Lanz2013}}
\end{deluxetable*}

\end{document}